\documentclass[aps,rmp,showpacs,amsmath,amssymb,amsfonts,lengthcheck,twocolumn,longbibliography,floatfix,superscriptaddress]{revtex4-2}

\usepackage[utf8]{inputenc}
\usepackage[top=0.75in,bottom=0.75in,left=0.75in,right=0.75in]{geometry}
\usepackage{graphicx}
\usepackage{subfigure}
\usepackage{amsthm}
\usepackage{soul}
\usepackage{verbatim}
\usepackage{dcolumn}
\usepackage{bm}
\usepackage{epsf}
\usepackage{color}
\usepackage[colorlinks=true,citecolor=blue,linkcolor=blue,urlcolor=blue]{hyperref}%
\usepackage{xcolor}
\usepackage{dsfont}
\usepackage{tabularray}
\usepackage{amsmath}

\newcommand{\ed}{e}
\newcommand{\qmark}[1]{``#1''}
\begin{document}

\newcommand{\Qenv}{\dot{Q}^{\mathrm{env}}}
\title{Physical Approaches to Metabolic Scaling in Living Systems}

\author{Efe Ilker}
\affiliation{Aix-Marseille Université,\unpenalty~INSERM,\unpenalty~DyNaMo,\unpenalty~Turing\unpenalty~Centre\unpenalty~for Living Systems,\unpenalty~Marseille 13009,\unpenalty~France}

\affiliation{Max Planck Institute for the Physics of Complex Systems,\unpenalty~01187 Dresden,\unpenalty~Germany}
\author{Michael Hinczewski}
\affiliation{Department of Physics,\unpenalty~Case Western Reserve University,\unpenalty~Cleveland OH 44106}

\author{Xingbo Yang}
\affiliation{Cluster of Excellence Physics of Life (PoL),\unpenalty~Dresden,\unpenalty~01062,\unpenalty~Germany}

\author{Frank Jülicher}
\affiliation{Max Planck Institute for the Physics of Complex Systems,\unpenalty~01187 Dresden,\unpenalty~Germany}
\affiliation{Center for Systems Biology Dresden,\unpenalty~01307 Dresden,\unpenalty~Germany}

\begin{abstract}
Living systems continuously transform matter and energy through the chemical processes that constitute their metabolism. The overall metabolic rate of an  organism correlates positively with its body mass, however both the exact scaling behavior and possible explanations for this behavior have been under intense debate for two centuries. This review synthesizes empirical findings and theoretical frameworks on the energetics of living systems from an interdisciplinary perspective, with a focus on physical concepts. A general thermodynamic framework to study metabolism is laid out, together with a coarse-grained description of metabolic biochemistry. The rich history of experimental work in this field is surveyed, revealing a variety of metabolic scaling patterns at different levels of biological organization, from individual cells to whole populations. Several biophysical models proposed to explain the sublinear scaling of metabolic rate with body mass are summarized. Though the traditional focus has been on adult organisms, the review also highlights recent advances that probe metabolism during development. Improvements in experimental techniques, extensive datasets, and a host of open questions, suggest the field will continue gaining momentum in the near term. The review concludes with an outlook for this future progress: an interdisciplinary approach to elucidate metabolic scaling across different developmental stages and organism sizes.

    
    
\end{abstract}

\maketitle

\tableofcontents



\section{Introduction}
Living organisms are chemically active systems that operate far from thermodynamic equilibrium. They
are open systems that take up energy and material resources from their environment and transduce them into other forms. Eventually, energy and matter leave the organism in the form of heat and waste products. This constant flux of energy and matter is essential to maintain life functions, even in the resting state of an organism. It also permits an organism to grow and to change its morphology during development or during ageing. Investigating how  energetic consumption varies in different organisms, and how the energy flux is maintained for life processes, are broad topics at the interface of biology, chemistry, and nonequilibrium physics.

The chemical processes in cells that underlie biosynthesis, energy conversion, and energy consumption constitute metabolism. The metabolic rate of an organism is often associated with energy flows, typically probed by the total rate of heat release or the consumption rate of chemical resources such as oxygen. Quantitative metabolic rate measurements started at the dawn of modern chemistry, with the seminal work of Lavoisier and Laplace in the 18th century \cite{lavoisier1783}. However, comparative studies across animal species surged in the 20th century, particularly with Max Kleiber’s pioneering studies \cite{kleiber1947body,kleiber1961fire}. Kleiber reported that resting metabolic rate scales sublinearly with body mass, approximately following a 3/4 power law across a wide range of species. This ``Kleiber's law'' is the most well known form of metabolic scaling, and suggests that the metabolic rate is not simply proportional to organism mass. Modern experimental surveys have now extended this work, compiling metabolic rates across all kingdoms of life and more than twenty orders of magnitude in mass~\cite{makarieva2008mean,delong2010shifts,hatton2019linking,hoehler2023metabolic}. Despite this wealth of data, both the physical basis and the statistical validity of Kleiber's law---versus other scaling forms within and between species---remain points of contention in animal physiology~\cite{hulbert2014sceptics}.

More generally, metabolic scaling raises interesting questions across multiple disciplines. It is a key thermodynamic property of organisms, inextricably tied to them being active, out-of-equilibrium systems governed by the laws of physics. At the same time, metabolic scaling was shaped by natural selection during the evolution of life on earth, and is only one example of the many allometric scaling behaviors that characterize the structure and dynamic organization of organisms~\cite{schmidt1984scaling}. The causal relationships underlying these various behaviors are complex and not fully clear. However we know that analogous scaling properties in inanimate matter are often fundamental to our understanding of those physical systems. In particular, metabolic scaling is reminiscent of the patterns exhibited by fluxes or the rate of entropy production in general active matter and nonequilibrium systems. 

While historically the main subject of scaling studies has been adult organisms, there are also interesting connections between metabolism and organism size during post-embryonic growth~\cite{murray1921normal,von1957quantitative,west2001general}. Recent evidence demonstrates links to even earlier stages of development, with metabolism playing a role in how internal structures such as organs and body plan are shaped and transformed from embryos onward~\cite{miyazawa2018revisiting,garfinkel2024historic,ghosh2023developmental,song2020chemical}. Theoretical biophysics potentially provides conceptual frameworks to understand morphogenesis through active physicochemical processes that can link these different aspects. Thus, there is a need to broaden the discussion on metabolic scaling to include all stages of organism development, and put it on the same footing as the more traditional comparisons between adult organisms at different sizes and between species.


Explaining metabolic scaling remains one of the great open questions of biology, with competing theories focused on different aspects of the structure, transport, energetics, and evolutionary history of organisms 
\cite{west1997general,mortola2013thermographic, banavar1999size, mcmahon1973size, kooijman2010dynamic, white2022metabolic}.
Together with the fact that the pertinent modeling and experimental literature is widely scattered over many disciplines, this makes the field challenging to review. However, advances in nonequilibrium thermodynamics, active matter theory, developmental biology, and experimental techniques provide new opportunities to elucidate these phenomena. An overview of the field from a physics perspective can thus serve as an entry point, highlighting connections between biological scaling, physical principles, and the conceptual ideas that may guide future research.

In this review, we discuss the thermodynamic basis and experimental results for metabolic scaling studies as well as the theoretical approaches that attempt to explain these observations. The purpose of the review is to present the material in a unified framework, treating living systems as physical systems and using physical observables.  The review is organized as follows. Section~\ref{section2} serves as an introduction to the problem of metabolism, including a thermodynamic framework to study spatially organized metabolic processes, a simplified description of the underlying biochemistry using macrochemical reactions,  and a discussion of experimental approaches to study metabolism. In Section~\ref{section3}, we review empirical observations on the scaling of metabolic rate with the mass of organisms. There are fascinating scaling relationships not just for whole animal metabolism, but across a variety of levels from cells to tissues to whole populations. We also discuss some of the implications of these empirical patterns in the context of post-embryonic growth. In Section \ref{section4}, we survey a range of biophysical models proposed to explain sublinear metabolic scaling in living systems. Section \ref{section5} discusses recent advances of the role of metabolism in developmental biology. Finally, in Section \ref{section6}, we summarize and provide an outlook on the future development of the field. We refer the reader to Table \ref{notations} for notation and abbreviations, and to Table \ref{concepts} for key concepts frequently used in the text.

\section{Fundamentals of metabolism}\label{section2}

This section covers the physical observables of metabolism, the coarse-grained biochemistry that governs their dynamics, as well as the experimental techniques that can be used to quantify metabolic rates. To ground this discussion, we start with a general framework for the thermodynamics of metabolic processes.

 \subsection{Thermodynamics of metabolic processes}\label{thermo}
 Thermodynamics of living systems is generally studied by a macro description of chemical transformations and energy balance \cite{blaxter1989energy,ghosh2023developmental,koojiman1995stoichiometry, harold1987vital,von1993thermodynamic}. Here, we present a systematic approach starting from a local thermodynamic description using the strategy developed for irreversible thermodynamics \cite{de2013non, prigogine1947etude}. This provides general relations for fluxes of energy and matter, and thermodynamic quantities in spatially heterogeneous systems. We then relate this approach to the macro description by coarse-graining the metabolic processes.

 \begin{table}[t]
 \def\arraystretch{1.25}
     \centering
     \begin{tblr}{c c p{1.5in} c}
        Notation & &Definition && Unit\\
        \hline
         & & Organism-level & & \\
        \hline
         $B$ & &Basal metabolic rate  & & W\\
         $b$ & &Mass-specific basal metabolic rate $= B/M$ & & W/g\\
        $M$ & & Total mass of organism (including water) & & g\\
        $\dot{Q}$ & & Heat release rate from organism into the environment & & W\\
        $\Phi_{\rm{O}_2}$ & & Oxygen consumption rate (OCR) && mol/s \\
        \hline
        & & Local (position-dependent)& & \\
        \hline
        $e,g,h,p$ & & Densities (per unit volume) of energy, Gibbs free energy, enthalpy, and pressure && kJ/L\\
        $s$ & & Density (per unit volume) of entropy && kJ/(K $\cdot$ L)\\
        $T$ & & Temperature  && K\\
        $n_i$ & & Density (per unit volume) of molecule $i$ && mol/L\\
        $s_i$ & & Entropy of molecule $i$ && kJ/ (K $\cdot$ mol)\\
        $h_i$ & & Enthalpy of molecule $i$ && kJ/mol\\
        $\Delta H_{\alpha}$ & & Enthalpy of reaction $\alpha$ && kJ/mol \\
        $\mu_i$ & & Chemical potential of molecule $i$ && kJ/mol\\
        $\Delta G_{\alpha}$ & & Gibbs free energy of reaction $\alpha$ && kJ/mol \\

        $r_{\alpha}$ & & Flux of reaction $\alpha$ per unit volume  && mol/(L $\cdot$ s) \\
        $r_i$ & & Net change of molecule $i$ per unit volume due to reactions  && mol/(L $\cdot$ s) \\

                        \hline
        & & Macrochemical & & \\
        \hline
        $\mathcal{R}_{\beta}$ & & Total flux of macrochemical reaction $\beta$  && mol/s \\

        $\Delta h_{\rm ox}$ & & Enthalpy change per oxygen consumed && kJ/mol \\
     \end{tblr}
     \caption{Notation and abbreviations used in the text.}\label{notations}
\end{table}
 
\subsubsection{Chemical processes and energy conservation}   The nonequilibrium thermodynamics of metabolic processes can be described by considering the spatiotemporal dynamics of energy and matter.  Energy and chemical components can either be transformed into different forms or transferred to their surroundings, hence their dynamics should obey conservation laws and balance equations. For a general description of metabolic processes, let us consider a system consisting of $K+1$ chemical components, i.e. $K$ chemical species in a solvent with $R$ chemical reactions 
\begin{equation}
    \sum_{i=0}^K \sigma_{i\alpha}^+ X_i \rightleftharpoons \sum_{i=0}^K \sigma_{i\alpha}^- X_i\label{eq:reactions}
\end{equation}
where $X_i$, with $i=0,\dots,K$ are the chemical species, $\alpha=1,...,R$ represent the reactions, and  $\sigma_{i\alpha}^{\mp}$ are the stoichiometric coefficients.  The solvent corresponds to $i=0$.
The spatiotemporal evolution of particle  concentrations $n_i$ defined as particle number per unit volume, obeys a local balance equation:
\begin{equation}
    \frac{\partial n_i}{\partial t}+ \nabla \cdot {\bf j}_i = r_i \ . \label{eqchm1}
\end{equation}
Here ${\bf j}_i$ are the spatial fluxes of chemical concentrations while the source term originates from chemical reactions, given by $r_i=\sum_{\alpha=1}^{R}\sigma _{i\alpha} r_{\alpha}$ where $\sigma_{i\alpha}=\sigma_{i\alpha}^- - \sigma_{i\alpha}^+$ are the net stoichiometric coefficients and $r_{\alpha}$ are  reaction fluxes per unit volume. Thus, the local concentrations may change either through reactions or transport of material.   

During metabolic processes, the energy may take different forms (thermal, chemical, mechanical, etc.), yet the total energy is conserved. Conservation of energy implies that the change in local energy density is balanced the amount of energy exchanged with the surroundings. Accordingly, the local energy density $\ed$, i.e. the energy per unit volume obeys:
\begin{equation}
    \frac{\partial \ed}{\partial t}+\nabla\cdot {\bf j}_{\ed}=0\label{eqe1}
\end{equation}
where ${\bf j}_{\ed}$ is the energy flux vector representing the transport of energy between volume elements.

 \begin{table}[t]
 \def\arraystretch{1.25}
     \centering
     \begin{tblr}{c c  p{1.5in} }
        Concept & &Summary \\
        \hline
        Allometric scaling & & Relationship of shape, anatomy, physiology to body size typically expressed as a power-law function of body mass\\
        Basal metabolism & & Metabolism at rest without locomotion, feeding, and other non-basal activities (Section \ref{section3})\\
        
        Macrochemical reactions & & Coarse-grained chemical reactions that represent the overall chemical transformation of metabolism (Section \ref{sec:macro})\\
        Metabolic rate & & Defined as heat release rate $\dot{Q}$, but see Section \ref{sec:iib} \\
         Metabolic scaling & & Scaling of metabolic rate with organism mass  \\
        Kleiber's law & & Empirical observation that for some species $B\sim M^{3/4}$\\
        Thornton's rule && Empirical observation that enthalpy change per oxygen consumed in full oxidation of an organic material is roughly $-450$ kJ/mol within 10$\%$ variation (Section \ref{thorntons_rule})  \\

        Warburg effect & &
        A metabolic phenomenon in which highly proliferative cells such as cancer cells prefer lactate fermentation for ATP production instead of respiration even when oxygen is abundant 
     \end{tblr}
     \caption{Summary of key concepts frequently used in the text.}\label{concepts}
 \end{table}

\subsubsection{Time evolution of thermodynamic quantities}
 We consider that local equilibration times are faster than the dynamics of hydrodynamic modes at large scales. Thus, we use the assumption of local equilibrium for small volume elements while the system can be out-of-equilibrium globally \cite{de2013non,bauermann_energy_2022,julicher2018hydrodynamic}. 
 This suggests that for a small volume element $V$ at local thermal equilibrium, we can define the Gibbs free energy $G=E-TS+pV$ as a thermodynamic potential.  The Gibbs free energy is a function $G(N_0,\dots,N_K,T,P)=\sum_{i=0}^K \mu_i N_i$ of particle numbers $N_i$, temperature $T$ and pressure $p$. Here, $E$ denotes internal energy, $S=-\partial G/\partial T$ entropy
 and chemical potentials are defined as
 $\mu_i=\partial G/\partial N_i$.
Finally, we define enthalpy as $H=E+pV$.
For  spatially extended and heterogeneous systems, we additionally define
densities of energy $e=E/V$, enthalpy $h=H/V$ and entropy $s=S/V$ as well as 
concentrations $n_i=N_i/V$.

Using this framework, we can write heat balance and entropy balance equations. The detailed derivations are given in Appendix \ref{appendixa}. The energy density can be written as  $e=h-p$.
In the following, we consider for simplicity systems that are maintained at constant pressure $p$. The rate of change of the energy density is then $\partial_t\ed =\partial_t h$ with 
\begin{equation}
 \partial_t h=\sum_i h_i\partial_t n_i+c_p\partial_t T \quad ,
 \label{eq:dth0}
\end{equation}
where we have defined the particle enthalpy $h_i= \mu_i-T\partial \mu_i/\partial T\vert_p$ and the volumetric heat capacity at constant pressure $c_p=T\partial s/\partial T\vert_p$ (see Appendix \ref{appendixa}).  
The energy flux can  be decomposed into heat flux and transport of particle enthalpy as 
\begin{equation}
    {\bf j}_{\ed}={\bf j}_{q}+\sum_{i=0}^K{\bf j}_{i}h_i \quad ,\label{eq:je}
\end{equation}
 which defines   the heat flux ${\bf j}_{q}$. As a result, using Eqs. \eqref{eqchm1} and \eqref{eqe1} and the thermodynamic relations, 
we obtain the heat balance equation
\begin{eqnarray}
       &&c_p\frac{\partial T}{\partial t}+\nabla\cdot {\bf j}_{q}={\dot{\theta}_{q} } \quad ,\nonumber\\ 
       &&\dot{\theta}_{q}=-\sum_{\alpha=1}^{R} r_\alpha \Delta H_\alpha - \sum_{i=0}^{K} {\bf j}_i \nabla h_i
       \label{eqheatbalance}
\end{eqnarray}
where $\dot{\theta}_{q}$ is the local rate of heat generation ($\dot{\theta}_{q}>0$) or absorption ($\dot{\theta}_{q}<0$), and $\Delta H_\alpha=
\sum_{i=0}^K\sigma_{i\alpha}h_i$ is the reaction enthalpy. 


 We can also express the entropy balance as
\begin{equation}
    \frac{\partial s}{\partial t}+\nabla\cdot {\bf j}_s=\dot{\theta}_s\geq 0 \label{eqs1}
\end{equation}
where $s$ is entropy density, ${\bf j}_{s}$ is the entropy flux and $\dot{\theta}_s$ is the local entropy production rate per volume describing irreversible processes.
The second law of thermodynamics  requires $\dot{\theta}_s\geq0$. 
The entropy flux can be decomposed into contributions associated with particle fluxes, and heat flux, 
\begin{equation}
{\bf j}_{s}=\frac{{\bf j}_{q}}{T}+\sum_{i=0}^K{\bf j}_{i}s_i \quad ,
\end{equation}
where particle entropy is defined as $s_i=-\partial \mu_i/\partial T\vert_P$. 
Using $\partial_t s =\sum_{i=0}^K s_i\partial_t n_i+T^{-1}c_p \partial_t T$ together with the heat balance in Eq.~\eqref{eqheatbalance}, we obtain the entropy production
rate per volume
\begin{equation}
    T\dot{\theta}_s=- \sum_{\alpha=1}^{R}r_\alpha \Delta G_\alpha - \sum_{i=0}^{K}{\bf j}_i \cdot \nabla \mu_i - {\bf j}_s\cdot\nabla T \label{epr}
\end{equation}
where $\Delta G_{\alpha}=\sum_{i=0}^K \sigma_{i\alpha}\mu_i$ are the reaction Gibbs free energies. We observe that $-\Delta G_{\alpha}, -\nabla\mu_i, -\nabla T$ are the thermodynamic driving forces for each process (respectively) whereas $r_{\alpha}, {\bf j}_i, {\bf j}_s$ are the associated fluxes.

\subsubsection{Total heat release rate}

The total energy of a system with volume $V_{\mathcal{S}}$ is given by $E_{\mathcal{S}}=\int_{V_{\mathcal{S}}}e \ dV$. The change in $E_{\mathcal{S}}$ can be expressed via changes in energy density and the volume of the system by writing the net energy flux vector at the system-environment interface. Thus, we can write the time evolution of $E_{\mathcal{S}}$ as
\begin{equation}
      \frac{d E_{\mathcal{S}}}{d t}= - \oint_{\partial V_{\mathcal{S}}}{\bf j}_e^{\rm int} \cdot d{\bf A}\label{dEdt1}
\end{equation}
where ${\bf j}_e^{\rm int}={\bf j}_e^{\rm in/out}-e^{\rm in/out}{\bf u}$ is the energy flux across the system-environment interface outwards from the system (see Appendix \ref{appendixb}). Here, ${\bf u}$ is the local interface velocity, $d{\bf A}$ is the outward directing vector at the system-environment interface with a magnitude of infinitesimal area element $dA$, and superscripts \qmark{in}, \qmark{out} represent the side of the interface that is inside or outside the system. 
We define the heat released by the system as the total heat flux
leaving the system
\begin{equation}
    \dot Q= \oint_{\partial V_{\mathcal{S}}}{\bf j}_q^{\rm out} \cdot d{\bf A}\label{dEdt2} \quad .
\end{equation}
With the definition of heat flux via Eq. (\ref{eq:je}) and applying ${\bf j}_i^{\rm int}= {\bf j}_i^{\rm in/out}-n_i^{\rm in/out}{\bf u}$ from Eq.~\eqref{eqjiint}, we have
\begin{equation}
    \frac{d E_{\mathcal{S}}}{d t}=-\dot{Q}
    -\oint_{\partial V_{\mathcal{S}}}\left (\sum_{i=0}^K h_i^{\rm out}{\bf j}_i^{\rm int}+p {\bf u}\right )\cdot d{\bf A} \label{eqetotmain}
\end{equation}
Now using $d E_{\mathcal{S}}/{d t}=\int_{V_\mathcal{S}}  \partial_t h \ dV +\oint_{\partial V_\mathcal{S}} e^{\rm in}{\bf u}\cdot {\bf dA}$, we can write
\begin{eqnarray}
\dot{Q}&=&-\int_{V_{\mathcal{S}}}\frac{\partial h}{\partial t}dV
\nonumber \\&-&\oint_{\partial V_{\mathcal{S}}}\left(h^{\rm in}{\bf u}+\sum_{i=0}^K h_i^{\rm out}{\bf j}_i^{\rm int}\right)\cdot d{\bf A} \quad .\label{calorimetry0main}
\end{eqnarray}
Using Eq. (\ref{eq:dth0}), we can write this in the form
\begin{eqnarray}
\dot{Q}+\int_{V_{\mathcal{S}}}c_p \frac{\partial T}{\partial t}dV=&-&\sum_{i=0}^K \int_{V_{\mathcal{S}}}h_i\frac{\partial n_i}{\partial t}dV\nonumber\\&-&\sum_{i=0}^K\oint_{\partial V_{\mathcal{S}}} h_i^{\rm out}{\bf j}_i^{\rm int}\cdot d{\bf A} \nonumber\\&-&\oint_{\partial V_{\mathcal{S}}}h^{\rm in}{\bf u}\cdot d{\bf A} \quad .\label{calorimetry1main}
\end{eqnarray}
The left-hand side of Eq.~\eqref{calorimetry1main} is measurable with a calorimeter, while the right-hand side shows the enthalpic changes due to particle fluxes and chemistry and hence is related to the biochemistry of the system. Moreover, the right hand side of Eq.~\eqref{calorimetry1main} has three contributions. The first term is due to concentration changes of chemical species inside the organism. The second term is due to exchanged molecules between organism and environment and the third term is due to growth/degrowth. 


\subsubsection{Coarse-grained metabolism}\label{sec:IIa4}
When discussing metabolism, it is often useful to reduce the number of variables and to focus on the key molecular species using coarse-graining methods. Here, we demonstrate an approach based on coarse-graining of each spatial region (system, environment, interface).

We can rewrite Eq.~\eqref{calorimetry1main} by integrating out the spatial information in each region inside, outside, and the interface.  We define the rate of change of number of molecules in the system $\Gamma_i=\int_{V_{\mathcal{S}}}\frac{\partial n_i}{\partial t}dV$, the net exchange fluxes as $\Phi_i=\oint_{\partial V_{\mathcal{S}}}{\bf j}_i^{\rm int}\cdot d{\bf A}$. We aim to relate the growth associated term $\oint_{\partial V_{\mathcal{S}}}h^{\rm in}{\bf u}\cdot d{\bf A}$ to biomass growth by defining biomass in terms of its elemental composition, such as the numbers of carbon, hydrogen, oxygen, and nitrogen atoms. The composition of biomass is defined by the fractions of molecular component $i$ present in a small volume with concentrations $n_i$. 
To express fraction of elements in the biomass, we define a molecular composition matrix $\eta_{ji}$ with entries that are the number of chemical element $j$ contained in molecule $i$. The  concentration of element $j$ (e.g., concentration of carbon atoms) in biomass is  $\epsilon_j=\sum_i \eta_{ji} n_i$.  The composition of biomass is then given by the fractions $\epsilon_j/ \epsilon_b$, where $\epsilon_{b}=n_b$ is the concentration of a reference element, which we chose to be carbon.  Thus $\epsilon_j/ \epsilon_b$ is the number of atoms of element $j$ in an
effective biomass molecule with one carbon atom and $n_b$ is the
concentration of this effective biomass molecule. 
We also define the enthalpy $h_b$
per biomass molecule, such that $h_b n_b=\sum_i h^{\rm in}_i n_i=h^{\rm in}$, or $h_b=h^{\rm in}/n_b$.
The rate of change of the number of biomass molecules in the organism due to growth is  $\Pi_b=\oint_{\partial V_{\mathcal{S}}} {n}_b {\bf u} \cdot d{\bf A}$. Thus finally, we rearrange Eq.~\eqref{calorimetry1main} as:
\begin{eqnarray}
\dot{Q}+\int_{V_{\mathcal{S}}}c_p\frac{\partial T}{\partial t}dV=&-&\sum_{i=0}^K \bar{h}_i^{\rm in}\Gamma_i-\sum_{i=0}^K \bar{h}_i^{\rm out}\Phi_i -\bar{h}_b \Pi_b\nonumber \\
&+&\delta \dot{Q} \quad .\label{calorimetry2main}
\end{eqnarray}
where we introduced volume averaged enthalpies 
$\bar{h}_i^{\rm in}=V_{\mathcal{S}}^{-1}\int_{V_{\mathcal{S}}} h_i dV$, $\bar{h}_i^{\rm out}=V_{\mathcal{E}}^{-1}\int_{V_{\mathcal{E}}} h_i dV$ in the system and its environment as well as the 
chacteristic molecular enthalpy representing the biomass that is formed during growth 
$\bar h_b = A^{-1}\oint_{\partial V_{\mathcal{S}}} h_b d{A}$.
The last term $\delta \dot{Q}$ is due to the spatial heterogeneities in enthalpy $\delta h_{i}^{\rm in/out}=h_{i}-\bar{h}_{i}^{\rm in/out}$, $\delta h_{b}=h_{b}-\bar{h}_{b}$ and it is given by:
\begin{eqnarray}
    \delta \dot{Q}=&-&\sum_{i=0}^K \int_{V_{\mathcal{S}}}\delta h_i^{\rm in}\frac{\partial n_i}{\partial t}dV-\sum_{i=0}^K\oint_{\partial V_{\mathcal{S}}} \delta h_i^{\rm out}{\bf j}_i^{\rm int}\cdot d{\bf A} \nonumber \\&-&\oint_{\partial V_{\mathcal{S}}} \delta h_b {n}_b {\bf u} \cdot d{\bf A}.
\end{eqnarray}
The first three terms on the right hand side of Eq.~\eqref{calorimetry2main} stem from the integration of the spatial information of concentrations within each region (inside, outside, interface) leading to net fluxes respectively $\Gamma_i$, $\Phi_i$, $\Pi_b$. The rearrangement in Eq.~\eqref{calorimetry2main} becomes powerful as we can consider that for many chemical species involved in metabolism $\Gamma_i= 0$, because they are to a good approximation stationary, and $\Phi_i= 0$, because they are not exchanged with the environment. This reduces the number of relevant fluxes. We therefore restrict our attention to the subset of species with nonzero fluxes and we define a reduced set of chemical species and an associated flux vector:
\begin{subequations}
\begin{align}
    \tilde{\mathbb{X}}=&\{{\bf \tilde{X}}^{\rm in},{\bf \tilde{X}}^{\rm out},X_b\} \quad \label{eq:xtilde},\\
    \mathcal{I}=&(\tilde{\bf \Gamma},\tilde{\bf \Phi},\Pi_b)^T \quad. \label{rvect}
\end{align}
\end{subequations}
We use a tilde to denote quantities in the reduced subset, with component indices relabeled. Here $\tilde{\bf \Gamma}$ and $\tilde{\bf \Phi}$  are vectors collecting the nonzero components of $\Gamma_i$ and $\Phi_i$, respectively,  
with dimensions $K_{\tilde{\Gamma}}\leq K+1$, $K_{\tilde{\Phi}}\leq K+1$. The vectors $\tilde{\bf X}^{\rm in
}$ and $\tilde{\bf X} ^{\rm out}$ collect the 
$K_{\tilde\Gamma}$ and $K_{\tilde\Phi}$ chemical species inside and outside the system,  respectively, and $ X_b$
denotes the effective biomass molecule. 
We define the molecular enthalpies 
\begin{equation}
   \tilde{{\bf h}}=(\tilde{\bf h}^{\rm in}, \tilde{\bf h}^{\rm out},\bar{h}_b) \label{eq:enthalpyarray}
\end{equation}
corresponding to the enthalpies of chemical species in  $\tilde{\mathbb{X}}$ defined in Eq.~\eqref{eq:xtilde}. 
We can thus write Eq.~\eqref{calorimetry2main} concisely as
\begin{eqnarray}
\dot{Q}+\int_{V_{\mathcal{S}}}c_p\frac{\partial T}{\partial t}dV&=&-\sum_{i=1}^{K_{\rm net}} \tilde{h}_i \mathcal{I}_i+\delta\dot{Q} .\label{qdotmacrochemical1}
\end{eqnarray}
where $K_{\rm net}=K_{\tilde\Gamma}+K_{\tilde\Phi}+1$ and $\tilde h_i$ and 
$\mathcal{I}_i$ denote the components of $\tilde {\bf h}$
and $\mathcal{I}$.
Thus, when ignoring $\delta \dot{Q}$ for simplicity, it is sufficient to know the net fluxes of the $K_{\rm net}$ relevant chemical species to estimate the total rate of heat release.  

\subsubsection{Macrochemical description}\label{sec:macro}

The coarse-grained fluxes $\mathcal{I}_i$ describing
metabolism are governed by chemical transformation 
processes. These chemical processes can be represented as 
net reactions at the macro level. These net macrochemical reactions 
do not occur microscopically in the 
system but are the effective outcomes of large reaction 
networks. Biochemists typically aim to represent macrochemical reactions in terms of known reaction pathways. Here, we present a formalized framework.

Starting from a set of $\tilde R$ linearly independent 
macro-chemical reactions, we can express the components of the flux $\mathcal{I}$ as linear combination of the fluxes $\mathcal{R}_\beta$ of $\tilde R$ net reactions  
\begin{equation}
\mathcal{I}_i=\sum_{\beta}^{\tilde{R}}\tilde{\sigma}_{i\beta}\mathcal{R}_{\beta} \label{eqnetreaction1} \quad .
\end{equation}
Here $\tilde{\sigma}_{i\beta}$ denote effective stoichiometric coefficients accounting for the net molecule number changes in macrochemical reactions. The effective stoichiometric matrix should satisfy element conservation. We detail the definition of element composition matrix $\tilde{\eta}_{ij}$ for the reduced chemical species array defined in Eq.~\eqref{eq:xtilde} and derive the conservation relation Eq.~\eqref{eqelement3} in Appendix \ref{app:c}.  Then, using Eq.~\eqref{eqnetreaction1} in Eq.~\eqref{eqelement3}, 
\begin{equation}
\sum_{i=1}^{K_{\rm net}}\tilde{\eta}_{ji}\tilde{\sigma}_{i\beta}=0\label{eqelement2}
\end{equation}
which should hold for each independent reaction. Thus for $\tilde{R}$ independent reactions, each column of $\tilde{\sigma}$ should be composed of linearly independent right null vectors of the matrix $\tilde{\eta}$. The number of linearly independent net reactions is the dimension of the right nullspace of matrix $\tilde{\eta}$, i.e., $\tilde{R}=\text{nullity}(\tilde\eta)=K_{\rm net}-\mathbb{E}_{\rm net}$ when $K_{\rm net}> \mathbb{E}_{\rm net}$ where $\mathbb{E}_{\rm net}$ is the number of elements. If $K_{\rm net} \leq \mathbb{E}_{\rm net}$, that means no reaction is possible with the given set of chemical species\footnote{In that case, one should consider adding chemical species to the set for balancing elements.}. Then, the macrochemical reactions can be defined analogously to Eq.~\eqref{eq:reactions}, but involving only the reduced set of chemical species in Eq.~\eqref{eq:xtilde}
\begin{equation}
    \sum_{i=1}^{K_{\rm net}} \tilde{\sigma}_{i\beta}^+ 
    \tilde{\mathbb{X}}_i \rightleftharpoons \sum_{i=1}^{K_{\rm net}} \tilde{\sigma}_{i\beta}^- \tilde{\mathbb{X}}_i\label{eq:netreactions}
\end{equation}
where $\tilde{\sigma}_{i\beta}^{\pm}$ are the stoichiometric coefficients, and the net stoichiometric coefficient is defined as before $\tilde{\sigma}_{i\beta}=\tilde{\sigma}_{i\beta}^{-}-\tilde{\sigma}_{i\beta}^{+}$. 

As an example, we illustrate the macrochemical reaction in aerobic metabolism by considering the full oxidation of glucose. We consider resting metabolism with no biomass growth, and  the glucose (carbon source) is used up from internal reservoirs. 
In this case, the molecules exchanged with the environment are $\tilde{\bf X}^{\rm out}=\{\text{O}_2, \text{CO}_2,\text{H}_2 \text{O}\}$ and glucose is considered inside the system $\tilde{\bf X}^{\rm in}=\{\text{C}_6 \text{H}_{12} \text{O}_6\}$ and we use $\Pi_b=0$ (no biomass growth) and hence ignore $X_b$ ($X_b=\emptyset$). Accordingly, the reduced chemical species array and corresponding element composition matrix are given by
\begin{eqnarray}
\tilde{\mathbb{X}} =\left(
\begin{array}{c}
\text{C}_6 \text{H}_{12} \text{O}_6 \\
\text{O}_2\\
\text{CO}_2\\
\text{H}_2 \text{O}
\end{array}
\right), \quad
       \tilde\eta =
\left(
\begin{array}{cccc}
6  & 0 & 1 & 0 \\
12 & 0 & 0 & 2 \\
6  & 2 & 2 & 1
\end{array}
\right) \label{eq:xtilde2}
\end{eqnarray}
where the rows of $\tilde{\eta}$ are respectively the number of C, H, O elements in molecules given in $\tilde{\mathbb{X}}$. The only null vector of $\tilde\eta$ is proportional to $\tilde{\sigma}=(-1,-6,6,6)^T$. Then, the macrochemical reaction can be written as:
\begin{equation}
 \mathrm{C_6H_{12}O_6+6O_2\rightleftharpoons 6CO_2 + 6H_2O} \quad .
 \label{eq:macroglucose}
\end{equation}
As a result, the rates of change of the molecular species in $\tilde{\mathbb{X}}$ can be obtained from the macrochemical reaction rates and the stoichiometric coefficients.

We now return to our general formulation and finally write Eq.~\eqref{qdotmacrochemical1} as:
\begin{eqnarray}
\dot{Q}+\int_{V_{\mathcal{S}}}c_p\frac{\partial T}{\partial t}dV&=&-\sum_{\beta=1}^{\tilde{R}} \mathcal{R}_{\beta}\Delta H_{\beta}  +\delta\dot{Q}\label{qdotmacrochemical2}
\end{eqnarray}
where $\Delta H_{\beta}=\sum_{i=1}^{K_{\rm net}} 
\tilde{\sigma}_{i\beta} \tilde{h}_i$. 

Summarizing this section, we have derived expressions for local heat balance (Eq.~\eqref{eqheatbalance}), local entropy production rate (Eq.~\eqref{epr}), and the total rate of heat release (Eqs. \eqref{calorimetry1main},~\eqref{qdotmacrochemical1}) by constructing the irreversible thermodynamics of metabolic processes. We related the heat release rate to the enthalpy change of macrochemical reactions in Eq.~\eqref{qdotmacrochemical2} from the knowledge of chemical species with non-zero net fluxes by using element conservation. This is a backward approach, which does not necessarily require knowledge of the entire chemical network, and is sometimes referred as ``black-box metabolism''.

\subsection{Metabolic rate} \label{sec:iib}

The metabolism of organisms reflects the fact that life operates far from
thermodynamic equilibrium.  The metabolic rate can be defined as the flux of energy associated with the metabolism of an organism. The energy flux is equivalent to the input power, however there is no way of directly measuring it. On the other hand, due to its nonequilibrium nature, metabolism is also a dissipative process. A key variable that is associated with both energy
flux and dissipation is the total rate of heat release by the organism ($\dot{Q}$). This heat release rate has been widely used as a proxy for metabolic rate, since it is well defined and can be measured by several techniques, as we will explain in Section \ref{exper_approaches}. 

We should note the caveats in using the heat release rate as a proxy for the metabolic rate. In energy metabolisms such as photosynthesis or endothermic methanogenesis \cite{liu2001microbial,von1999does}, heat can be absorbed rather than released, thereby reducing $\dot{Q}$. An alternative proxy (given that metabolism is dissipative) is the entropy production rate, which is always positive. Consequently, entropy production can provide a consistent measure of metabolic activity, making it a compelling substitute for the traditional metabolic rate based on $\dot{Q}$. We give in Eq.~\eqref{eq:totaleprv2} an expression for the total entropy production rate that is analogous to heat release described by Eq.~\eqref{qdotmacrochemical2}. However, a key challenge is the lack of a direct measurement technique for the entropy production rate, which makes it much more difficult to estimate than the rate of heat release. Estimating entropy production rate would require quantification of fluxes of chemical species and their associated chemical potentials as well as heat. Recent works explore such approaches in order to determine the entropy production rate in unicellular growth \cite{cossetto2025thermodynamic,cossetto2025charting}, but the available data is still limited compared to traditional $\dot{Q}$ measurements.

Consequently in this review we will define the metabolic rate as the heat release rate $\dot{Q}$. This is a common proxy for aerobic organisms, and as we will see below it can also be related to other metabolic observables (like the oxygen consumption rate). With this choice of definition, Eqs.~\eqref{qdotmacrochemical1} and \eqref{qdotmacrochemical2} are the key expressions. These equations are also useful when converting between different experimental quantities, such as the oxygen consumption rate or nutrient intake rates. Throughout this paper, we present specialized forms of Eqs.~\eqref{qdotmacrochemical1} and \eqref{qdotmacrochemical2}  under different metabolic rate measurement conditions. 

\subsection{Cellular metabolism} 
 \begin{figure*}[t]
 \includegraphics[width=\textwidth]{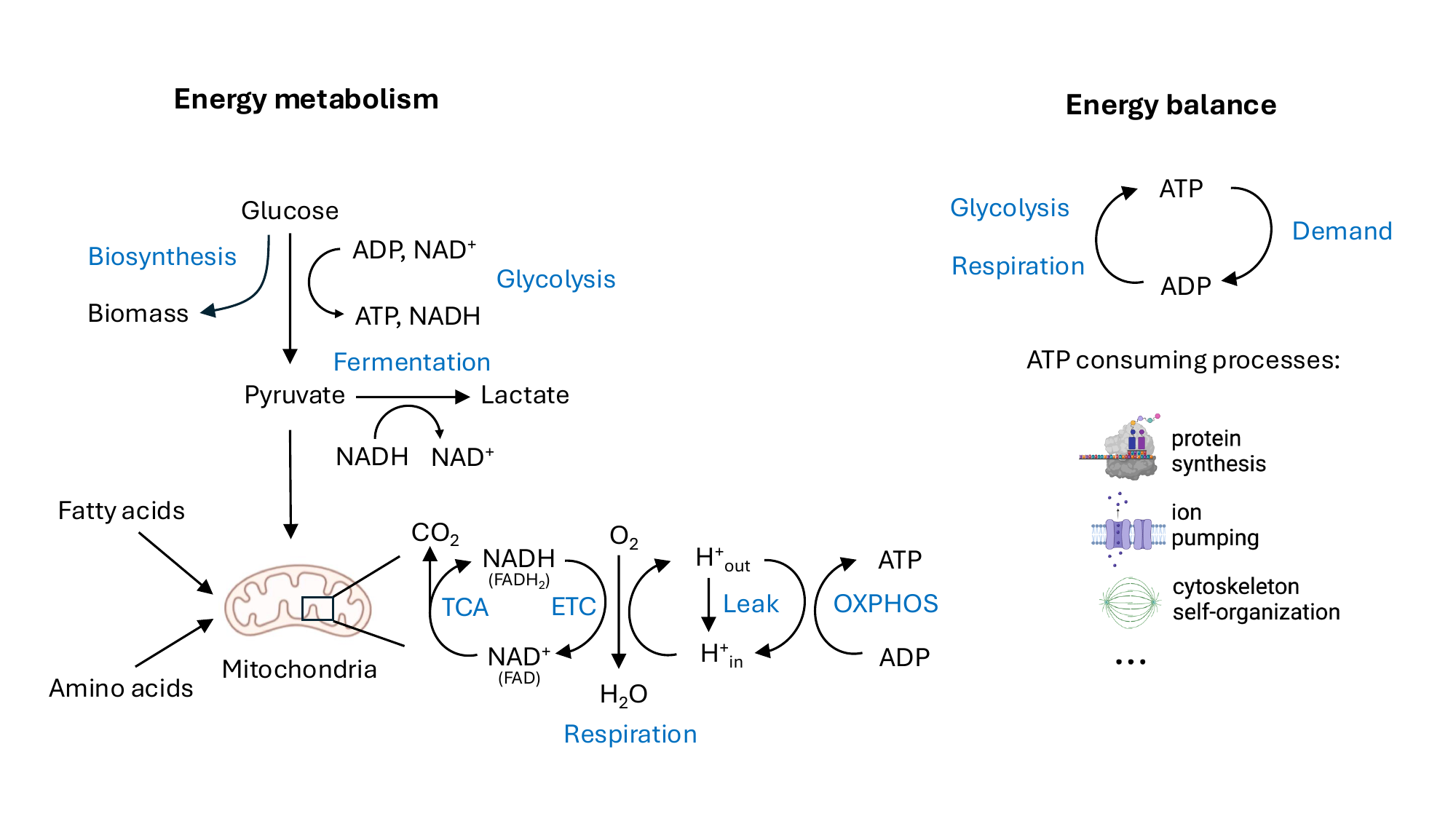}
 \caption{{\bf Coarse-grained cellular energy metabolism.} Left panel: energy metabolism composed of glycolysis, fermentation, mitochondrial respiration and biosynthesis. Glycolysis and respiration are the two major pathways that produce ATP, the energy currency of the cell. The tricarboxylic acid (TCA) cycle, the electron transport chain (ETC), the proton leak and oxidative phosphorylation (OXPHOS) are indicated. Each arrow represents a net chemical reaction, as defined in Eq.~\eqref{eq:netreactions}, or transport across cellular compartments. Right panel: energy balance of ATP production and hydrolysis. ATP demand represents all the ATP consuming processes in the cell.}
 \label{fig:cell_metabolism}
 \end{figure*}
 \begin{figure*}[t]
 \includegraphics[width=0.9\textwidth]{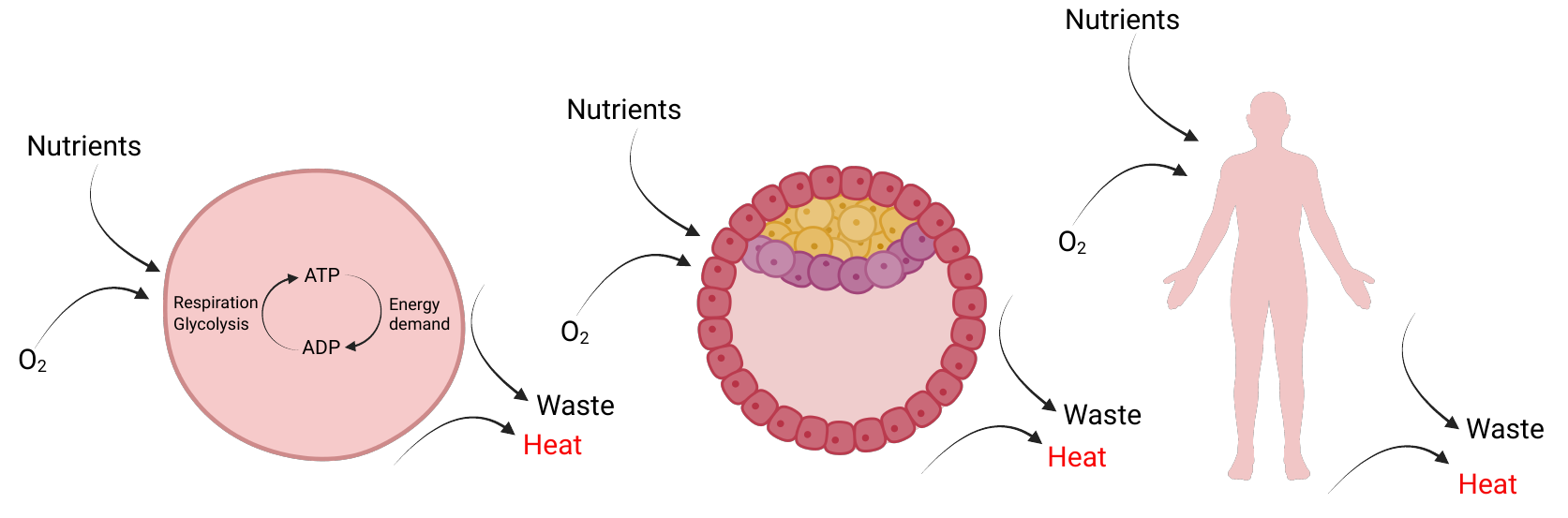}
 \caption{{\bf Net fluxes into and out of living systems}. Coarse-grained metabolism at the whole system level for cells, tissues, and the entire animal can be described via net input (nutrients, oxygen) and output (waste, heat, and internal compositional change) fluxes. These fluxes can be linked by  macrochemical reactions, Eq.~\eqref{eq:netreactions}. The metabolic rate is defined as the rate of heat release of the organism to the environment. Eq.~\eqref{qdotmacrochemical2} relates the heat release rate ($\dot{Q}$) to macrochemical reaction rates $\mathcal{R}_{\beta}$ and associated enthalpy changes $\Delta H_{\beta}$.}
 \label{fig:metabolic_rate}
 \end{figure*}
In this section, we provide a coarse-grained view of energy metabolism in cells (Fig.~\ref{fig:cell_metabolism}) and discuss different observables related to the metabolic rate (Fig.~\ref{fig:metabolic_rate}). In cells, the energy required for biological processes is predominantly supplied through the hydrolysis of adenosine triphosphate (ATP) to adenosine diphosphate (ADP) and inorganic phosphate (Pi).  In biology, several distinct metabolic pathways contribute to ATP production from ADP.
Glycolysis and mitochondrial respiration represent the two major ATP production pathways in the cell. 

Glycolysis is composed of 10 enzymatic steps that collectively convert one glucose molecule ($\rm{C_6H_{12}O_6}$) to two pyruvate molecules ($\rm{C_3H_4O_3}$), generating two ATP and two NADH in the process. Nicotinamide adenine dinucleotide (NAD) is an electron carrier, existing in the reduced (NADH) or oxidized (NAD+) state. Since glycolysis requires NAD+ to proceed, the cell needs to regenerate NAD+ from NADH. This can be achieved through fermentation, which converts pyruvate to lactate ($\rm{C_3H_6O_3}$), generating NAD+ in the process. Glycolysis plus fermentation therefore represents an anaerobic metabolic pathway that produces ATP by converting glucose to lactate. While lactate is the main fermentation product in mammalian cells, organisms such as yeast and some prokaryotes produce other fermentation products, such as ethanol and acetate.

In the presence of oxygen, pyruvate can be transported into mitochondria to be fully oxidized to carbon dioxide and produce a great amount of ATP. This process is termed mitochondrial respiration, representing the aerobic energy metabolism of the cell. Mitochondrial respiration is composed of a myriad of chemical steps, but it can be seen in a coarse-grained view as three coupled cycles: the $\rm{NADH/NAD^+}$ cycle (or $\rm FADH_2/FAD$ cycle), the proton cycle and the ATP/ADP cycle. The $\rm{NADH/NAD^+}$ cycle uses free energy from nutrients, provided by pyruvate for example, to pump protons across the the mitochondrial inner membrane, creating a proton gradient across the membrane. This proton gradient drives the synthesis of ATP by flowing protons through ATP synthase, a rotary molecular complex embedded in the mitochondrial inner membrane, a mechanism much like how flowing water drives a water wheel. This process is termed the oxidative phosphorylation. Since not all protons will go through ATP synthase, those protons that flow back into the mitochondrial matrix without driving ATP synthesis constitute what is termed the proton leak. In prokaryotic cells, which lacks mitochondria, oxidative phosphorylation occurs across the plasma membrane. 

To capture these metabolic pathways quantitatively, we introduce net chemical reactions, as defined in Eq.~\eqref{eq:netreactions}, for the labeled arrows in Figure \ref{fig:cell_metabolism} that represent net reactions involved in glycolysis, fermentation and mitochondrial respiration. At steady state when ATP production balances ATP consumption, we can sum up the net reactions to obtain the macrochemical reactions for the cell that represent the net carbon flow that drives the cell out of equilibrium. Based on the macrochemical reactions, we quantify the total metabolic rate of the organism in terms of the rate of heat release (Figure \ref{fig:metabolic_rate}). We further show that the oxygen consumption rate (OCR), denoted as $\Phi_{\text{O}_2}$, is a proxy for the total metabolic rate.

We first write down the net reaction for glycolysis: 
\begin{equation}
\begin{aligned}
\mathrm{C_6H_{12}O_6 + 2ADP + 2Pi + 2NAD^+}
\rightleftharpoons \\
\mathrm{2C_3H_{4}O_3+2ATP+2NADH+2H_2O}
 \label{eq:glyc}
 \end{aligned}
\end{equation}
Two ATP molecules are produced per glucose from glycolysis if the glucose is converted all the way to pyruvate. Here, the phosphate group $\mathrm{Pi}$ denotes hydrogen phosphate $\mathrm{HPO_4^{2-}}$, which is the dominant form at cellular pH. Alternatively, the glycolysis intermediates can be directed as building blocks for biosynthesis at the expense of ATP production. Two molecules of the high energy electron carrier $\rm{NADH}$ are produced from glycolysis, which can be oxidized back to $\rm{NAD^+}$ through fermentation. The net reaction for fermentation is
\begin{equation}
\begin{aligned}
\mathrm{C_3H_{4}O_3 + NADH + H^+}
\rightleftharpoons
\mathrm{C_3H_{6}O_3+NAD^+}
 \label{eq:ferm}
 \end{aligned}
\end{equation}
Summing up net reactions, Eqs.~\eqref{eq:glyc} and  \eqref{eq:ferm}, we arrive at the net reaction of glycolysis plus fermentation
\begin{equation}
\begin{aligned}
\mathrm{C_6H_{12}O_6 + 2ADP + 2Pi + 2H^+}
\rightleftharpoons \\ 
\mathrm{2C_3H_{6}O_3+2ATP+2H_2O}
 \label{eq:glyc_ferm}
  \end{aligned}
\end{equation}
This net reaction does not require oxygen, hence is an anaerobic ATP production pathway. In the presence of oxygen, the pyruvate can enter mitochondria to be fully oxidized to produce more ATP through mitochondrial respiration. Once pyruvate enters the mitochondria, it drives the production of $\rm{NADH}$, $\rm{FADH_2}$ and guanosine triphosphate ($\rm{GTP}$) through the tricarboxylic acid (TCA) cycle. $\rm{GTP}$ can be converted to $\rm{ATP}$ through the nucleoside diphosphate kinase (NDPK). The net reaction of the TCA cycle plus the NDPK reaction is
\begin{equation}
\begin{aligned}
\mathrm{C_3H_{4}O_3 + 4NAD^+ + FAD + ADP + Pi + 2H_2O}
\rightleftharpoons \\
\mathrm{4NADH+FADH_2+3H^++ATP+3CO_2}
 \label{eq:tca}
 \end{aligned}
\end{equation}
Each $\rm{NADH}$ or $\rm{FADH_2}$ donates two electrons to one oxygen atom through the mitochondrial Electron Transport Chain (ETC) and pumps protons across the mitochondrial inner membrane. The net reaction of the ETC with $\rm{NADH}$ or $\rm{FADH_2}$ as the respective electron donors can be written as
\begin{equation}
\begin{aligned}
\mathrm{NADH + H^+ +\frac{1}{2}O_2}
\rightleftharpoons
\mathrm{NAD^++H_2O+10\Delta H^+}
 \label{eq:etc_nadh}
 \end{aligned}
\end{equation}
and
\begin{equation}
\begin{aligned}
\mathrm{FADH_2 + \frac{1}{2}O_2}
\rightleftharpoons
\mathrm{FAD+H_2O+6\Delta H^+}
 \label{eq:etc_fad}
 \end{aligned}
\end{equation}
where $\Delta \rm{H^+}$ denotes the pumped proton across the inner membrane of the mitochondria from the matrix to the intermembrane space. Every oxidation of $\rm{NADH}$ ($\rm{FADH_2}$) pumps a total of 10 (6) protons across the mitochondrial inner membrane, thus generating a proton gradient across the membrane as well as an electric potential difference. This electrochemical potential gradient drives protons to go through the transmembrane ${\rm F_0}$ subunit of the 
${\rm F_0F_1}$-ATP synthase complex embedded in the mitochondrial inner membrane, turning the rotary ${\rm F_1}$ subunit to make ATP from ADP and Pi, with the reaction 
\begin{equation}
\begin{aligned}
\mathrm{ADP+Pi+H^++4\Delta H^+}
\rightleftharpoons
\mathrm{ATP+H_2O} \quad .
 \label{eq:atpsyn}
 \end{aligned}
\end{equation}
Here $\Delta \rm{H}^+$ denotes the proton flow through the ATP synthase. 
The stoichiometry coefficient 4 derives from the fact that approximately 3 protons are required to synthesize 1 ATP through the ATP synthase, and it takes an additional 1 proton to transport to import one phosphate via the proton coupled mitochondrial phosphate carrier (PiC). The ATP produced from the mitochondria is exchanged with ADP in the cytoplasm through the adenine nucleotide translocator (ANT). ATP production through ETC and ATP synthase constitutes the oxidative phosphorylation process. Meanwhile, protons can also leak through the mitochondrial inner membrane via other proton channels such as mitochondrial uncoupling protein (UCP), producing heat rather than ATP. 

Mitochondrial respiration can be defined as the net reaction that converts nutrients to produce ATP in the presence of oxygen. Pyruvate as discussed is a major nutrient that supports mitochondrial respiration. Since glycolysis produces pyruvate, it is convenient to combine glycolysis with mitochondrial respiration by summing up the reactions in Eqs.~\eqref{eq:glyc} and \eqref{eq:tca}-\eqref{eq:atpsyn} to arrive at the net reaction for glycolysis plus mitochondrial respiration
\begin{equation}
\begin{aligned}
\mathrm{C_6H_{12}O_6+6O_2+32ADP+32Pi+32H^+}
\rightleftharpoons \\
\mathrm{6CO_2+32ATP+38H_2O}
 \label{eq:resp_glyc}
  \end{aligned}
\end{equation}
Notice that mitochondrial respiration produces 32 ATP molecules per glucose as compared to 2 ATP molecule per glucose from fermentation (reaction Eq.~\eqref{eq:glyc_ferm}). Mitochondrial respiration is therefore a more efficient ATP production pathway compared to fermentation in terms of the yield of ATP per glucose. Proton leak and diversion of glucose for biosynthesis can both decrease the ATP yield per glucose, hence reaction Eq.~\eqref{eq:resp_glyc} represents an upper limit of ATP yield per glucose. In addition to pyruvate, amino acids and fatty acids can also serve as nutrients for mitochondrial respiration, leading to different yields of ATP per molecule. It is worth noting that despite the lower yield of ATP per glucose for fermentation compared to respiration, highly proliferative cells such as cancer cells heavily use fermentation even in the presence of abundant oxygen, a phenomenon termed the ``Warburg effect'', first discovered by Otto Warburg in the 1920s \cite{warburg24,vander2009understanding}. Recent studies have shown that the Warburg effect also occurs in developing embryos \cite{krisher2012role}.
%

The resulting ATP produced from glycolysis and respiration powers all ATP consuming processes in the cell. This is achieved by enzymes coupling the ATP hydrolysis reaction with ATP-driven reactions that do not happen spontaneously, such as for example protein synthesis, ion pumping and polymerization of actin filaments and microtubules (Figure \ref{fig:metabolic_rate}). The ATP hydrolysis reaction is given by
\begin{equation}
\begin{aligned}
\mathrm{ATP+H_2O}
\rightleftharpoons
\mathrm{ADP+Pi+H^+}
 \label{eq:atp_hydr}
  \end{aligned}
\end{equation}

ATP is constantly turning over due to the rapid dynamics of ATP consumption and production. At steady state, when ATP production balances ATP consumption, we can eliminate ATP, ADP, and Pi from the reactions. Combining reaction Eq.~\eqref{eq:glyc_ferm} with Eq.~\eqref{eq:atp_hydr}, we obtain the net reaction for anaerobic metabolism
\begin{equation}
 \mathrm{C_6H_{12}O_6\rightleftharpoons 2C_3H_{6}O_3}\quad .
 \label{eq:glyc_ferm_steady} 
\end{equation}
Combining Eqs.~\eqref{eq:resp_glyc} and \eqref{eq:atp_hydr}, we obtain the net reaction for aerobic metabolism
\begin{equation}
 \mathrm{C_6H_{12}O_6+6O_2\rightleftharpoons 6CO_2 + 6H_2O} \quad .
 \label{eq:glyc_oxphos_steady}
\end{equation}
These net reactions, representing the overall chemical transformation of metabolism, are called macrochemical reactions
\cite{ghosh2023developmental}. Thus, with steady-state assumptions the macrochemical reaction of aerobic metabolism appears as a combustion reaction, Eq.~\eqref{eq:macroglucose}. The activity of each pathway is quantified by its metabolic flux, which is defined as the turnover rate of molecules through the metabolic reaction. Fermentation activity is quantified by the lactate production rate, while respiration activity is quantified by the oxygen consumption rate. 

In addition to constantly taking in nutrients from the environment, cells can store nutrients for usage during starvation. Typical internal nutrient stores are glycogen and fatty acids. Glycogen is a multibranched polysaccharide of glucose stored in glycogen granules. It can be mobilized to fuel glycolysis. Fatty acids are usually stored as lipid droplets in the cell and can be mobilized to fuel mitochondrial respiration.

Photosynthesis by plants, cyanobacteria and algae is another important energy metabolic pathway that harness energy from light to convert carbon dioxide and water to glucose and oxygen, which as a net macrochemical reaction amounts to the reverse of Eq.~\eqref{eq:glyc_oxphos_steady}.
\\

\subsection{Heat release per oxygen consumed}
\label{thorntons_rule}
For aerobic metabolism, the oxygen consumption rate (OCR denoted by $\Phi_{\text{O}_2}$) is proportional to the heat release rate $\dot{Q}$. To show this, let us take the example of respiration by using glucose (Eq.~\eqref{eq:macroglucose}). The glucose is used up from internal stores, while the $\text{O}_2$, $\text{CO}_2$, $\text{H}_2 \text{O}$ is exchanged with the environment, and hence $\tilde{{\bf h}}=(h_{\rm glu}^{\rm in},h_{\rm O_2}^{\rm out},h_{\rm CO_2}^{\rm out},h_{\text{H}_2\text{O}}^{\rm out})^T$ corresponding to chemical species array $\tilde{\mathbb{X}}$ in Eq.~\eqref{eq:xtilde2}. The macrochemical equation, Eq.~\eqref{eq:macroglucose}, would have  $\Delta H_{}=\sum_{i=1}^{K_{\rm net}} 
\tilde{\sigma}_{i\beta} \tilde{h}_i=-h_{\rm glu}^{\rm in}-6h_{\rm O_2}^{\rm out}+6h_{\rm CO_2}^{\rm out}+6h_{\text{H}_2\text{O}}^{\rm out}$. 
Accordingly, we write the enthalpy change per oxygen consumed as $\Delta h_{\rm ox}=\Delta H/6$. Considering $\partial _t T=0$ in the organism and neglecting $\delta \dot{Q}$ in Eq.~\eqref{qdotmacrochemical2},  we have
\begin{equation}
   \dot{Q}=-\Phi_{{\rm O}_2} \Delta h_{\rm ox} \label{qox}\quad .
\end{equation}
To estimate $\Delta h_{\rm ox}=\Delta H/6$, we can take the standard values of enthalpy of formation. At $T=298 ^{\circ}$K, the enthalpy of formation values are given as $h_{{\rm glu}}=-1264.2 \text{ kJ/mol}$ (value for glucose in an aqueous environment), $h_{{\rm CO}_2}=-393.5 \text{ kJ/mol}$, $h_{{\rm H}_2{\rm O}}=-285.8 \text{ kJ/mol}$, and  $h_{{\rm O}_2}=0$ by definition \cite{heijnen1999bioenergetics}. This leads to $\Delta h_{\rm ox}^{\rm glu}=-468.6 \text{ kJ/mol}$. More generally, for different carbon sources (glucose, fatty acid, amino acid) used in energy metabolism, the enthalpy change per oxygen consumed remains within a narrow range of $\Delta h_{\rm ox}=-433\text{ to }-468.6 \text{ kJ/mol}$ (see Appendix \ref{appendixd}). Thus, typical energy conversions of oxygen consumption rates assume an average value of $\Delta h_{\rm ox}=-450 \text{ kJ/mol}$ which is known as \qmark{Thornton's rule} \cite{thornton1917xv}. As a result, Thornton's rule often allows us to use $\Phi_{{\rm O}_2}$ and $\dot{Q}$ measurements interchangeably by assuming Eq.~\eqref{qox} with an approximate $\Delta h_{\rm ox}=-450 \text{ kJ/mol}$ irrespective of the carbon source used up by the organism. Thornton's rule is quite general: combustion of organic compounds and combustion of microbial biomass also approximately follow this relation ~\cite{cordier1987relationship,von1993thermodynamic}. 
If one needs higher accuracy, there are empirical formulas correcting for different oxidation states of carbons in the biomass, improving the prediction of heat released per oxygen consumed in the context of biomass combustion \cite{ho1979assimilation}.

Strong deviations from the Thornton value that can be captured by comparative analysis of $\Phi_{{\rm O}_2}$ and $\dot{Q}$ may indicate alterations in energy metabolism. For instance, a value of $\dot{Q}/\Phi_{{\rm O}_2}>500 \text{ kJ/mol}$ would indicate the use of anaerobic metabolic pathways. In particular, this ratio can reach $\dot{Q}/\Phi_{{\rm O}_2}>600 \text{ kJ/mol}$ in cancer cells that primarily use fermentation for energy metabolism \cite{gnaiger1990anaerobic}.


 %

 \vspace{-1em}

\subsection{Experimental approaches to measure metabolic rate}\label{exper_approaches}

We have defined metabolic rate as the rate of heat release of organisms to the environment (Section \ref{sec:iib}). Hence the direct way to measure metabolic rate is through a calorimeter. Isothermal calorimetry is a standard technique to measure the heat release rate of a sample with a constant temperature reference. Appendix \ref{appendixc} provides a discussion relating the heat release rate of the sample to the temperature difference between the sample and the reference (Eq.~\eqref{eq:qnewton}). The best devices can measure a heat release rate on the order of 200 pW \cite{bae2021micromachined}, which corresponds to the heat release rate from a {\it C. elegans} or {\it Drosophila} embryo. Since calorimetry is a bulk technique, it is unable to measure heat release rate with spatial resolution.

Based on the thermodynamic framework, Eq.~\eqref{qdotmacrochemical1} relates the heat release rate of the organism $\dot{Q}$ to the metabolic fluxes $\mathcal{I}$ involved in the net reactions, allowing researchers to leverage the extensive array of metabolic flux measurement techniques to estimate the heat release rate.
If we know the enthalpy changes associated with the net reactions, the heat release rate of the organism can be 
equivalently calculated from Eq.~\eqref{qdotmacrochemical2}. One of the most frequently measured metabolic fluxes is the oxygen consumption rate $\Phi_{\text{O}_2}$. Due to Thornton's rule (Section \ref{exper_approaches}), $\dot{Q}$ can be reliably inferred for aerobic organisms from $\Phi_{\text{O}_2}$, and the latter can be measured through respirometry. Hence respirometry is considered a form of indirect calorimetry.

Multiple techniques exist to measure $\Phi_{\text{O}_2}$ across different scales. The oxygen consumption rate 
$\Phi_{\text{O}_2}$ of a population of cells can be measured using sealed-chamber respirometry, where
$\Phi_{\text{O}_2}$ is measured by the oxygen depletion rate within the chamber hosting the cells \cite{ferrick2008advances}. The advantage of this technique is that the measurement is fast, on the order of minutes, and metabolic perturbations can be performed during the measurement through the injection of drugs. To measure $\Phi_{\text{O}_2}$ of single cells, such as oocytes, open-chamber respirometry is implemented \cite{lopes2005respiration}. The sample is placed at the bottom of a thin capillary well with an open top where oxygen can diffuse in. Due to a balance between diffusion and oxygen consumption by the sample, a linear steady-state oxygen gradient is established within the well that can be measured by a motor-controlled oxygen sensor. The consumption rate $\Phi_{\text{O}_2}$ is then proportional to the oxygen concentration gradient. The advantage of this technique is that it can be used to measure $\Phi_{\text{O}_2}$ down to fmol/s, on the order of the rate of a single mouse oocyte \cite{ottosen2007murine}. A disadvantage is that it takes hours for the oxygen to reach steady state. Recent technology development has enabled the measurement of $\Phi_{\text{O}_2}$ with subcellular resolution. Fluorescence lifetime imaging (FLIM) 
is used to measure the fluorescence lifetime of NADH with optical resolution. Because NADH fluorescence lifetime changes drastically when NADH binds to an enzyme compared to free NADH, FLIM of NADH can be used to infer the concentrations of bound and free NADH quantitatively \cite{sharick2018protein}. A coarse-grained NADH redox model was developed to interpret the changes of free and bound NADH concentrations in terms of the change of the oxidation flux of NADH through the mitochondrial electron transport chain (ETC flux), which is proportional to $\Phi_{\text{O}_2}$ normalized by mitochondrial volume. This technique has enabled the discovery of subcellular spatial gradients of ETC flux within a single mouse oocyte \cite{yang2021coarse}. The limitation of this technique is that it measures relative changes of the ETC flux. To obtain absolute measurements, a calibration with a respirometer is needed.

 The macrochemical description, Eq.~\eqref{eq:macroglucose}, relates the nutrient uptake rate and the waste secretion rate to the oxygen consumption rate through stoichiometry. Therefore if the macrochemical reaction is known, the heat release rate can be inferred by measuring the nutrient uptake rate or the waste secretion rate. For organisms taking up glucose from the environment, the glucose uptake rate can be measured using mass spectrometry \cite{paczia2012extensive}, nuclear magnetic resonance spectroscopy \cite{arunachalam2024robustness} or fluorescence-based biochemical assays on spent media by the sample \cite{martin1995role}. In short, the concentration of glucose in the media culturing the sample is measured at different times, and the glucose uptake rate is determined via the rate of concentration change. Using the macrochemical equation, Eq.~\eqref{eq:macroglucose}, the oxygen consumption rate can be calculated from the glucose uptake rate. Finally, using the Thornton's rule, the heat release rate can be inferred. 
 
 The nutrient uptake and waste secretion rate measurements also enable inference of the heat release rate for an anaerobic organism. For anaerobic organisms converting glucose to lactate according to Eq.~\eqref{eq:glyc_ferm_steady}, the heat release rate can be calculated from the reaction enthalpy change and the glucose uptake rate or the
 lactate secretion rate according to Eq.~\eqref{qdotmacrochemical2}. The lactate secretion rate can be inferred by measuring the extracellular acidification rate (ECAR) of the media through pH measurement \cite{plitzko2018measurement, mookerjee2017quantifying}. The enthalpy change of the macrochemical reaction, Eq.~\eqref{eq:glyc_ferm_steady}, can be calculated from the enthalpy of formation of glucose and lactate at standard conditions using literature values \cite{hammes2015physical}.

For growing organisms, the relation between $\Phi_{\text{O}_2}$ and $\dot{Q}$ needs to be corrected to account for the accumulation of biomass. The concept of biomass has been defined 
in Section \ref{sec:IIa4}. Eq.~\eqref{calorimetry2main} also relates the heat release rate to the flux of biomass accumulation ($\Pi_b$), which is in turn related to the growth rate of the organism for balanced growth. As an example of growth rate measurement, a spectrophotometer can be 
used to measure the growth of a 
bacterial cell population by measuring the decrease of transmitted light intensity through the sample \cite{mira2022estimating}. Cell counting or direct mass measurement can be used to measure the growth rate of tissue culture cells or of larger organisms \cite{vembadi2019cell}. The enthalpy of formation of the biomass can be estimated from bomb calorimetry \cite{ho1979assimilation}.

 \vspace{1em}
\section{Scaling of Metabolic Rate}\label{section3}

   The heat release rate $\dot{Q}$ of an organism can depend on a wide variety of internal and external factors, including developmental stage, the level of activity, time post-digestion, pregnancy, and ambient temperature. Over the last century, there has been a special focus on the basal value of $\dot{Q}$, which reflects a ``minimal'' metabolic rate measured under a specific set of conditions that depends on the class of organism (as described in more detail in Sec.~\ref{defb} below). We will denote this basal rate as $B\equiv \dot{Q}$. Comparing values of $B$  
   within and across different species reveals complex scaling relationships that have intrigued researchers for decades. The most famous of these is Kleiber's law, $B \sim M^{3/4}$~\cite{kleiber1932body,kleiber1947body,kleiber1961fire},  where $M$ is the total (wet) mass of the organism. This scaling, first observed in a dataset of mammals and birds, has been the subject of intense debate as to its biological origins~\cite{harrison2022white} and the extent to which it is truly a universal ``law'' or more of an ``empirical approximation'' valid in certain regimes~\cite{hulbert2014sceptics}.

   Here we review the evidence of power-law scaling of $B$, and explore the consequences of this scaling for organismal growth. Sec.~\ref{hist} starts with a short history of metabolic rate measurements, followed by a discussion of the technical challenges in defining and estimating $B$ across many different species in Sec.~\ref{defb}. In the modern era of large-scale $B$ data sets covering all domains of life~\cite{makarieva2008mean,delong2010shifts,hatton2019linking,hoehler2023metabolic}, we now have a more nuanced picture of interspecies scaling of $B$, described in Sec.~\ref{inter}. The precise form of the power laws depends on the biological scales and categories of interest. Kleiber's law still approximately holds within similar classes of multicellular organisms, but we see substantial deviations in unicellular eukaryotes and prokaryotes. On the largest scales, from bacteria to whales, there is an approximately linear $B \sim M$ relation. As explained in Sec.~\ref{univ}, this leads to several interesting consequences: a universal metabolic time scale and bounds on the metabolic rate per unit mass. Sec.~\ref{intra} looks at how $B$ varies within a species, and even within a single organism during growth. The picture here is similarly complex: organisms with relatively simple growth plans have been found to exhibit Kleiber's law, i.e. the nearly perfect 3/4 scaling in planaria across six decades of mass~\cite{thommen2019body}. On the other hand, arthropods can pass through radically different developmental stages known as instars, with each instar possibly exhibiting distinct non-Kleiber $B$ scaling~\cite{glazier2024multiphasic}. In Sec.~\ref{sec:growth} we describe the consequences of sublinear scaling of $B$ as the organism matures, where the ratio of $B$ to mass decreases over time. This fact may be connected to one of the fundamental features of animal growth---that mass tends to asymptotically approach a limiting value with age. Finally, in the last two sections we look at the ramifications of metabolic scaling at biological levels both smaller and larger than the organism: Sec.~\ref{sec:tissue} describes the heterogeneous ways tissue and cellular-level metabolism scale with organism mass $M$, while Sec.~\ref{sec:popscaling} considers the combined metabolism of whole populations, and connections to ecological scaling relationships.
   


\subsection{History of metabolic rate measurements and scaling}\label{hist}

\begin{figure}
    \includegraphics[width=\columnwidth]{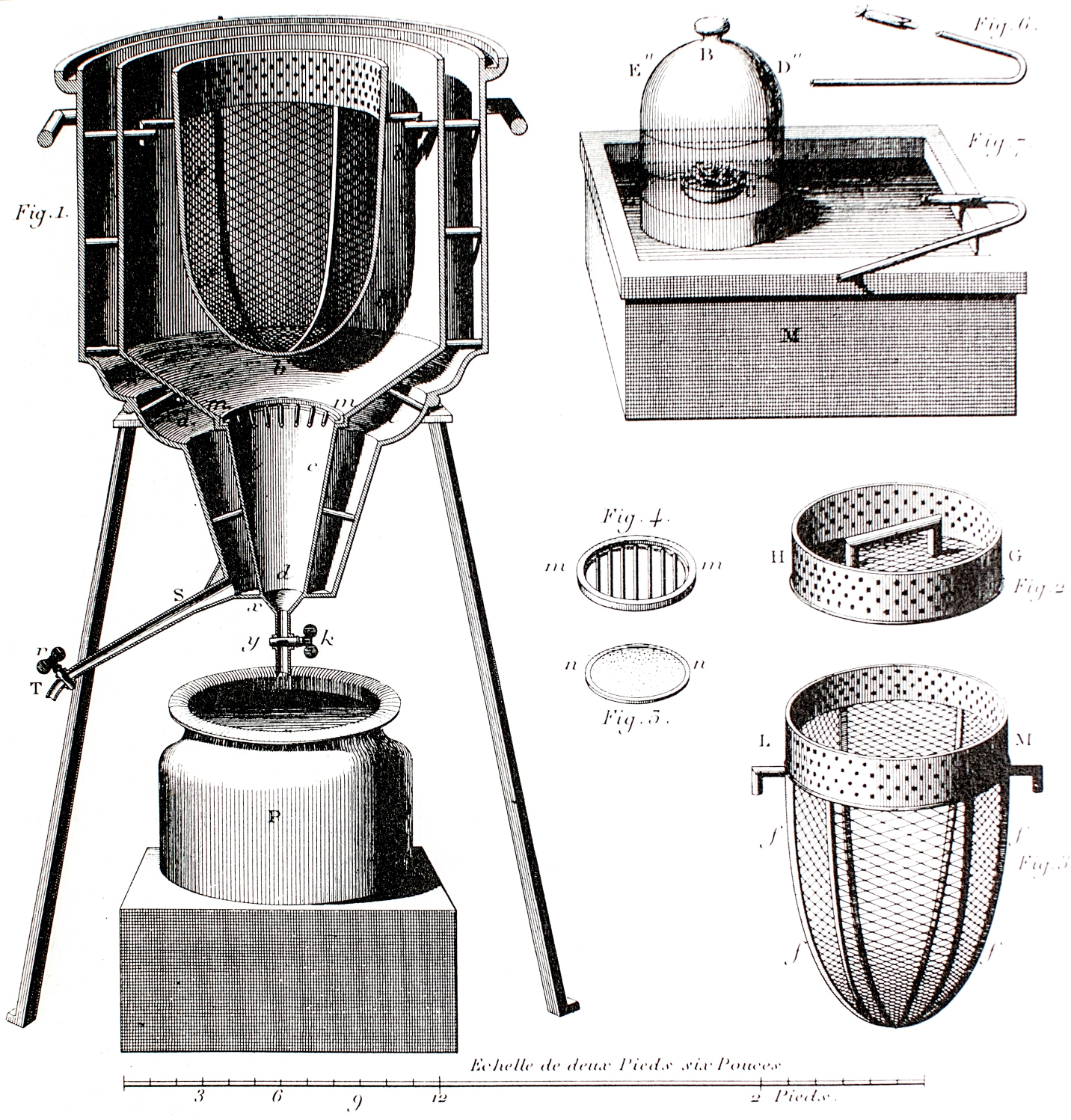}
    \caption{Schematic of Lavoisier and Laplace's 1783 guinea pig experimental setup, the first direct measurement of the metabolic rate $\dot{Q}$ of a living organism. Left: the calorimeter where the animal was placed in a wire basket surrounded by ice. The water from the melted ice, collected in the vessel below, was used to estimate $\dot{Q}$. Bottom right: detail of the wire basket. Top right: the bell jar apparatus used for the follow-up respirometry measurements. After spending up to ten hours in the calorimeter, the guinea pig was placed for the same duration inside the bell jar in a basin containing a shallow pool of mercury. Gas volume changes due to oxygen consumption and carbon dioxide production could be measured by tracking the height of the mercury within the jar. Image credit: {\it Oeuvres de Lavoisier}, vol. II, plate II (public domain).}\label{lavoisier}
\end{figure}

Attempts to quantify metabolic rates of animals date back to the very beginnings of modern chemistry and thermodynamics~\cite{holmes1985lavoisier,hulbert2004basal,hulbert2014sceptics}. The first $\dot{Q}$ measurement was for a guinea pig in 1783, among the first direct measurements of heat dissipation in any physical system. This was part of a series of landmark experiments carried out by Antoine Lavoisier and Pierre-Simon Laplace using their newly invented ice calorimeter, depicted in Fig.~\ref{lavoisier}.  The test subject was placed inside a wire basket surrounded by ice within an insulated vessel for up to ten hours, and $\dot{Q}$ was estimated from the quantity of ice melted per unit time. The same guinea pig was then moved to an air chamber---formed by a bell jar immersed in a basin of mercury---for a similar number of hours. This enabled quantifying respiratory consumption and production rates, $\Phi_{\text{O}_2}$ and $\Phi_{\text{CO}_2}$, using a combination of gas volume changes (differences in the height of mercury) and weighing the amount of CO$_2$ absorbed by a caustic alkali solution. Needless to say, these marathon sessions were not always tolerated well by the guinea pigs. Lavoisier reported that one suffered ``malaise derived from the fact that it was half immersed in the mercury. It died a half hour after it was taken out of the bell jar''~\cite{holmes1985lavoisier}.

Lavoisier and Laplace's experimental protocol simultaneously gave birth to both direct and indirect calorimetry, whose modern descendants we described in Sec.~\ref{exper_approaches}. They were the first to empirically demonstrate the revolutionary implications of Eq.~\eqref{qdotmacrochemical2}: that $\dot{Q}$ solely depends on the macrochemical reactions describing the net transformation and exchange of chemical species, regardless of the complexity of what is happening inside the system. They did this by showing that $\dot{Q}/\Phi_{\text{CO}_2}$ was similar for a guinea pig and burning charcoal, allowing Lavoisier to declare that animal respiration is just a slow type of combustion. The experimental ingenuity and attention to detail needed to achieve this result are remarkable, even by modern standards. If one converts from French pre-metric units\footnote{Lavoisier reported that on average a guinea pig would produce 224 {\it grains} of CO$_2$ and melt 13 {\it onces} of ice over 10 hours~\cite{holmes1985lavoisier}. Using the Parisian pre-metric to metric conversion factors of 1 {\it once} = 576 {\it grains} and 1 {\it grain} = 0.053 g~\cite{encyclopedie1842} this corresponds to 12 g = 0.27 mol of CO$_2$ and 397 g of ice. The mean CO$_2$ rate (negative because it is produced rather than consumed) is $\Phi_{\text{CO}_2} = -7.5\times 10^{-6}$ mol/s. Given the latent heat of fusion of water (333.55 J/g), the amount of melted ice implies a total heat release of 132 kJ over the experiment, at a mean rate of $\dot{Q} = 3.68$ W. We thus get a value of $\dot{Q}/\Phi_{\text{CO}_2} = -491$ kJ/mol, which can be converted to $\dot{Q}/\Phi_{\text{O}_2}$ as described in the text.}, and uses the measured respiratory ratio $\Phi_{\text{CO}_2}/\Phi_{\text{O}_2} = -0.95$ for guinea pigs at rest~\cite{yilmaz2008assessing}, Lavoisier's reported value for $\dot{Q}/\Phi_{\text{CO}_2}$ is equivalent to $\dot{Q}/\Phi_{\text{O}_2} \approx 466$ kJ/mol. This is strikingly similar to the $\approx 450$ kJ/mol value of Thornton's rule, described in Sec.~\ref{thermo}, though it predates Thornton's result by more than 130 years.

Respirometry advanced rapidly after this period, in forms that became less excruciating for the test subjects. Lavoisier himself supervised the first measurements of $\Phi_{\text{O}_2}$ in a human in 1790, revealing that metabolic rates depend on activity levels, digestion, and temperature~\cite{holmes1985lavoisier}. The 19th century saw a proliferation of $\dot{Q}$ estimates in animals using respirometry---mammals, birds, amphibians, and reptiles---as well as the first scaling hypothesis for $\dot{Q}$~\cite{hulbert2014sceptics,schmidt1984scaling}. Sarrus and Rameaux posited on theoretical grounds in 1838 that for endothermic animals $\dot{Q} \sim A$, the surface area~\cite{sarrus1839application}. Their plausible argument was that since heat was dissipated through the surface, the production of heat should scale with $A$ to compensate for the loss. Because volume $V \sim A^{3/2}$ and $M \sim V$, this predicts that $\dot{Q} \sim M^{2/3}$. After $\dot{Q}$ measurements in dogs and rabbits~\cite{rubner1883einfluss,richet1889chaleur} seemed to fit the $\dot{Q} \sim A$ scaling, the Sarrus-Rameaux ``surface law'' became increasingly accepted as fact. By the turn of the 20th century, Voit compiled a survey of metabolic data in a variety of species, claiming a univeral ratio $\dot{Q} / A \approx 50$ W/m$^2$~\cite{voit1901grosse}.

It was against this surface law orthodoxy that Max Kleiber pushed back in his classic 1932 paper~\cite{kleiber1932body}, stating in his introduction that the ``law is not at all so clear today as it appeared to its early discoverers.'' What had occurred in the intervening years, enabling Kleiber's analysis, was a growing consensus on the definition and protocol for measuring the basal metabolic rate $B$ in birds and mammals.  This was spurred in part by the increasing sophistication of agricultural science, and the need for precise comparisons among farm animals. Kleiber himself was a researcher in the Animal Husbandry department of the institution that later became UC Davis. By ensuring that measurements obeyed certain criteria to count as basal (described in detail in the next section) many of the confounding factors that influence $\dot{Q}$ could be controlled for. The result was a set of 13 $B$ vs. $M$ data points---ranging four orders of magnitude in mass from a dove to steer---that Kleiber found could be best-fit by a power law with exponent 0.74. Inspired by Kleiber's work, Brody and Procter published a larger analysis of over 50 species later the same year~\cite{brody1932relation}. They found exponents of 0.64 for birds and 0.734 for mammals, values which are nearly identical to the best-fit exponents (0.66 and 0.73 respectively) calculated from more comprehensive modern surveys, as shown in Fig.~\ref{interspecies}. Despite these early indications that scaling could vary between different classes of organisms, Kleiber in the years that followed popularized a simplified universal exponent of 3/4, which became his eponymous ``law''. As he explained in his monograph {\it The fire of life}~\cite{kleiber1961fire}, ``there is not much point in discussing whether the 0.734 power or the three-quarter power of body weight fits the metabolic results better''. He preferred 3/4 in part because it was easy to calculate $M^{3/4}$ on a slide rule, without referring to a table of logarithms---a practical measure in the pre-calculator era when estimating farm animal metabolism to determine dosages of drugs or dietary supplements~\cite{hulbert2014sceptics}.

The idea of Kleiber's law as more than just an agricultural convenience, but rather something fundamental and applicable across biological scales, found support in the work of Hemmingsen~\cite{hemmingsen1960energy}. By correcting for the thermal dependence of metabolism in ecotherms and unicellular organisms (as described in Sec.~\ref{defb}), he mapped experimental measurements for these classes to a reference temperature of 20$^\circ$C. A quasi-universal relation $B \sim c M^{3/4}$ provided a reasonable fit to endotherms, ectotherms, and unicellular species, but requiring a different (and progressively smaller) prefactor $c$ for each of the three categories. A similar feature is still present in recent analyses~\cite{delong2010shifts,hatton2019linking,hoehler2023metabolic}, though a more general scaling form $c M^\alpha$ is used, with both $c$ and $\alpha$ varying between groups (see Fig.~\ref{interspecies} for an example).

While using different scaling forms for different groups adds more flexibility and provides better fits to experimental data, it opens up questions of which groups to choose. For example does one use the broad categories of Fig.~\ref{interspecies} (mammals, birds, plants, ectotherms, etc.) or subdivide these into smaller classes? Critics of Kleiber's law, particularly those who favor the older 2/3 surface law, have used this ambiguity to cast doubt on the universality of the 3/4 exponent, even for a single category like mammals. \cite{heusner1982energy} argued that it is better to split mammals into groups of related species (i.e. cattle, sheep, rats, mice), described by  $B \approx c M^{2/3}$ relations with different $c$ for each group. However subdividing at this fine level makes any power scaling difficult to justify, since the species in each group do not differ by more than an order of magnitude in their adult masses. \cite{dodds2001re} showed that smaller mammals (with masses $<$ 10 kg) seem to fit the 2/3 law, while deviations at larger masses could reflect either a true biological difference or be an artifact of scarcer data in that range. Fitting the mammal data set in Fig.~\ref{interspecies}(a) with the restriction $M < 10$ kg gives an exponent of 0.66, consistent with this claim. Other critics have focused on the conditions in which the data are gathered and how it is subsequently processed. For example most modern datasets, like those of \cite{makarieva2008mean} and \cite{hoehler2023metabolic}, do not normalize temperatures for endotherms, as is typically done for ectotherms since Hemmingsen. Instead they use measurements at thermoneutrality (described in the next section), which require different temperatures for different species. In contrast, \cite{white2003mammalian} opted to map mammal data to a mean body temperature of 36.2$^\circ$C, and used $M$ and $B$ points that are averages over entire mammalian orders, rather than individual species. This led to a scaling that was closer to the 2/3 exponent. Alternatively, one can reject pure power law scaling altogether, and look for evidence of curvature in log-log plots of $B$ versus $M$~\cite{kolokotrones2010curvature}, indicative perhaps of cross-over between power laws that become dominant in different $M$ regimes.

Thus, nearly two centuries after the idea of metabolic scaling was introduced, the debate about its details and scope continues. However several conclusions can be safely drawn. Given the noise in the biological data (either real variation or measurement-induced) and questions about which categories of organisms to analyze, a single power law applied across all of biology is too reductive. On the other hand, when focusing on particular groups or scales of organization, metabolic power laws can indeed be useful ``empirical approximations''~\cite{hulbert2014sceptics}, highlighting underlying biological trends whose origins and mechanisms we do not fully understand. In the next sections we delve into the specifics of the measurements, the scaling results, and their implications.

\subsection{Defining and measuring $B$ across different organisms}\label{defb}

\begin{figure}[t]
    \centering\includegraphics[width=\columnwidth]{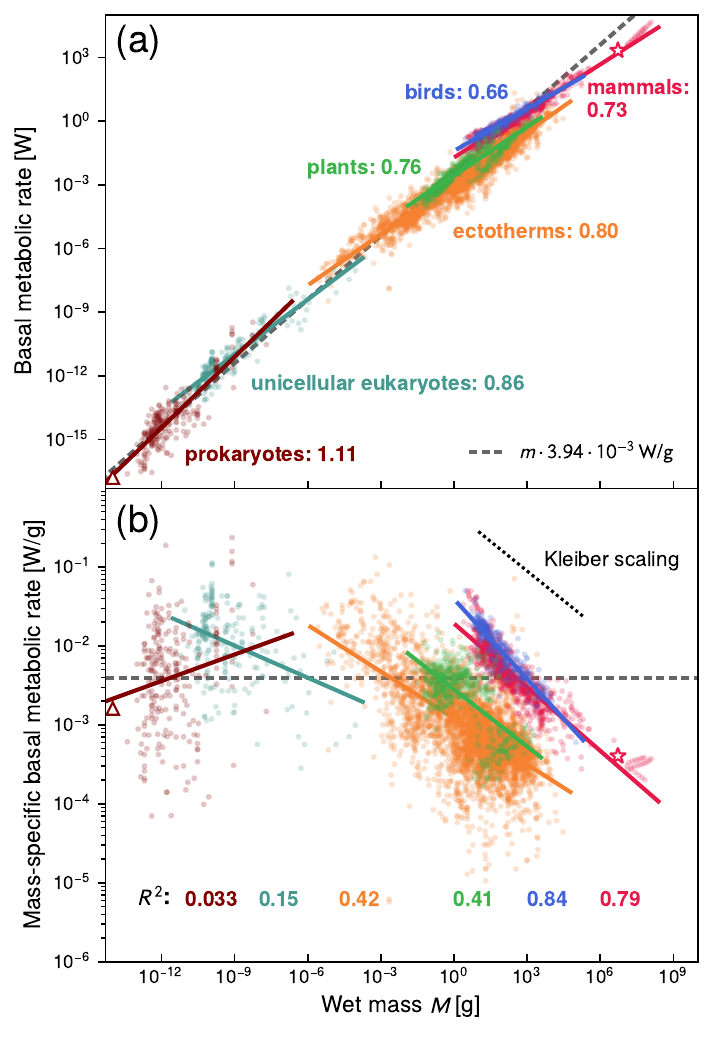}
    \caption{(a) Basal metabolic rate $B$ versus wet mass $M$ for organisms across various domains of life, adapted from the data sets collected in \cite{hoehler2023metabolic}. For each class of organisms we show the best-fit exponent $\alpha$ for the scaling $B \sim M^{\alpha}$ within the class. The dashed line represents a global fit to linear scaling, $B \sim M$. (b) Same as panel (a), except showing $b = B/M$, the mass-specific basal metabolic rate. The dashed line corresponds to the mean mass-specific value, $\bar{b} = 3.94\times 10^{-3}$ W/g. The dotted line represents Kleiber law scaling, $B \sim M^{3/4}$, or equivalently $b \sim M^{-1/4}$. $R^2$ values are shown from linear regression on the log-log plotted $b$ data (solid lines), color coded by class. In both panels the star represents the largest organism for which $B$ was directly measured (an orca), while the triangle represents the smallest organism (the bacterium {\it F. tularensis}).
    }\label{interspecies}
\end{figure}

In the decades since Kleiber's first study~\cite{kleiber1932body}, researchers have surveyed the relationship between $B$ and $M$ across a remarkable range of organismal scales. Fig.~\ref{interspecies}(a) shows part of a data set ($7095$ points) taken from \cite{hoehler2023metabolic}, which augmented earlier surveys such as \cite{makarieva2008mean,savage2004predominance,rubalcaba2020oxygen}. For $B$ data across 22 orders of magnitude of mass to be meaningfully compared on the same plot, it is worth first discussing some of the technical challenges in defining and measuring a ``basal'' metabolic rate among such a diverse set of organisms. The vast majority of existing $B$ data in the literature (including at least 85\% of the points in Fig.~\ref{interspecies}) are from respirometry measurements, typically using OCR to estimate $B$ via Thornton's rule. Direct calorimetry (such as recently used in \cite{thommen2019body} to validate Kleiber's law in planarians) is far less common, though in principle it provides more reliable measurements of $B$. Where comparisons have been made, the differences between direct and indirect methods range from a couple of percent for an ectotherm~\cite{walsberg2006using} up to $\sim 40\%$ for mammals/birds~\cite{walsberg2005direct}. These errors are significant in some contexts (like tracking of metabolism in individual animals over time) but not large enough to substantially affect scaling in broad inter-species comparisons such as Fig.~\ref{interspecies}.

To qualify as a ``basal'' metabolic rate, different criteria are used among different classes of organisms. For animals~\cite{mcnab1997utility,makarieva2008mean}, the ideal $B$ measurement is done for an individual that is an adult (to minimize energetic expenditures on growth), non-moving, and fasting (or ``post-absorptive'').  The latter condition entails that a certain time interval (that varies with species) has passed since the last meal, to ensure the body is relying on stored energy reserves. The organism should also not be in gestation, since a significant portion of the metabolism is then diverted to reproductive processes. For endothermic animals like mammals and birds (red and blue points in Fig.~\ref{interspecies}, respectively), there is an additional criterion of thermoneutrality: the measurement should be done at the range of temperatures where the metabolic rate is near minimum. Below this range endotherms need to expend excess energy to maintain core body temperature, while above it they expend energy to facilitate heat loss to the environment (i.e. via water loss and other active cooling processes)~\cite{porter2009size}. Further details about the thermoneutral range can be found in Sec.~\ref{thermoregulation}.

Fulfilling all these criteria for a particular organism may require some experimental ingenuity, particularly in cases like aquatic animals where the ``non-moving'' state could be hard to achieve. This is illustrated nicely by one study~\cite{worthy2013basal} which carried out a $B$ measurement for one of the largest organisms in the Fig.~\ref{interspecies} data set: an orca whale, marked by a red star in the figure at $M = 5\times 10^6$ g. (All data points with larger $M$ are also for cetaceans, like the blue whale at the extremum, but these are estimated rather than measured $B$ values, extrapolated from smaller species~\cite{lockyer1981growth}.) The orca in question was a single adult male who had been trained over a several month period to be able to rest under a hood placed directly over its blowhole for the duration (up to 30 min) of the respirometry measurement. The orca had to fast overnight before the measurement, and be kept at water temperatures (13$^\circ$C) presumed to be thermoneutral for the species.

For ectotherms, on the other hand, there is no concept of thermoneutrality, since metabolic rate is typically a monotonically increasing function of ambient temperature~\cite{nespolo2011using}. The basal metabolic rate, which is often called the standard metabolic rate in the ectothermic context~\cite{chabot2016determination}, has to come with a temperature specification. Comparing ectothermic rates across species on the same graph (i.e. the orange points in Fig.~\ref{interspecies}) thus only makes sense if all the values correspond to the same temperature. Unfortunately measurements are carried out at a range of temperatures, and require some way of normalizing to a single reference temperature, which is chosen to be 25$^\circ$C for the ectothermic points in Fig.~\ref{interspecies}~\cite{makarieva2008mean}.  There are two main phenomenological approaches to quantify the metabolic temperature dependence $B(T)$, inspired by the functional forms used to fit the variation of chemical reaction rates with temperature in physical chemistry~\cite{white2006scaling}.

The first approach assumes Arrhenius-like behavior of $B(T)$~\cite{gillooly2001effects} to relate metabolic rates at the reference and measurement temperatures $T_\text{ref}$ and $T_\text{meas}$,
\begin{equation}\label{arr}
B(T_\text{ref}) = B(T_\text{meas}) e^{-E_b/(k_B T_\text{ref})+E_b/(k_B T_\text{meas})},
\end{equation}
where $k_B$ is the Boltzmann constant, temperature is measured in K, and there is some phenomenological ``barrier energy'' parameter $E_b$. The motivation of Eq.~\eqref{arr} comes from treating the rate $\Phi_{\text{O}_2}$ in Eq.~\eqref{qox}, $\dot{Q} = -\Phi_{\text{O}_2} \Delta h_\text{ox}$, derived from a macrochemical net reaction, as if it was an ordinary chemical reaction rate, with an Arrhenius dependence $\Phi_{\text{O}_2} \propto \exp(-E_b/(k_B T))$. Since the temperature dependence of $\Delta h_\text{ox}$ is presumed to be far more gradual over the biologically relevant temperature range (and thus $\Delta h_\text{ox}$ can be approximated as a constant), this leads to the exponential relationship in Eq.~\eqref{arr}, assuming  basal conditions where $\dot{Q} = B$. The analysis in \cite{gillooly2001effects} claimed that $E_b$ is roughly a universal constant among different domains of life, approximately $E_b \approx 60$ kJ/mol, reflecting an effective activation energy scale shared by organisms with similar underlying biochemistries. This energy scale is comparable to that of ATP hydrolysis under physiological conditions, $\approx 50-70$ kJ/mol~\cite{milo2010bionumbers}.

The second approach, more commonly used among biologists, is expressed in terms of a van 't Hoff $Q_{10}$ coefficient,
\begin{equation}\label{q10}
B(T_\text{ref}) = B(T_\text{meas}) Q_{10}^{(T_\text{ref}-T_\text{meas})/10\:\text{K}}.
\end{equation}
$Q_{10}$ quantifies the factor by which $B(T)$ increases for every 10 K increase in ambient temperature. This is the approach used to normalize the ectothermic data in Fig.~\ref{interspecies}, with slightly different values of $Q_{10}$ in the range 1.4--2.5 chosen for different groups (reptiles, fish, amphibians, etc.), based on best-fits to experimental data~\cite{white2006scaling,makarieva2008mean}. Under the typical case where $|T_\text{ref} - T_\text{meas}| \ll T_\text{ref}$, the two approaches in Eqs.~\eqref{arr} and \eqref{q10} can be mapped onto each other, with
\begin{equation}
    Q_{10} \approx e^{\frac{E_b \cdot 10\:\text{K}}{k_B T_\text{ref}^2}}.
\end{equation}
For $T_\text{ref} = 298$ K and $E_b = 60$ kJ/mol this yields $Q_{10} = 2.3$, compatible with the experimental fitting range.

Though neither of the above functional forms for $B(T_\text{ref})$ are strictly derived from first principles, they can reasonably describe experimental data over the limited temperature ranges available for measurement. Normalization using either approach yields similar scaling behaviors when doing interspecies comparisons~\cite{white2006scaling}.

The entire conceptual framework of basal metabolic rates can be extended to non-animal groups as well~\cite{makarieva2008mean}. For photosynthetic cases (most plants, cyanobacteria, algae) this requires measuring metabolic rates in the dark, when the organism is relying on respiration and internal energy stores. For larger plants, like trees, there are typically size constraints on the measurements, which means either using smaller stages of growth (seedlings or saplings) or measuring rates in constituent parts like leaves. Finally, for unicellular organisms $B$ is usually taken to be the endogenous metabolic rate, defined as $\dot{Q}$ for a non-growing cell in a nutrient-free medium. All non-animal groups in Fig.~\ref{interspecies} (plants, unicellular eukaryotes, prokaryotes) have their metabolic rates normalized to 25$^\circ$C using Eq.~\eqref{q10}.

\subsection{Interspecies scaling relationships}\label{inter}

A survey of $B$ data across the spectrum of life, as seen in Fig.~\ref{interspecies}, reveals a variety of interesting relationships depending on the scale and group of interest. Despite the historical focus on sublinear scaling relationships of $B$ versus $M$ in individual groups, there is a striking global trend that is worth discussing first: as seen in Fig.~\ref{interspecies}(a), a linear fit $B \sim M$ roughly captures the behavior of $B$ across 22 orders of magnitude in mass. This is also reflected in Fig.~\ref{interspecies}(b), which shows the same data points but in terms of mass-specific basal metabolic rate $b = B/M$. As noted in earlier studies~\cite{makarieva2008mean,hoehler2023metabolic}, the results cluster remarkably close to the mean value, estimated as $\bar{b} = 3.94 \times 10^{-3}$ W/g for the organisms surveyed in the figure. In fact, 86\% of these organisms have $b$ values within an order of magnitude of the mean. This includes both the smallest directly measured organism in the data set (the bacterium {\it Francisella tularensis}, just 0.2--0.7 $\mu$m in diameter~\cite{jones2005infectious}, marked as a triangle in the figure) and the largest one (the orca, marked by a star). In Sec.~\ref{univ} below we will explore one consequence of the global linear scaling: a universal metabolic time scale that can be expressed in terms of an ATP turnover rate.

If one partitions the data into classes (mammals, birds, ectotherms, plants, unicellular eukaryotes, and prokaryotes, highlighted by distinctly colored points) one can fit power law relationships $B \sim M^\alpha$ to each class. The resulting best-fit exponents $\alpha$ are shown next to each class in Fig.~\ref{interspecies}(a). Despite the accumulation of data over the decades, there is broad historical consistency: as mentioned earlier, the mammal (0.73) and bird (0.66) exponents are very close to those of \cite{brody1932relation}, while the ectotherm (0.8) and unicellular eukaryote (0.86) exponents are slightly higher than those first reported by \cite{hemmingsen1960energy} for those classes (0.74 and 0.77 respectively). Given the scatter in the metabolic data, one might reasonably ask whether these class-based best-fit exponents have any explanatory power over and beyond the universal $B \sim M$ trend. One simple way to answer this question is to factor out this global trend, by looking at $b$ versus $M$ as in Fig.~\ref{interspecies}(b). The lines in this panel correspond to a scaling $b \sim M^{\alpha -1}$ for each class, so Kleiber's law would be a slope of $-1/4$. We can quantify what fraction of the variance in $b$ is explained by the power-law scaling in each class via the coefficient of determination, $R^2$, calculated from linear regression to the log-log plotted data points. The $R^2$ values for the different classes are shown at the bottom of the figure. Mammals and birds have the highest $R^2$ (0.79 and 0.84 respectively), reflecting the robust nature of the sublinear scaling laws for these groups. The data for ectotherms and plants is far noisier ($R^2 = 0.42$ and $0.41$). On the other end of the spectrum, prokaryotes and unicellular eukaryotes have $R^2 = 0.033$ and 0.15, indicating that the concept of separate power laws in these cases (outside of the global $B \sim M$) is tenuous. Unsurprisingly, their fitted $\alpha$ values are closest to one (1.11 for prokaryotes, 0.86 for unicellular eukaryotes), but we  should be wary of assigning significance to the difference of these exponents from 1.

\subsection{Universal metabolic time scale and bounds on mass-specific metabolism}\label{univ}

This global trend of $B \sim M$, from the smallest bacteria to the largest animals, allows us to extract a universal metabolic time scale. To derive this scale, we start with the basic assumption of resting metabolic rate experiments, $B \approx \dot{Q}$, and use Eq.~\eqref{qox}, which relates $\dot{Q}$ to the oxygen consumption rate $\Phi_{{\rm O}_2}$. We can thus express the linear scaling of $B$ as
\begin{equation}\label{u1}
   B \approx \bar{b} M \approx \dot{Q} = -\Phi_{{\rm O}_2}\Delta h_{\rm ox},
\end{equation}
valid for large-scale variation in $M$. Here $\Delta h_{\rm ox}=-450$ kJ/mol from Thornton's rule. From Eq.~\eqref{u1} we can extract a mass-specific oxygen consumption rate,
\begin{equation}
    \label{u3}
    \frac{\Phi_{{\rm O}_2}}{M} = -\frac{\bar{b}}{\Delta h_{\rm ox}} = 5.3 \times 10^{15}\:\text{s}^{-1}\text{g}^{-1}.
\end{equation}
For typical humans with $M = 60$ kg, this predicts $3 \times 10^{20}$ oxygen molecules per second, which is the correct order of magnitude: the true number is closer to $10^{20}$ s$^{-1}$~\cite{flamholz2014quantified}, since the $b$ value for humans (along with other larger mammals) is lower than $\bar{b}$, as seen in Fig.~\ref{interspecies}(b).

Using glucose respiration as a reference, we can express Eq.~\eqref{u3} in terms of ATP turnover. We assume a steady state concentration of ATP at a typical scale of $\sim 1$ mM, characteristic of cells in a variety of organisms~\cite{milo2010bionumbers}. Maintenance of this concentration requires consumption to be balanced by production, and since approximately 5 ATP are produced for every O$_2$, we have the rate of ATP consumption $\Phi_\text{ATP} \approx 5 \Phi_{{\rm O}_2}$. The 1 mM ATP concentration can be translated to the number of ATP molecules per unit mass, $N_\text{ATP}/M = 5.5 \times 10^{17}$ g$^{-1}$, where we have used a typical cell density of 1.1 g/mL~\cite{milo2010bionumbers}. We thus get:
\begin{equation}
    \label{u4}
    \frac{\Phi_\text{ATP}}{N_\text{ATP}} = \frac{5 \Phi_{{\rm O}_2}}{M} \frac{M}{N_\text{ATP}} = 4.8 \times 10^{-2}\:\text{s}^{-1},
\end{equation}
which corresponds to an average ATP lifetime of about 21 seconds. This gives a fundamental metabolic time scale of ATP turnover that roughly holds across all domains of life. Given the molecular mass of ATP, $M_\text{ATP} = 8.4 \times 10^{-22}$ g, for humans this corresponds to a turnover of $\sim 10^2$ kg of ATP per day, similar to our total body weight.

It is interesting to compare the ATP turnover rate from Eq.~\eqref{u4} to two extreme scenarios. Consider a ``minimal'' metabolism, where ATP production is just large enough to compensate for spontaneous ATP hydrolysis. This limit is biologically unrealistic (since no ATPase enzymes are present to catalyze hydrolysis), but it can serve as a lower bound. Using a spontaneous ATP hydrolysis rate value extrapolated from experiments, $r^\text{spont}_\text{ATP}/N_\text{ATP} = 1.7 \times 10^{-8}$ s$^{-1}$, corresponding to pH 7 and 25$^\circ$C~\cite{moeller2022hydrolysis}, we get a minimal $b$ value of $b_\text{min} = 1.4 \times 10^{-9}$ W/g. Comparing $\bar{b}$ to $b_\text{min}$, the six orders of magnitude speed-up of typical biological systems reflects the efficacy of enzymes. On the other end of the scale, we can imagine another unrealistic scenario as an upper bound: all ATPase enzymes match the rate $r^\text{max}_\text{ATP}/N_\text{ATP} = 390$ s$^{-1}$ achieved by the fastest known motor protein, myosin XI from the alga {\it Chara corallina}~\cite{ito2007kinetic}. This would correspond a maximal value $b_\text{max} = 33$ W/g. All the $b$ values in Fig.~\ref{interspecies}(b) easily fall between these two limits.

\subsection{Intraspecies metabolic scaling}\label{intra}

\begin{figure}[t]
    \centering\includegraphics[width=\columnwidth]{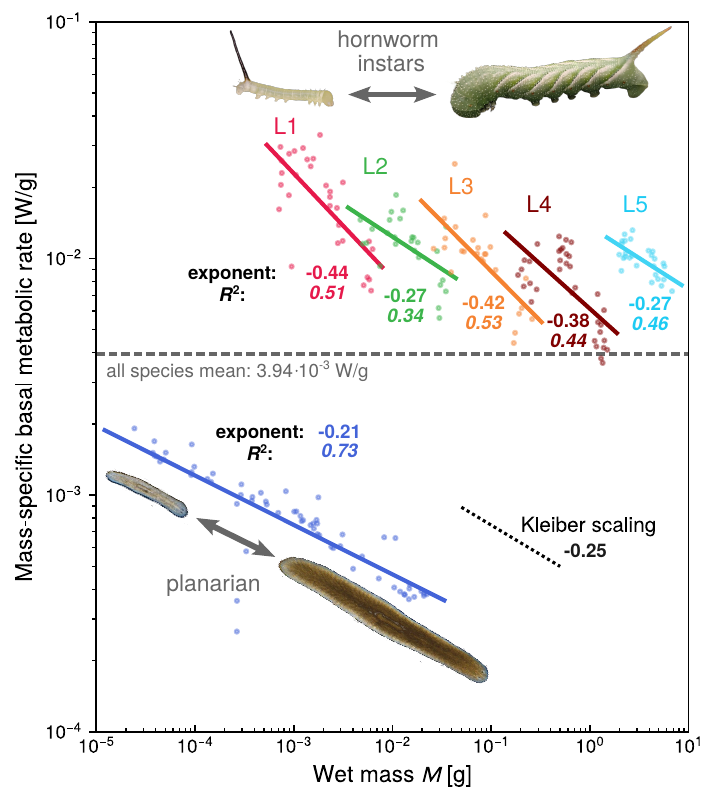}
    \caption{Examples of mass-specific basal metabolic rates $b=B/M$ versus $M$ within a single species. Top: five instars (developmental stages) L1 to L5 of the tobacco hornworm, the larva of the moth {\it Manduca sexta}~\cite{sears2012ontogenetic}. For each instar, the best fit exponent $\gamma$ for the scaling $b \sim M^\gamma$ is shown, along with the $R^2$ value for linear regression on the log-log plot. Kleiber scaling corresponds to $\gamma = -1/4$. The dashed line indicates the mean $\bar{b}$ for all species, taken from Fig.~\ref{interspecies}(b). Bottom: the planarian {\it Schmidtea mediterranea} for a range of body sizes induced by feeding or starvation~\cite{thommen2019body}. In both species, the data was converted from dry mass to wet mass as described in the references, and for the hornworm case the $B$ was converted from units of $\mu$L CO$_2$ hr$^{-1}$ to W using the measured {\it M. sexta} respiratory ratio $\Phi_{\text{CO}_2}/\Phi_{\text{O}_2} = -0.88$~\cite{alleyne1997effects} and the Thornton's rule value $\Delta h_{\rm ox} = -450$ kJ/mol.}\label{intraspecies}
\end{figure}

Though the modern consensus for defining the basal metabolic rate requires adult organisms, it is interesting to relax this assumption and explore how $B$ scales across sizes within a single species~\cite{glazier2005beyond}---historically the first kind of scaling to be systematically studied, as in the early measurements on dogs and rabbits~\cite{rubner1883einfluss,richet1889chaleur}. In order to reliably identify power-law scaling, we need $M$ to vary by orders of magnitude, which typically requires measurements at different developmental stages. Here we focus in particular on post-embryonic development (for example larval / juvenile forms), leaving the discussion of metabolic changes during embryogenesis to Sec.~\ref{sec_embryo}.

Fig.~\ref{intraspecies} shows two representative examples of intraspecies scaling, expressed in terms of the mass-specific rate $b$ versus $M$, which illustrate different limits of complexity. On the bottom is data from the planarian {\it Schmidtea mediterranea}~\cite{thommen2019body}, which exhibits single power-law scaling across three decades of mass, with an exponent very close to the Kleiber mass-specific value of $-1/4$. The scaling is robust ($R^2 = 0.73$) and the $b$ values lie close in magnitude to the all-species mean $\bar{b}$ from Fig.~\ref{interspecies}(b) (dashed line). The broad range of $M$ in this case is not due to distinct developmental stages, but to the relatively unique ability of planarians to reversibly and dramatically change body size in response to feeding or starvation (ranging from $\sim$ 0.5 to 20 mm for {\it S. mediterranea}). The simplicity of the scaling---Kleiber's law in miniature---suggests that the underlying metabolic mechanisms here are the same across sizes. For {\it S. mediterranea} the physiological basis of the scaling is in fact better understood than in other organisms: it is caused by size-dependent energy storage scaling that leads to increased mass per cell as $M$ gets larger (while $B$ per cell remains relatively fixed)~\cite{thommen2019body}.

In contrast, when an organism undergoes significant developmental changes during its life history, the scaling can become quite complex. The top of Fig.~\ref{intraspecies} shows results for tobacco hornworm larvae, caterpillars of the moth {\it Manduca sexta}~\cite{sears2012ontogenetic}. These larvae go through five instars---developmental periods between molts---that span a remarkable four decades of mass. Each instar appears to have its own scaling with distinct exponents, staggered relative to one another in a similar way to the scaling of different organism classes in Fig.~\ref{interspecies}(b). However the extent to which these correspond to true power laws is debatable: the mass range within each instar is roughly a single decade, which combined with the noise in the data means the power law fits exhibit lower $R^2$ values of $\sim 0.3 - 0.5$. This has led to an ongoing discussion on whether the scaling actually has multiple developmental phases~\cite{glazier2024multiphasic}, or is better described via a single overall power law with $\alpha \sim 0.9$~\cite{packard2025commentary}. Assuming the apparent discontinuities in $b$ versus $M$ at the instar boundaries are real, $B$ can suddenly jump up while $M$ remains temporarily fixed---a phenomenon also seen during embryonic development (Sec.~\ref{sec_embryo}).

\subsection{Metabolic scaling and organismal growth}\label{sec:growth}

Power-law scaling of metabolic rate within a single species, described in the previous section, has interesting consequences when considered at the level of a single organism. In particular, it can lead to predictions for the growth of the organismal mass $M(t)$ over time $t$~\cite{west2001general}. To understand this connection, let us start with a simplified version of Eq.~\eqref{qdotmacrochemical2}, which relates $\dot{Q}$ to macrochemical reactions:
\begin{equation}
\begin{split}
\dot{Q}&=-\sum_{\beta = 1}^{\tilde{R}} \mathcal{R}_\beta \Delta H_\beta.
\end{split}\label{qdot_simple}
\end{equation}
Here we have assumed a constant temperature $T$ and ignored spatial inhomogeneities in enthalpy that lead to variations $\delta \dot{Q}$.  The $\beta$th macrochemical reaction is characterized by a flux $\mathcal{R}_\beta$ and enthalpy change $\Delta H_\beta$. We restrict ourselves to considering only the basal metabolism, so $\dot{Q} = B$, and the terms on the right do not include contributions from locomotion, feeding, and other non-basal activities~\cite{hou2008energy}. However growth---synthesis of new biomass---has a non-zero contribution even at rest, and may be a significant component of the total $B$ during juvenile stages. To distinguish this from the strict definition of basal metabolism (valid for non-growing adults) the rate in this context is sometimes referred to as the {\it resting} metabolic rate. However for simplicity we will continue to denote it as the basal rate $B$, like we did in Sec.~\ref{intra}.

Let us separate out two important macrochemical reactions from the sum in Eq.~\eqref{qdot_simple}: the net respiration reaction (for example Eq.~\eqref{eq:resp_glyc}) and a net reaction for biomass synthesis, which we will label with subscripts resp and syn respectively. Then Eq.~\eqref{qdot_simple} becomes:
\begin{equation}
    B = -\mathcal{R}_\text{resp}\Delta H_\text{resp} - \mathcal{R}_\text{syn} \Delta H_\text{syn} -\sum_{\beta \in {\cal M}} \mathcal{R}_\beta \Delta H_\beta.\label{qdot_detailed}
\end{equation}
The remaining reactions in the sum on the right form a set ${\cal M}$ we will call the ``maintenance'' reactions, which describe all the other cellular processes involved in the resting metabolism. The reaction rates are related to each other via the balance of ATP production and consumption: in particular if $\gamma_\text{resp}$ is the ATP produced in the respiration reaction, $\gamma_\text{syn}$ is the ATP consumed in the biosynthesis reaction, and $\gamma_\beta$ is the ATP consumed in one of the remaining reactions $\beta \in {\cal M}$, then balanced ATP turnover requires:
\begin{equation}\label{atp_balance}
\gamma_\text{resp} \mathcal{R}_\text{resp} = \gamma_\text{syn} \mathcal{R}_\text{syn} + \sum_{\beta \in {\cal M}} \gamma_\beta \mathcal{R}_\beta.
\end{equation}
Solving for $\mathcal{R}_\text{resp}$ from Eq.~\eqref{atp_balance} and plugging into Eq.~\eqref{qdot_detailed}, we find
\begin{equation}
\begin{split}
    B &= -\left(\Delta H_\text{syn} +\frac{\gamma_\text{syn}}{\gamma_\text{resp}}\Delta H_\text{resp}\right) \mathcal{R}_\text{syn}\\
    &\quad -\sum_{\beta \in {\cal M}} \left(\Delta H_\beta
    + \frac{\gamma_\beta}{\gamma_\text{resp}} \Delta H_\text{resp}\right) \mathcal{R}_\beta.
\end{split}\label{b_split}
\end{equation}
This equation groups the contributions to $B$ into two types: (i) terms involving maintenance reactions whose rates $\mathcal{R}_\beta \propto M$ approximately scale with the overall mass of the organism; (ii) the synthesis contribution proportional to $\mathcal{R}_\text{syn}$, which scales like the mass growth rate $\mathcal{R}_\text{syn} \propto dM/dt$. If we make the time dependence of $M(t)$ and $B(t)$ explicit, we can rewrite the equation for basal metabolism as:
\begin{equation}
    \label{growth}
    B(t) = B_m M(t) + E_m \frac{dM(t)}{dt},
\end{equation}
where $B_m M(t)$ represents all the type (i) terms grouped together (the maintenance contribution) and $E_m dM/dt$ represents the synthesis contribution.

In the simplest version of the model, the two coefficients---$B_m$ [units: W/g], the mass-specific maintenance cost, and $E_m$ [units: J/g], the mass-specific synthesis cost---are taken to be constants. Because the summed  enthalpy change terms in parentheses in Eq.~\eqref{b_split} typically turn out to be negative, the negative signs in front cancel and $B_m$ and $E_m$ in Eq.~\eqref{growth} are positive. If we further assume a single power-law scaling $B(t) = B_0 M^\alpha(t)$ holds for the basal metabolic rate across the entire lifetime of the organism, then we obtain a growth equation that determines the dynamics of $M(t)$,
\begin{equation}
    \label{wbe}
    B_0 M^\alpha(t) = B_m M(t) + E_m \frac{dM(t)}{dt}.
\end{equation}
This growth equation, with $\alpha$ set to 3/4 following Kleiber's law, was first formulated by West, Brown, and Enquist, and is known as the WBE ontogenic growth model~\cite{west2001general}. It was later considered for general $\alpha$ in both biological~\cite{kempes2012growth} and non-biological~\cite{bettencourt2007growth} contexts. In mathematical form, this growth equation has several predecessors in early literature, posited by Murray (with $\alpha = 2/3$) as a growth model for farm animals~\cite{murray1921normal}, and more famously by von Bertalanffy with a general exponent $\alpha$ as well as separate power-law scaling of the maintenance term with mass~\cite{von1957quantitative}. The interpretation of the terms in the equation differed in these earlier models, with the metabolic rate term taken to reflect anabolism (the building up of biomolecules) and the maintenance term identified with catabolism (the consumption of biomolecules that fuels anabolism). In the original WBE derivation, the term on the left of Eq.~\eqref{wbe} was identified as the power input into the resting metabolism, and the equation had an intuitive interpretation as a statement of energy conservation: the power input was expended either on maintenance or growth. However it is worth noting that the experimental evidence for scaling is in the context of dissipated heat $B$, not power input, so starting with Eq.~\eqref{qdot_simple} is more thermodynamically justified. Hence the energy expenditures on the right in Eq.~\eqref{wbe} are dissipated heat costs associated with the maintenance and growth reactions, as was acknowledged in follow-up work~\cite{hou2008energy}. 

Several generalizations of Eq.~\eqref{wbe} are possible~\cite{ilker2019modeling}: the simple power law $B_0 M^\alpha(t)$ may be replaced by a more general functional form $B(M(t))$ that reflects changes in metabolic scaling across developmental stages, like in the hornworm example of Fig.~\ref{intraspecies}(a). There may be analogous complexity in the mass-specific cost coefficients, with $B_m$ and $E_m$ replaced by $B_m(M(t))$ and $E_m(M(t))$. For example maintenance costs per unit mass in juvenile birds are lower than in adult birds, before endothermic heat production has attained its mature levels~\cite{werner2018energy}. Consistent with this idea, \cite{ilker2019modeling} showed that fitting the generalized growth model (with constant $E_m$) to experimental data of $B(M(t))$ and $M(t)$ for turkeys and Japanese quail~\cite{dietz1997effect} yielded a maintenance cost $B_m(M) \sim M^\zeta$ with $\zeta \approx 0.5 - 1$ for birds in the earliest growth stages.

In contrast to a dependence $B_m(M)$, a constant mass-specific synthesis cost $E_m$ is found to be remarkably consistent not just within the lifetime of individual organisms, but across species. It exhibits a universal order of magnitude $E_m \sim {\cal O}(10^3)$ J/g. For example in an analysis of data from prokaryotes and unicellular eukaryotes tabulated in \cite{lynch2015bioenergetic}, the interspecies average is $E_m = 2,600$ J/g~\cite{ilker2019modeling}. At much larger size scales, $E_m = 2,700-9,500$ J/g for embryonic and juvenile mammals, $800-2,800$ J/g for embryonic birds and fishes, and $1,400-7,500$ J/g for juvenile birds~\cite{hou2008energy}. 

One can derive the rough scale of $E_m$ using a simple model of biomass production. Let us assume the biomass macrochemical reaction can be expressed as synthesis of a hypothetical molecule CH$_{x}$O$_{y}$N$_{z}$ representing average composition ratios for C, H, O, N. For example in {\it E. coli} typical values are $x=1.77$, $y=0.49$, $z=0.24$~\cite{milo2010bionumbers}. We denote the corresponding molecular mass as $m_\text{biom} = 12 + x+ 16 y+14z$ g/mol. The summed enthalpy change $\Delta H_\text{syn} + (\gamma_\text{syn}/\gamma_\text{resp}) \Delta H_\text{resp}$ in Eq.~\eqref{b_split} can be interpreted as the enthalpy change for a net reaction that combines both respiration and biomass synthesis. A simple example of such a combined reaction, based on conversion of glucose and ammonia to biomass, would be
\begin{equation}
\begin{split}
    &\mathrm{C_6 H_{12} O_6} + \sigma \mathrm{O_2} + \chi \mathrm{NH_3}\\
    &\qquad\rightleftharpoons \xi \mathrm{CH}_x\mathrm{O}_y \mathrm{N}_z + \epsilon \mathrm{CO_2}+\nu \mathrm{H_2 O}.
\end{split}\label{combo_reaction}
\end{equation}
Here the ATP, ADP, and Pi terms on both sides have canceled, leaving a net reaction for production of $\xi$ molecules of biomass for each molecule of glucose. The remaining constants are related through mass balance as:
\begin{equation}\label{stoich}
\begin{split}
    \sigma &= 6 - \frac{1}{4}\xi(4+x-2y-3z),\\
    \chi &= \xi z, \quad \epsilon = 6-\xi,\\
    \nu&= 6 - \frac{1}{2}\xi(x-3z).
\end{split}
\end{equation}
To estimate $\xi$, one can use information about the typical biomass yield per glucose molecule, expressed as $\alpha = \xi m_\text{biom}/m_\text{gluc}$, where $m_\text{gluc} = 180$ g/mol is the molecular mass of glucose. A representative value for ${\it E. coli}$ is $\alpha = 0.5$~\cite{milo2010bionumbers}, and we can then express $\xi = m_\text{gluc} \alpha/m_\text{biom}$. If the enthalpy change associated with the combined reaction approximately obeys Thornton's rule for the consumption of $\sigma$ O$_2$ molecules, then 
\begin{equation}\label{enth_approx}
    \Delta H_\text{syn} + (\gamma_\text{syn}/\gamma_\text{resp}) \Delta H_\text{resp} \approx \sigma \Delta h_\text{ox},
\end{equation}
where $\Delta h_\text{ox} \approx -450$ kJ/mol. This approximation follows from the fact that Eq.~\eqref{combo_reaction} can be thought of as a partial combustion of glucose (using $\sigma$ oxygen molecules), with the rest of the glucose converted into biomass. Since biomass generally has a similar enthalpic content to the glucose (as supported by the fact that biomass combustion also approximately follows Thornton's rule~\cite{cordier1987relationship,von1993thermodynamic}), the main contribution to the enthalpy change is the partial combustion.

Given the form of the biomass reaction, we can express the overall mass increase as $dM/dt = w \xi m_\text{biom} \mathcal{R}_\text{syn}$, where the factor of $w$ is the ratio of wet-to-dry mass, accounting for the fact that $M$ is wet mass but $m_\text{biom}$ is a unit of dry biomass. For bacteria a typical value is $w = 4.5$~\cite{milo2010bionumbers}. This relation for $dM/dt$ allows us to find an expression for $E_m$ using Eqs.~\eqref{b_split}-\eqref{growth}:
\begin{equation}\label{em}
    E_m = -\frac{\Delta H_\text{syn} + \frac{\gamma_\text{syn}}{\gamma_\text{resp}} \Delta H_\text{resp}}{w\xi m_\text{biom}}.
\end{equation}
Combining this with Eqs.~\eqref{stoich} and \eqref{enth_approx} and using $\xi m_\text{biom} = \alpha m_\text{gluc}$, we finally get
\begin{equation}
\begin{split}
    E_m &\approx-\left(\frac{1}{30\alpha} - \frac{4+x-2y-3z}{4(12+x+16y+14 z)}\right)\frac{w^{-1}\Delta h_\text{ox}}{\text{g/mol}}\\
    &= 2,600\:\text{J/g},
\end{split}
\end{equation}
with the second line using the parameter values estimated for ${\it E. coli}$. The 
result is consistent with the observed $E_m$ scales quoted above.

Returning to Eq.~\eqref{wbe}, there are a variety of interesting properties of the growth curves $M(t)$ that are solutions of the WBE growth model with fixed $B_m$ and $E_m$. For $\alpha <1$ the curve $M(t)$ tends toward an asymptotic value $M(t) \to M_a$ as $t \to \infty$, where the growth rate vanishes, $dM/dt \to 0$. This naturally accounts for the existence of a saturating ``adult'' mass for organisms, with
\begin{equation}
    M_a = \left(\frac{B_0}{B_m}\right)^{\frac{1}{1-\alpha}}.
\end{equation}
The saturating behavior no longer exists for $\alpha \ge 1$, and in the special case of $\alpha = 1$ we have pure exponential growth for $M(t)$. Notably prokaryotes have $\alpha \approx 1$, as seen in Fig.~\ref{interspecies}, which is consistent with roughly exponential growth behavior. In this case the largest organism size is not defined by an asymptotic limit, but by the criteria for cell division, i.e. if the mass has doubled~\cite{kempes2012growth}.

\begin{figure}[h!]
    \centering\includegraphics[width=\columnwidth]{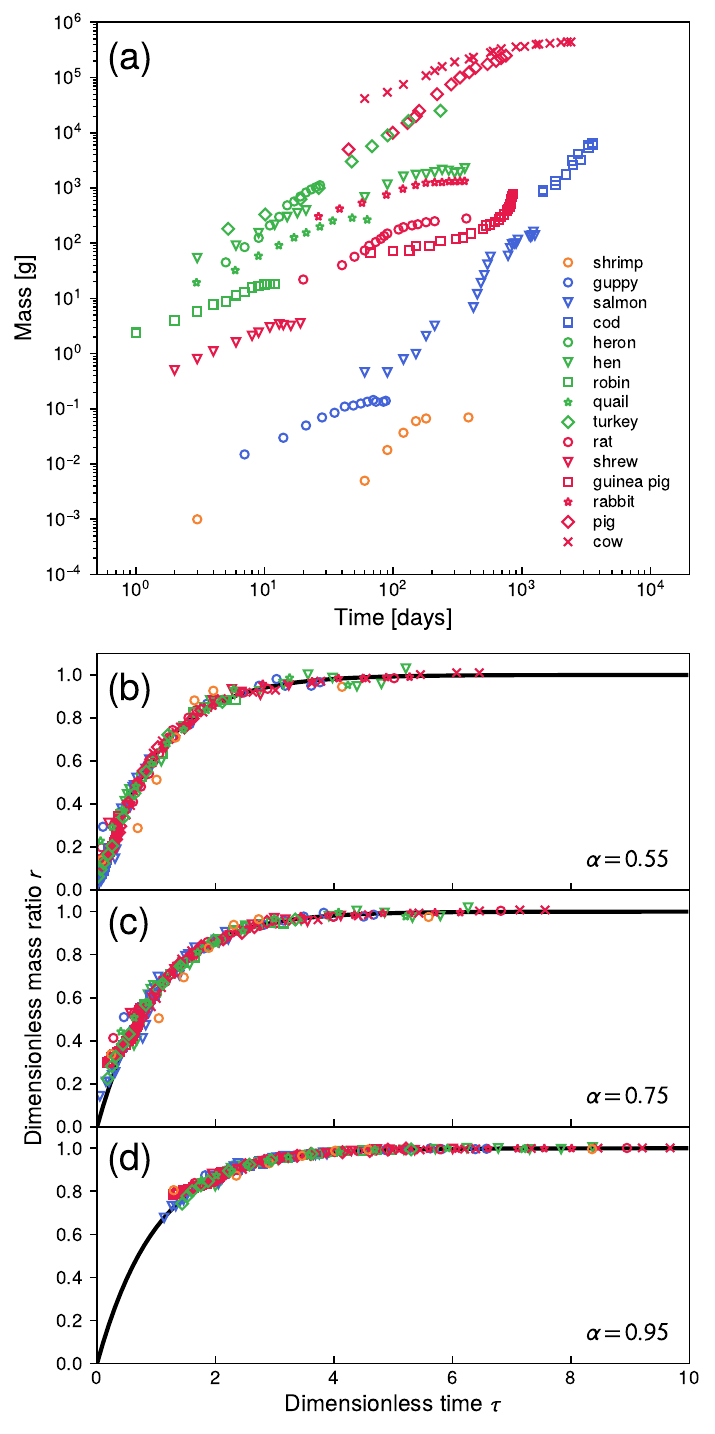}
    \caption{(a) Measured data points for mass $M(t)$ versus time since birth $t$ for a variety of animals. All measurements are taken from \cite{west2001general} except for quail and turkey \cite{dietz1997effect}. (b-d) The same trajectories, but now expressed in terms of the dimensionless mass ratio $r$ and dimensionless time $\tau$ defined in Eq.~\eqref{dimen}. The three different panels correspond to three different choices for the metabolic scaling exponent: $\alpha = 0.55$, $0.75$, $0.95$. The black curve is the universal growth curve in Eq.~\eqref{coll}.  The collapse of the data onto this universal curve is not particularly sensitive to the value of $\alpha$.}\label{animal_growth}
\end{figure}

For $\alpha <1$ we can analytically solve for $M(t)$ from Eq.~\eqref{wbe} and rewrite the growth trajectory elegantly as
\begin{equation}\label{coll}
    r(\tau) = 1-e^{-\tau},
\end{equation}
via the dimensionless mass ratio $r$ and dimensionless time $\tau$,
\begin{equation}\label{dimen}
\begin{split}
    r &= \left(\frac{M}{M_a}\right)^{1-\alpha},\\
    \tau &= \frac{B_0}{E_m}(1-\alpha)M_a^{\alpha-1} t-\ln \left[1-\left(\frac{M_0}{M_a}\right)^{1-\alpha}\right].
\end{split}
\end{equation}
Here $M_0 = M(0)$ is the mass at birth, $t=0$. A striking consequence of this, first shown by \cite{west2001general} for the Kleiber's value $\alpha=3/4$, is that $M(t)$ curves from very different organisms (from a cow to a guppy) can all be collapsed onto a universal growth curve defined by Eq.~\eqref{coll}. Fig.~\ref{animal_growth}(a) shows the thirteen $M(t)$ trajectories for various organisms analyzed in \cite{west2001general}, plus two more (quail and turkey) from \cite{dietz1997effect}. Fig.~\ref{animal_growth}(c) shows the corresponding collapse using $\alpha = 3/4$. This collapse is one of the most compelling aspects of the original WBE growth model, and in itself might be taken as evidence that Kleiber's law with $\alpha = 3/4$ holds for all these organisms. However it might be more accurate to say that the collapse is indicative of general sublinear scaling of $B(t)$, without pointing to a precise exponent. In panels (b) and (d) we see a similar collapse when choosing different values $\alpha = 0.55$ or 0.95, showing that the existence of the collapse is not particularly sensitive to the value of $\alpha$.  In fact for the quail and turkey cases the experimental data indicated that the metabolic exponent $\alpha(t)$ itself was time-dependent, with values $\alpha(t) > 1$ for small times transitioning to $\alpha(t) <1$ at large times. Despite this added complexity, the dimensionless data points for quail and turkey still lie near the universal curves, though the mapping is based on choosing fixed values of $\alpha$ in Fig.~\ref{animal_growth}(b-d). Hence we have to be careful in drawing conclusions about the form of metabolic scaling from the behavior of the dimensionless time and mass ratio.

\subsection{Tissue and cellular-level metabolic scaling}\label{sec:tissue}

For organisms with complex internal anatomies there is another aspect of metabolic scaling that merits attention: the fact that heat dissipation is highly heterogeneous in space, with some regions accounting for a disproportionate amount of the total $B$. This has been most extensively studied across different tissues and organs in mammals~\cite{schmidt1984scaling}. For humans, a small set of metabolically significant organs (heart, kidney, brain, lungs, and abdominal organs like the liver) comprising $\sim 8\%$ of an individual's mass account for $\sim 72\%$ of $B$~\cite{aschoff1971}. For rats the discrepancy is smaller, but the numbers are still lopsided: $\sim 14\%$ versus $\sim 39\%$~\cite{rolfe1997cellular}.

The relative metabolic activity of different tissues is commonly quantified via a mass-specific metabolic rate $b_i$ for the $i$th tissue type. If $m_i$ is the corresponding mass of that tissue in the organism, such that summing over all types we have $M = \sum_i m_i$, then $B$ can be approximated as $B \approx \sum_i m_i b_i$~\cite{wang2001reconstruction}. The sum is typically restricted to major organs/tissues, with the remaining metabolic activity accounted for by a catch-all ``residual'' type. This simple model can produce reasonably accurate predictions of whole organism $B$ values. For example estimates of $b_i$ for human tissues~\cite{elia1992organ} were used in conjunction with $m_i$ measurements via magnetic resonance imaging to predict $B$ values in a diverse cohort of individuals, closely agreeing with values from respirometry~\cite{wang2010specific}.

To experimentally estimate $b_i$ there are two major approaches---{\it in vivo} methods and tissue slices---both with technical limitations. {\it In vivo} approaches, like those of \cite{elia1992organ} in humans, rely on measuring O$_2$ concentration differences between arterial blood carrying oxygen into an organ relative to venous blood carrying it away, for example using catheter-drawn blood samples. Combined with separate estimates of blood flow through the organ, this gives an approximate oxygen consumption rate, which is converted to units of W using Thornton's rule, and divided by the organ mass to get $b_i$ in units of W/g. More recently, positron emission tomography with specially designed tracers has been used to measure oxygen consumption in certain organs like the brain~\cite{wang2012organ}. Given the complexity of these {\it in vivo} approaches, we have relatively few mammalian $b_i$ values estimated in this manner. There is a more common, long-standing {\it in vitro} alternative: respirometry measurements on tissue slices taken out of organs and immersed in ionic solutions designed to mimic physiological conditions. The $b_i$ value here is proportional to the oxygen consumption rate per mass of the tissue slice. These measurements have been conducted for almost a century~\cite{schmidt1984scaling}, but it was recognized early that they sensitively depend on the solution composition~\cite{krebs1950body}, and other factors like tissue thickness (which affects oxygen diffusion into the cells) and the fraction of cells damaged during the slicing process~\cite{couture1995relationship}. Hence there is a general expectation that $b_i$ values from tissue slices may only be rough approximations, often underestimating the true values.

\begin{figure}[t]
\centering\includegraphics[width=\columnwidth]{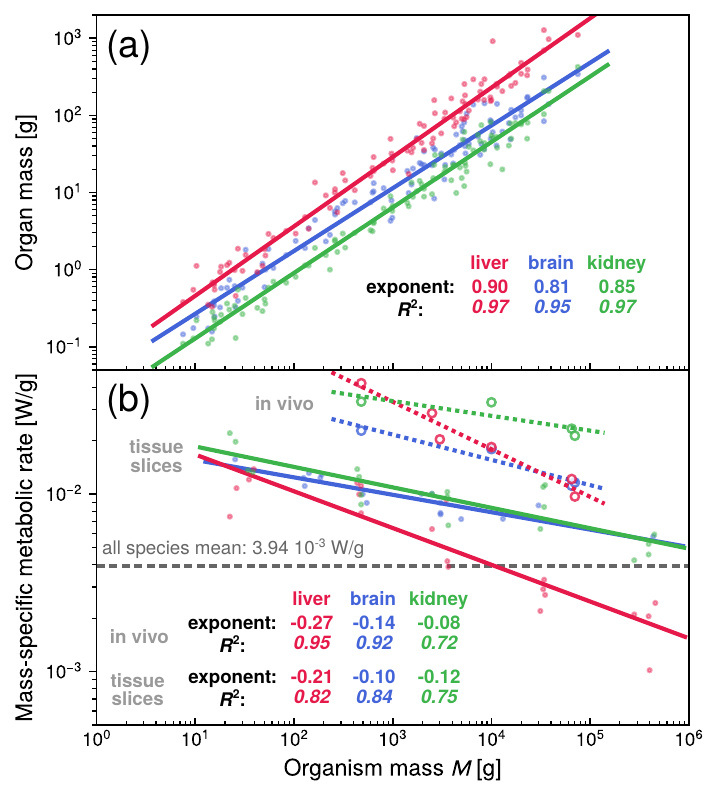}
\caption{(a) Organ masses for liver (red dots), brain (blue dots), and kidney (green dots) for 100 mammal species, as a function of total organism mass $M$~\cite{navarrete2011energetics}. The three solid lines are corresponding power law fits, with exponents and $R^2$ values indicated in the figure. (b) Mass-specific tissue metabolic rates from the same organs (liver, brain, kidney) in a subset of mammalian species as a function of organism mass $M$. Solid dots are estimates from tissue slice experiments (liver and kidney from \cite{couture1995relationship}, brain from \cite{elliott1948metabolism}), while open circles are {\it in vivo} data~\cite{wang2012organ}. The horizontal dashed line is the whole-organism average value $\bar{b}$ across all species from Fig.~\ref{interspecies}(b) for comparison. Solid and dashed lines represent power law fits to the tissue slice and {\it in vivo} data respectively, with the corresponding exponents and $R^2$ values indicated in the figure.}\label{tissue}
\end{figure}

Fig.~\ref{tissue} shows representative data from a variety of mammalian species. There are two separate scaling relationships to consider: how $m_i$ for the $i$th organ scales with $M$, shown in Fig.~\ref{tissue}(a) for liver, brain, and kidney; and also how $b_i$ scales with $M$, shown in Fig.~\ref{tissue}(b) for the same organs, with both tissue slice and {\it in vivo} data. Both organ masses and mass-specific metabolic rates exhibit relatively robust power law scaling with $M$ over several decades, but the patterns vary with organ. As expected, tissue slice $b_i$ values are lower than {\it in vivo} estimates, but the power law trends are roughly consistent between the two approaches: liver $b_i$ has a steeper decline with $M$ than brain or kidney in both cases. If $m_i \sim M^{f_i}$ and $b_i \sim M^{g_i}$ for some exponents $f_i$ and $g_i$, then the corresponding term contributing to the whole-organism rate $B \approx \sum_i m_i b_i$ would scale like $m_i b_i \sim M^{f_i+g_i}$. In the simplest scenario all the $f_i+g_i$ values would be the same, and approximately equal to the Kleiber's value of 3/4 to be consistent with the known $B$ scaling for mammals as seen in Fig.~\ref{interspecies}(a). However we get variability in $f_i+g_i$, so metabolism is not partitioned proportionately among different organs at different mammal sizes: $f_i+g_i$ is in the range 0.63-0.69 for liver, 0.67-0.71 for brain, and 0.73-0.77 for kidney, depending on whether the {\it in vivo} or tissue slice exponents are used. These values are not too far from Kleiber scaling (and the data is noisy), but it does highlight potentially heterogeneous scaling behaviors among different organs. Though data from only three organs are shown here, the variability in scaling extends to other tissue types as well~\cite{schmidt1984scaling}.

\begin{figure}[t]
\centering\includegraphics[width=\columnwidth]{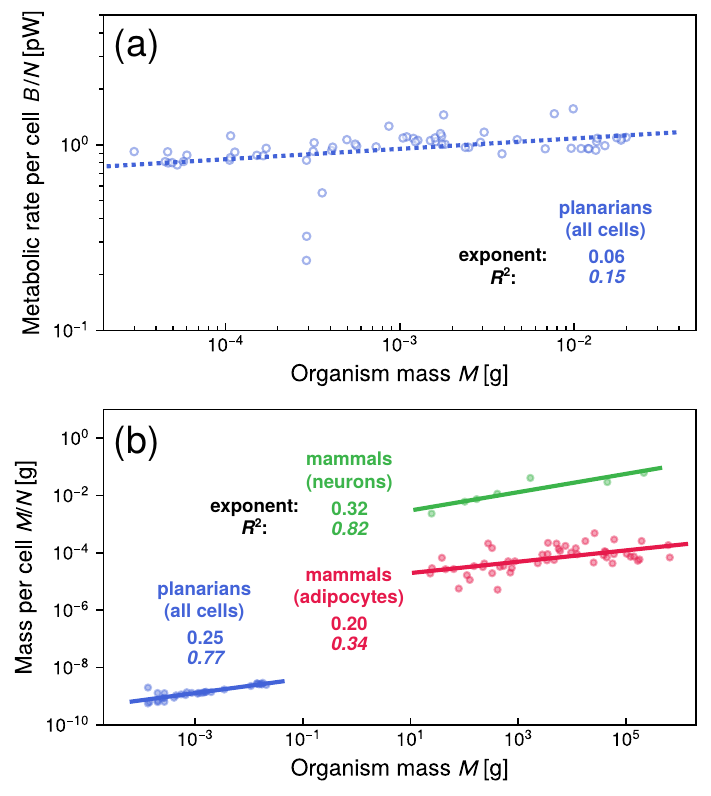}
\caption{(a) Basal metabolic rate per cell $B/N$ (blue dots) for planarians versus organism mass $M$~\cite{thommen2019body}. This is an average over all cells (and hence cell types) in the organism. The dashed line is a power law fit, with the exponent and $R^2$ value shown. The original data was in terms of $B/N$ versus cell number $N$. The latter was converted to $M$ using the power law fit from the planarian data shown in the bottom panel. (b) Mass per cell $M/N$ versus organism mass $M$. The blue dots are for planarians, averaged across all cells~\cite{thommen2019body}. The red dots are for mammalian species~\cite{savage2007scaling}, but for only two specific cell types: superior cervical ganglion neurons (green) and adipocytes from the dorsal wall of the abdomen (red). Solid lines are power law fits, with the exponents and $R^2$ values indicated in the figure.}\label{percell}
\end{figure}

Going below the tissue and organ level, it is worth considering how metabolic scaling plays out at the level of individual cells. Here the traditional view was that cell mass stays roughly the same across different animal species, independent of organism mass $M$~\cite{schmidt1984scaling}. In other words $M/N$, where $N$ is the total number of cells in the organism, was believed to be largely constant across different $M$. If this were true, and if the basal metabolic rate $B \sim M^\alpha$ for some exponent $\alpha$, this would imply a certain scaling of the metabolic rate per cell $B/N$. To see this, let us rewrite $B \sim M^\alpha$ as
\begin{equation}\label{percelleq}
    \frac{B}{N} \sim M^{\alpha-1} \frac{M}{N}.
\end{equation}
When $M/N$ is independent of $M$, then $B/N \sim M^{\alpha -1}$, which would translate to a decreasing rate $B/N \sim M^{-1/4}$ with increasing $M$ for Kleiber scaling $\alpha = 3/4$. Note that this is the metabolic rate for the whole organism averaged over the total number of cells, in contrast to the mass-specific rate $b_i$ for a given organ discussed earlier, which is defined in terms of the metabolic rate contribution from the organ, divided by the mass of the organ. In the scenario where mass and cell number have a fixed proportionality, $b_i$ is also proportional to the organ metabolic rate per cell. The decreasing power law trends for $b_i$ with $M$ in Fig.~\ref{tissue} for mammalian species, while not exactly exhibiting exponents of $-1/4$, make it plausible that $B/N \sim M^{-1/4}$ could be possible when averaging over the the whole organism.

Recent experimental evidence, however, complicates this picture. In mammals for example, we now know that $M/N$ is approximately constant for certain cell types, but not in others. There is a spectrum of scaling depending on cell type, going all the way to the opposite limit where $B/N$ is nearly constant and $M/N$ has power law scaling with $M$. Based on Eq.~\eqref{percelleq}, we see that $M/N \sim M^{1-\alpha}$ if $B/N$ is independent of $M$. These various scaling scenarios were first explored by \cite{savage2007scaling}, who analyzed data from 18 different cell types in mammals, either looking at direct measurements of $M/N$ values, or instead using average cellular volume as a proxy. While many cell types with relatively fast division times (i.e. erythrocytes, hepatocytes, fibroblasts, epithelial cells) fit the constant $M/N$ scenario, some slowly dividing cells were closer to the other limit, where $M/N \sim M^{1-\alpha}$. Fig.~\ref{percell}(b) shows $M/N$ versus $M$ data for two such cell types: superior cervical ganglion neurons (green) and abdominal adipocytes (red). Though the exponents (0.32 and 0.20 respectively) differ from the Kleiber expectation $1-\alpha = 1/4$, they do indicate that cellular-level mass scaling (just like the organ level) is heterogeneous in mammals.

Planarians provide a fascinating contrast: in this case the $M/N \sim M^{1-\alpha}$ scaling holds for the average over all cells, and $B/N$ is approximately constant~\cite{thommen2019body}. Fig.~\ref{percell}(b) (blue dots) shows the planarian $M/N$ data across two decades of $M$, scaling like $M^{1/4}$. Fig.~\ref{percell}(a) plots $B/N$, which is nearly flat as $M$ varies. A power law fit of $B/N$ gives a weak scaling of $M^{0.06}$, but the $R^2 = 0.15$ is quite low. In the planarian case the increase in $M/N$ with $M$ is directly related to energy storage molecules: cells in the largest planarians had 8 times more glycogen and 88 times more triglycerides than cells in the smallest individuals. One explanation is a strategy to manage different extremes of food availability, since planarians have a unique ability to grow in size under high food supply and shrink when faced with starvation. The larger animals have the ability to store away some of the energy, bulking up cellular mass, while the smaller, hungrier animals deplete these stores---compensating at either extreme to maintain a nearly constant metabolic expenditure per cell. The net effect of this is a very clean example of Kleiber's law scaling, as shown in Fig.~\ref{intraspecies}(b). Thus planarians provide perhaps the strongest connection discovered to date between Kleiber's law and a biochemical mechanism.  While the mechanism is clearly not universal across all animals (as evidenced by many mammalian cell types with non-constant $B/N$), the role of energy storage merits further exploration. It is probably not a coincidence that one of the cell types in mammals which shows power law scaling of $M/N$ is an adipocyte, a repository for fats like triglycerides.  

\subsection{Population-level scaling}\label{sec:popscaling}

To conclude our survey of metabolic scaling and its implications at different biological levels, we switch to the opposite end of the spectrum: the cumulative metabolism of whole populations. A full account of ecological scaling laws is beyond the scope of this review, but there are interesting analogues of metabolic scaling relations in ecological quantities. For example the population density $\rho$, the mean number of individuals per unit area, exhibits a nearly reciprocal behavior relative to organism mass $M$ compared to the basal metabolic rate results of Fig.~\ref{interspecies}~\cite{hatton2019linking}. Across the entire size range, from prokaryotes to mammals, there is an approximate scaling of $\rho \sim M^{-1}$, but individual classes (i.e. mammals, ectotherms, etc.) exhibit shallower slopes with $\rho \sim M^{-\alpha}$ and $\alpha < 1$. The individual class scaling results are staggered relative to one another, to be compatible with the overall $\rho \sim M^{-1}$ trend, just like the classes in Fig.~\ref{interspecies} relative to the $B \sim M$ trend. The value of $\alpha$ in certain cases is close to 3/4, an observation known as Damuth's law, after the study which first noted this relation for mammals in a variety of terrestrial habitats~\cite{damuth1981population}. The close reciprocity in how $B$ and $\rho$ scale with $M$ means that $\rho B$, the total population metabolism per unit area, is roughly independent of organism mass: values of $\rho B$ across 20 orders of magnitude in $M$ cluster around a mean scale of $10^{-3}$ W/m$^2$~\cite{hatton2019linking}. This has been interpreted as an ``energy equivalence rule''~\cite{brown2004toward}, with the approximate constancy of $\rho B$ suggesting that different populations of organisms are using roughly the same proportion of available energy resources regardless of body size. However it is worth noting that population density data is noisier than metabolic rates, resulting in a spread of $\rho B$ values over about six order of magnitude across all organisms. Thus whether energy equivalence is a meaningful ``rule'' is up for debate~\cite{isaac2013paradox}, in this way also paralleling the discussions around metabolic scaling ``laws''.

In addition to $\rho B$, one can consider the total population biomass density, $m_\text{tot} = \rho M$, or just total biomass $M_\text{tot}$. The maximum growth rate of $M_\text{tot}$ in a given population has a remarkably clean Kleiber-like scaling $\sim M^{3/4}$ with individual body mass $M$, valid across kingdoms of life~\cite{hatton2015predator,hatton2019linking}. Given a model of interactions between organisms (i.e. predator-prey dynamics or more generalized multi-species interactions) one can translate this scaling dependence into dynamical equations for $m_\text{tot}$, with a variety of implications. These equations are analogues of individual growth models like Eq.~\eqref{wbe}, with biomass density production rates that scale like $\sim m_\text{tot}^{3/4}$. For example, a simple model of predator-prey dynamics along these lines predicts that steady-state predator biomass density should scale with prey biomass density using the same exponent, $m^\text{pred}_\text{tot} \sim (m^\text{prey}_\text{tot})^{3/4}$, a relation that has been empirically validated~\cite{hatton2015predator}. More generally, ecological models with competitive interactions and production rates that have this 3/4 scaling (or any sublinear exponents $\alpha<1$) exhibit a striking feature: such models are more likely to have stable, coexisting populations as the diversity of species increases~\cite{hatton2024diversity}.

The ecological scaling laws described above, even when only approximate, raise interesting questions about the prevalence of sublinear scaling exponents, in particular those around the Kleiber value of 3/4. Does ecology inherit these exponents from the underlying metabolism of organisms, or is the causal connection reversed, with ecological constraints (translated to evolutionary pressures) driving metabolic scaling? Or are the mechanisms that underlie organism and population-level scaling completely independent, with the similarity in exponents a mathematical coincidence, aided in part by researchers seeking out the universal simplicity of 3/4 scaling even where the data is noisy and heterogeneous? All of these remain open questions so long as we do not have clear explanations of metabolic scaling at any level of biological organization. The next section describes some of the attempts to date to provide such explanations.

\section{Biophysical Models of Metabolic Scaling}\label{section4}

As described in Sec.~\ref{section3}, metabolic scaling is quite complex, characterized by a variety of exponents depending on the class of organisms and biological level of interest. However given the historical prominence of Kleiber's exponent of 3/4, and its relevance to mammals, several theoretical works have aimed to explain this value. In this section we will summarize some of these attempts, as well approaches providing possible explanations for a more general sublinear scaling. Living organisms are shaped by evolutionary optimization within structural or energetic constraints, and the proposed explanations typically emphasize either one or the other.  Accordingly, we can categorize models that account for scaling into two types, which we describe below: those based on constraints in body plan / anatomy (Sec.~\ref{sec:bodyplan}), and those based on arguments about the flow and allocation of energy (Sec.~\ref{sec:flow}).

 \vspace{-1em}
\subsection{Models based on constraints in body plan and anatomy}\label{sec:bodyplan}
In organisms, differential growth across cells, tissues, and systems leads to non-isometric changes in anatomical and physiological traits as body size increases \cite{d1917growth}. The intricate morphologies within animals, such as branched tissues and transport networks, often exhibit fractal self-similarity and power law scaling \cite{schmidt1984scaling}. In this spirit, significant efforts have been made to obtain the 3/4 exponent of metabolic scaling from fractal anatomy. Another  hypothesis of models in this category relates metabolic rate to muscle power, and states that mechanical stability limits the scaling of muscles and hence the power dissipated by muscles. In the following sections we present these models.

\subsubsection{Transport efficiency constraints}
\begin{figure*}[t] 
    \centering
\includegraphics[width=\textwidth]{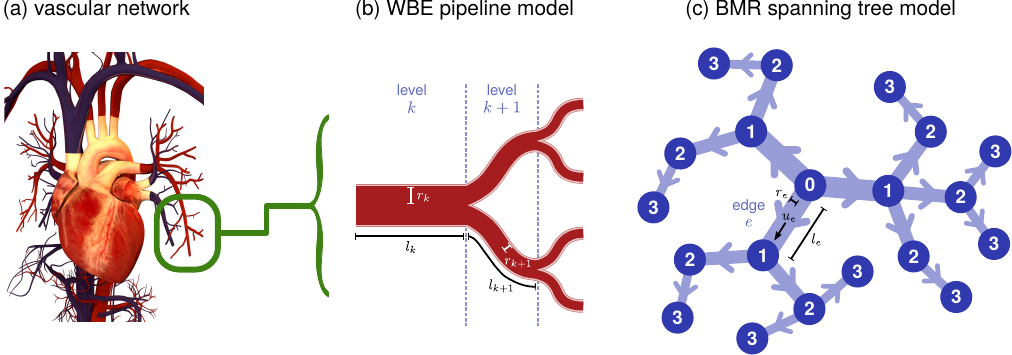} 
    \caption{(a) Example of a vascular network circulating the metabolic resources in an organism. Image credit: Bryan Brandenburg, Creative Commons CC BY-SA 3.0. (b) WBE hierarchical pipeline model~\cite{west1997general}, where the radius $r_k$ and length $l_k$ scales across hierarchy levels $k$. (c) BMR model of a spanning tree network \cite{banavar1999size}, transporting metabolites from source node 0 to other nodes via fluxes along the edges. The original model considers only the topology of the network, but here we lay out also the geometry that is required to understand the assumption \eqref{eq:vb_bmr}. Each edge $e$ is characterized by radius $r_e$, length $l_e$, and flow rate $u_e$. The integer labeling each node is its hierarchy level $k$.} 
    \label{fig:wbe}
\end{figure*}

Metabolic resources need to be at sufficient concentrations in the entire organism. In particular, oxygen is taken from the environment via inhalation or absorption of air from a surface, which then needs to be distributed in the body. 
For diffusive transport, the penetration depth of oxygen can be estimated  by  $\lambda_{\text{O}_2}=\sqrt{D_{{\rm O}_2}\tau_{{\rm O}_2}}$ where $D_{{\rm O}_2}$ and $\tau_{{\rm O}_2}$ are the diffusion coefficient and lifetime of $\text{O}_2$. 
Assuming first-order reaction kinetics, the lifetime of an oxygen molecule is $\tau_{{\rm O}_2}=n_{{\rm O}_2}V/\Phi_{{\rm O}_2}$ where $\Phi_{{\rm O}_2}$ is the total oxygen consumption rate $n_{{\rm O}_2}$ is the concentration of oxygen and $V$ is the volume of the organism. Considering the mass density of the organism to be $\rho=1$ kg/L and substituting $M=\rho V$ in Eq.~\eqref{u3}, we obtain the volume-specific oxygen consumption $\Phi_{{\rm O}_2}/V=8.75$  $\mu$mol/L/s. For $\text{O}_2$ concentration, we can use the solubility of oxygen in water at $298^\circ \text{K}$ allowing $n_{\text{O}_2}\lesssim 250$ $\mu$mol/L, which leads to
$\tau_{{\rm O}_2}=28.6\;$s.
Finally, using the diffusion coefficient at the same temperature $D_{{\rm O}_2}=2.4\times 10^{-3}$ $\text{mm}^2/$s \cite{ramsing2011seawater} we obtain $\lambda_{\text{O}_2}\simeq 0.25 \text{ mm}$. Beyond this size, diffusive transport would be insufficient in supplying oxygen reliably. Thus, in many animal species, the distribution of resources such as oxygen and nutrients to the entire organism is provided by a circulatory system. The properties of the circulatory system should be compatible with the metabolic demand in an organism. 
A set of hypotheses proposes a relation between the non-isometric scaling of metabolic rates to transport capacity limitations of fractal networks such as vascular systems (Fig.~\ref{fig:wbe}(a)). 



\vspace{1em}

\textit{West-Brown-Enquist model:} A prominent model was proposed by West, Brown, and Enquist \cite{west1997general}, known generally as the WBE model. Note that this is distinct from the WBE growth model, Eq.~\eqref{wbe}, by the same authors. Suppose the transport network is a  branched rigid pipeline network in 3D with a number of branches $N_k$, length of branches $l_k$ and radius of branches $r_k$ at the $k^{\rm th}$ hierarchy (Fig.~\ref{fig:wbe}(b)). Here, the total number of hierarchy levels is $c$ while $k=0$ and $k=c$ correspond to the primary branch (hence $N_0=1$) and terminal branch (corresponding to capillaries) respectively. In this network, the capillary  values of radius $r_c$ and length $l_c$ as well as the flow rate $u_c$ are assumed to be constant and independent of the organism size. This can be motivated by the fact that capillaries serve cells whose sizes do not vary as dramatically with the body size. Moreover, the fluid transfer rate across the total cross-sectional area of the capillaries is given by $J_c=N_c\pi r_c^2u_c$ and the metabolic rate is assumed to be proportional to $J_c$, such that $B=\gamma J_c$ with size-independent proportionality constant $\gamma$. As a result, 
\begin{equation}
    B= N_c\gamma \pi r_c^2u_c\,. \label{eq:bnc}
\end{equation}
The size independence of $r_c$, $l_c$, $u_c$, and $\gamma$ then suggests that the metabolic scaling is determined by the number of capillaries, i.e., $B\sim N_c$. The WBE model assumes a self-similar hierarchical network with a branching rate at each hierarchy given by $\alpha=N_{k+1}/N_{k}>1$, and the length and radius follow
\begin{equation}
   l_{k+1}/l_k=\alpha^a\,,\quad r_{k+1}/r_k=\alpha^b\,.
\end{equation}
The total volume of the network is $V_b=\sum_{k=0}^{c} N_k \pi r_k^2 l_k$.  Using the above scaling relations for $N_k, r_k, l_k$ across hierarchies one obtains $N_k=N_c \alpha^{k-c}$, $l_k=l_c\alpha^{(k-c)a_k}$, and $r_k=r_c\alpha^{(k-c)b_k}$ where $a_k$ and $b_k$ are hierarchy-dependent scaling coefficients. As a result, we can write
\begin{equation}
V_b=N_c \pi r_c^2 l_c\sum_{k=0}^c \alpha^{(k-c)\beta_k} \label{eq:vb}
\end{equation}
with $\beta_k=1+a_k+2b_k$. The WBE model then makes the following key assumptions. The first is that $a_k=-1/3$ at all hierarchies,  which is linked to a volume-preserving fractal network at each hierarchy, i.e., $\alpha (4\pi/3)(l_{k+1}/2)^3=(4\pi/3)(l_{k}/2)^3$. The second assumption is that the value of $b_k$ is determined by minimization of energy dissipation rate in the circulatory system. The WBE argument is that for pulsatile flows this is achieved for an area-preserving network, $\alpha \pi r_{k+1}^2=\pi r_k^2$, leading to $b=-1/2$ whereas for laminar flow the minimization produces the well-known Murray’s law \cite{murray1926physiological}, $\alpha \pi r_{k+1}^3=\pi r_k^3$ leading to $b=-1/3$. Their argument is that the mammalian circulatory system contains both flow regimes, and a transition from $b=-1/2$ to $b=-1/3$ should occur as one approaches the capillaries, at a $k>k^*$ where laminar flow dominates. We can now rewrite the total volume of the network from Eq.~\eqref{eq:vb} as
\begin{eqnarray}
    V_b=N_c \pi r_c^2 l_c \left(\sum_{k=0}^{k^*}\alpha^{(c-k)\beta_1}+\sum_{k=k^*+1}^{c}\alpha^{(c-k)\beta_2}\right)\label{eq:vb2}
\end{eqnarray}
where $\beta_{1,2}=1+a+b_{1,2}$. Applying $a=-1/3$, $b_1=-1/2$ for $k\leq k^*$ and $b_2=-1/3$ for $k> k^*$ yields $\beta_1=-1/3$ $\beta_2=0$. Performing the sums in Eq.~\eqref{eq:vb2} results in
\begin{equation}
       V_b=N_c \pi r_c^2 l_c \left(s_1\alpha^{c/3} +s_2\right)
\end{equation}
where $s_1=(1-\alpha^{-(k^*+1)/3})/(1-\alpha^{-1/3})$ and $s_2=c-k^*$,
with the second line obtained by performing the geometric sum. Using $\alpha_c=N_c$, we get $V_b=\pi r_c^2 l_c (s_1N_c^{4/3}+s_2N_c)$. Final piece is to introduce a proportionality of the network volume with organism mass,
\begin{equation}
    M= \zeta V_b \label{eq:mvb}
\end{equation}
which is empirically supported for blood volume in mammals \cite{stahl1967scaling}. 
Then, using the relation Eq.~\eqref{eq:bnc}, we obtain 
\begin{equation}\label{wbe43}
       M= \bar{s}_1B^{4/3}+\bar{s}_2B 
\end{equation}
where $\bar{s}_1=\zeta l_c(\gamma^{4}\pi r_c^2)^{-1/3}s_1$ and $\bar{s}_2=\zeta l_c\gamma^{-1}s_2$. For $B\gg (\bar{s}_1/\bar{s}_2)^{-1/3}$, Kleiber's law emerges from $M\sim B^{4/3}$.

However if one considers the full prediction of Eq.~\eqref{wbe43}, not just the asymptotic limit, there is a discrepancy with empirical data, as first noted by \cite{savage2008sizing}. Inverting Eq.~\eqref{wbe43} to find $B$ as a function of $M$, the predicted curvature in a $\log B$ versus $\log M$ graph would be concave down. In other words the apparent scaling exponent becomes larger as $M$ decreases. The experimental data for mammals, on the other hand, indicates a concave up curvature, with the exponent decreasing with smaller $M$~\cite{dodds2001re,kolokotrones2010curvature}.



\vspace{1em}

\textit{Banavar-Maritan-Rinaldo model:} A few years after the WBE model, Banavar-Maritan-Rinaldo (BMR) proposed an alternative formulation for transport constraints \cite{banavar1999size}. In this approach, it is argued that $B\propto V_b ^{3/4}$ emerges as a result of minimization of the circulatory volume $V_b$. The model considers a fully-connected network with $N+1$ nodes and $E$ edges where the center node is the source of fluid which supply nutrients, and the rest of the nodes are sinks with local consumption rate $b_{i}$ while the fluid is transported through the edges across the network (Fig.~\ref{fig:wbe}(c)). Then, the total consumption rate is $B=\sum_{i=1}^N b_i$. BMR assume that the circulatory volume is 
\begin{equation}
    V_b\propto \sum_{e}|I_e| \label{eq:vb_bmr}
\end{equation}
 where $I_e$ is the metabolite flux along edge $e$. However, for this assumption to hold true, there should be implicit assumptions about the volumetric flow as we show below. Let us consider a cylindrical geometry of each edge as in the pipeline model. Then the circulatory volume is a sum over edges $V_b=\sum_e \pi r_e^2 l_e$ where $r_e$ and $l_e$ are the radius and length of edge $e$. The volumetric flow rate along each edge is $I_e=\pi r_e^2 u_e$ where $u_e$ is the flow rate at edge $e$. Then, the assumption $V_b\propto \sum_{e}|I_e|$ implies that $|u_e|/l_e$ is constant across the network. 
 
 The sum $\sum_{e}|I_e|$ is minimized if the edges form a spanning tree covering all nodes such that there are no loops in the network. A spanning tree can be described as a hierarchical structure in which the level hierarchy $k$ is the number of edges from a given node to the source node with a maximal hierarchy level $\kappa$. The total consumption rate of nodes at hierarchy $k$ is $\tilde{b}_k=\sum_{i\in i_k} b_i$, where $i_k$ denotes nodes $k$-edge distant from the source node. Similarly, we define total flux at $k^{\rm th}$ hierarchy level as $\tilde{I}_k=\sum_{e\in e_k}I_e$ where $e_k$ denotes edges connecting nodes from $i_{k-1}$ and $i_{k}$ in the network. Then, the total consumption rate can be expressed as a sum over $\kappa$ hierarchies, 
\begin{eqnarray}
    B=\sum_{k=1}^{\kappa} \tilde{b}_k \label{eq:bbmr}
\end{eqnarray}
 while at steady state we have $\tilde{I}_k=\sum_{j=k}^{\kappa}\tilde{b}_{j}$. The latter leads to 
\begin{eqnarray}
    \sum_{e}I_e&=&\sum_{k}\tilde{I}_k \nonumber \\
    &=&\sum_{k=1}^\kappa k\tilde{b}_{k} \label{eq:ie}
\end{eqnarray} 
where we used $\sum_{k=1}^\kappa \sum_{j=k}^{\kappa}\tilde{b}_{j}=\sum_{k=1}^\kappa k\tilde{b}_{k}$ in the second line. 
Now, assuming that consumption rate at each node is same such that $b_k=\bar{b} n_k$ where $n_k$ is the number of nodes at $k^{\rm th}$ hierarchy, one gets $\sum_{k=1}^\kappa k\tilde{b}_{k}=\bar{b}\sum_{k=1}^\kappa n_k$ while $B=\bar{b}\sum_{k=1}^{\kappa} n_k$ using Eq.~\eqref{eq:bbmr}. Finally, taking a uniform node density per unit volume suggests $n_k\propto k^2$ and hence $\sum_{k=1}^\kappa k n_k \propto \kappa^4 $ and  $\sum_{k=1}^\kappa n_k \propto \kappa^3 $ for $\kappa\gg 1$. Combining these results yields $B\propto \kappa^3$ and $\sum_{k=1}^\kappa k\tilde{b}_{k}\propto \kappa^4$. Substituting the latter into Eq.~\eqref{eq:ie} and using Eq.~\eqref{eq:vb_bmr} would lead to $V_b\propto \kappa^4$. Consequently, this would imply a scaling relation $B\sim V_b^{3/4}$. Then, the relation Eq.~\eqref{eq:mvb} leads to $B\sim M^{3/4}$. A consequence of such scaling is that size of the organism $M\sim \kappa^4$. This implies that the node density, although uniform, decreases as organism size increases, $n_k/M\sim M^{-1/4}$. 

On the one hand, this approach proposes a general principle that could govern different transport systems. On the other hand, not having an explicit physiological transport network model as a basis makes   difficult to capture the biological implications and the actual geometrical scaling of the system which is part of criticisms of this model \cite{dodds2001re}.

In summary, both the WBE and BMR models focus on the efficiency of the supply network, assuming it meets the metabolic demand. In this view, supply efficiency constrains the metabolic rate. More generally, metabolic demand also depends on a variety of factors intrinsic to the living organism. In the rest of Section \ref{section4}, we revisit models that consider the determinants of metabolic demand.




\subsubsection{Mechanical stability constraints}
\vspace{1em}
Another hypothesis is based on a mechanical similarity principle  \cite{mcmahon1973size}. This approach has two major assumptions: i) the metabolic rate $B$ and the muscle power output $P_m$ are linearly related, ii) the muscle system should follow a similarity relation, preserving elastic instability. 

Let us first consider the second assumption. Take the muscle system as a mechanical structure with a geometry approximated by a vertical rod with length $l$, circular cross-section radius $r$, and elastic modulus $Y$. The total mass is given by $M=\rho \pi r^2 l$ where $\rho$ is the mass per unit volume (independent of size and species). The limit of stability of the rod under its own weight suggests $l_{\rm cr}\propto (Y/\rho)^{1/3}r^{2/3}$ (see Section II.21 of \cite{landau2012theory}). Thus, McMahon's proposal is that in order to maintain mechanical stability as the size of the rod grows, the scaling relation $l\propto r^{2/3}$ should hold. Thus, we can write a mechanical similarity relation $r\propto M^{3/8}$. The power output by a contracting muscle typically follows $P^{\rm m}\propto \sigma A_{\rm cs} \dot l$ where $\sigma$ is the tensile stress, $A_{\rm cs}$ is the cross-sectional area of the muscle, and $\dot l$ is the speed of shortening. McMahon further assumes that $\sigma$, and $\dot l$ are scale-independent and similar across different species, based on Hill's earlier work on muscle physiology \cite{hill1950dimensions}. Since $A_{\rm cs}\propto r^2$, using the above mechanical similarity relation, i.e., $r\propto M^{3/8}$, we reach $P^{\rm m}\propto M^{3/4}$. Incorporating this with assumption i, $B\propto P^{\rm m}$, we obtain 3/4 scaling of the metabolic rate with mass.

This model is an early attempt to relate structural anatomy to metabolic scaling. However, it lost attention in recent discussions of metabolic scaling, primarily because its connection to metabolic rate relies on power output in muscle dynamics which is not significant for basal metabolism and is irrelevant for organisms lacking muscles \cite{agutter2004metabolic}.


\subsection{Models based on flow and allocation of energy}\label{sec:flow}

Another class of approaches to modeling metabolic scaling ignores the structural details of internal anatomy and instead looks at how energy is transduced through the organism. In some models the focus is on energy outflow from the animal's surface, related to the regulation of body temperature, and in others on the relative allocation of energy to different internal processes. These approaches are often compatible with a variety of effective sublinear scaling exponents for $B$, depending on the weights of the different contributions, rather than pinpointing a unique Kleiber value of 3/4.

 
\subsubsection{Thermoregulation}\label{thermoregulation}

In endothermic animals $B$ has a dual role: the heat release is a byproduct of all the necessary biochemical processes occurring in resting metabolism, but it also helps maintain the animal at a near constant body temperature. For mammals, both core and surface temperatures remain remarkably consistent over six orders of magnitude of $M$, clustering within a few degrees of 37.4$^\circ$C for core temperatures, and the range 28-30$^\circ$C for surface temperatures~\cite{mortola2013thermographic}. Since a constant temperature reflects a balance between heat generation and heat outflow into the environment, the scaling of one should reflect the scaling of the other. It is not a priori clear whether there are stronger physical constraints on the generation or the outflow, but one can make different models based on focusing in varying degrees on one or the other.

In the simplest model, one implicitly assumes the overriding physical constraint on the scaling of the metabolic rate $B$ is heat outflow, which must be tailored such that constant body temperature is maintained at all $M$ for endothermic animals. The full thermodynamic details of how animal body temperature homeostasis are quite complex, but can be roughly captured in the following manner. By definition the basal metabolism for endotherms is measured within the thermoneutral zone, where the ambient environmental temperature $T_{\rm env}$ is in the range $T_{\rm lc} \le T_{\rm env} \le T_{\rm uc}$. The boundaries of the zone are dubbed the lower ($T_{\rm lc}$) and upper ($T_{\rm uc}$) critical temperatures, which vary between species~\cite{schmidt1984scaling,riek2013allometry}. Within the zone the organism does not need to expend significant additional metabolic resources to maintain temperature, so $\dot{Q} \approx B$.

To connect $B$ to heat outflow, one can approximate heat transfer from the animal body to the environment using Newton's law of cooling. This states that heat loss is proportional to the difference between body surface temperature $T_{\rm surf}$ and $T_{\rm env}$:
\begin{equation}\label{heatloss}
    B \approx h A (T_{\rm surf}-T_{\rm env}).
\end{equation}
Here $A$ is the surface area of the animal, and $h$ the heat transfer coefficient, which reflects the combined effects of all the heat loss processes that are occurring (primarily air convection and radiation~\cite{mortola2013thermographic}). The law is an approximation to linear order in the temperature difference $\Delta T = T_{\rm surf}-T_{\rm env}$, and is valid when $\Delta T / T_{\rm surf} \ll 1$, as is typically the case in biological contexts (with temperatures measured in Kelvin).

In order for Eq.~\eqref{heatloss} to be satisfied for all $T_{\rm env}$ in the thermoneutral zone (assuming $B$ and $T_{\rm surf}$ fixed for a given organism), the animal has to vary $h$ as $T_{\rm env}$ changes. For example blood flow to the surface can be increased, pushing up $h$, in response to increases in $T_{\rm env}$~\cite{schmidt1984scaling}. As $T_{\rm env}$ decreases, $h$ can be made smaller by decreasing blood flow (or fluffing up feathers / fur) until it reaches a lower value of $h_{\rm lc}$ at $T_{\rm lc}$. At this limit we have
\begin{equation}\label{heatlosslc}
    B \approx h_{\rm lc} A (T_{\rm surf}-T_{\rm lc}).
\end{equation}
If $h_{\rm lc}$ were universal, for example a minimum heat transfer coefficient set by the constraints of animal surface physiology, and if all animals shared the same $\Delta T_{\rm lc} = T_{\rm surf} - T_{\rm lc}$, this would predict that $B \sim A$. Since surface area for a constant density object scales like $A \sim M^{2/3}$, we would then get $B \sim M^{2/3}$, which is just a modern version of the Sarrus-Rameaux argument for a 2/3 scaling exponent~\cite{sarrus1839application} mentioned in Sec.~\ref{hist}.

Unfortunately real animal physiology contradicts these idealized assumptions, and as a result Eq.~\eqref{heatlosslc} is actually compatible with a variety of scaling exponents for $B$. The heat transfer coefficient $h_{\rm lc}$ can systematically vary with body size due to changes in surface insulation (like thickness and composition of fur), and there is evidence for example that the insulation value of fur increases for mammals with increasing $M$ up to about $M \sim 10$ kg~\cite{hart1956seasonal}. We thus expect a decreasing value of $h_{\rm lc}$ with $M$ over this mass range. On the other hand we see a scaling $\Delta T_{\rm lc} \sim M^\delta$, with exponent $\delta \approx 0.2$~\cite{riek2013allometry}, driven by decreasing $T_{\rm lc}$ as $M$ increases. The decrease in $T_{\rm lc}$ allows larger animals like polar bears (whose $T_{\rm lc}$ value is a remarkable $-30^\circ$C~\cite{blix1979modes}) to tolerate colder temperatures at rest without expending additional metabolic resources. In contrast $T_{\rm uc}$ shows only very weak scaling with $M$~\cite{riek2013allometry}. Given $\delta \approx 0.2$, and the fact that $A \sim M^{2/3}$, we see from Eq.~\eqref{heatlosslc} that $h_{\rm lc} \sim M^{-0.12}$ would be compatible with Kleiber scaling for $B$. More generally, without knowing the precise dependence of both $h_{\rm lc}$ and $\Delta T_{\rm lc}$ on mass we cannot use Eq.~\eqref{heatlosslc} to predict a unique scaling exponent for $B$.

\subsubsection{Models involving energy allocations with different scaling exponents}

Given the noisiness of metabolic rate data, and indications of curvature in log-log plots of $B$ versus $M$~\cite{kolokotrones2010curvature}, a reasonable hypothesis might be that the apparent 3/4 scaling is an approximation to more complicated behavior that cannot be described by a single power law. Several models posit that $B$ may be a linear combination of terms with different scaling exponents. At one end of the complexity spectrum is the allometric cascade model of \cite{darveau2002allometric}, where $B$ is taken to be a multi-term sum,
\begin{equation}\label{cascade}
B = \sum_{i \in {\cal P}} c_i M^{\alpha_i},
\end{equation}
where the label $i$ runs over the set ${\cal P}$ of all relevant metabolic processes, with each process characterized by an exponent $\alpha_i$ and coefficient $c_i$. Given the experimental evidence for the heterogeneity of metabolic scaling among different organs and tissue types (Sec.~\ref{sec:tissue}), Eq.~\eqref{cascade} is very plausible, but it leads to a model with a large number of difficult-to-measure parameters. For example to model $B$ in mammals, \cite{darveau2002allometric} used ten different metabolic processes (among them sodium-potassium pumps, protein turnover, cardiac work, etc.) requiring twenty parameters.

At the other limit of complexity is assuming just two dominant terms in Eq.~\eqref{cascade}. For example a term that scales like $M$ combined in the right proportion with a term that scales like $M^{2/3}$ could make $B$ look like it scales with an intermediate exponent (i.e. 3/4) over a broad range of $M$. The model of \cite{ballesteros2018thermodynamic}, building on predecessors in the earlier literature~\cite{swan1974thermoregulation,yates1981comparative}, makes precisely this assumption, writing $B$ as 
\begin{equation}\label{twoexp}
B = c_1 M + c_2 M^{2/3}
\end{equation}
with two parameters $c_1$ and $c_2$ that can be fit to data. This functional form is given some support by empirical $B$ results, in particular the observation by \cite{dodds2001re} that mammals in the small $M$ limit seem to exhibit 2/3 rather than 3/4 scaling. The two terms are interpreted as energy consumed by the organism in doing useful metabolic work ($\sim M$, since this consumption is spread among the entire organism) and energy dissipated as heat through the surface ($\sim M^{2/3}$) because of the inefficiency in the conversion of input power into work. In this picture $B$ plays the role of input power, which is then allocated to either work or dissipation. The problem with such an interpretation is that $B$ as measured by experiments is the total heat dissipated by the organism per unit time, and not necessarily proportional to energy intake. Moreover we know from the discussion in Sec.~\ref{thermoregulation} that though heat is dissipated through the animal surface, this does not inevitably imply that the dissipation scales with the surface area and hence $\sim M^{2/3}$.

Eqs.~\eqref{cascade}-\eqref{twoexp} predict different scaling of $B$ at small and large $M$ depending on which power law terms dominate. One can arrive at a similar result (though with distinct exponents) using a completely different approach known as dynamical energy budget (DEB) theory~\cite{sousa2010dynamic,kooijman2010dynamic,maino2014reconciling}, though it requires making assumptions about the scaling of energy reserves with organism size. The main argument of DEB related to metabolic scaling can be summarized as follows. Imagine the organism mass $M$ partitioned into two compartments $M = M_s + M_r$. The reserve mass $M_r$ consists of pools of resource molecules (lipids, carbohydrates, proteins) that are produced via nutrient assimilation and serve as the fuel for the basal metabolism. The remaining mass $M_s$ is the structural mass, and is where all the metabolic processes occur. The reserves are harnessed for the maintenance of the structure via enzymes that operate at the interface of the reserve pools. This is inspired by biological energy reservoirs like lipid droplets, which participate in metabolism via surface-associated proteins~\cite{walther2012lipid}. One assumes individual pools have a mean mass $m_r$, and there are $n_r$ such pools in a given organism, so $M_r = n_r m_r$. In principle both $n_r$ and $m_r$ could vary with organism size. DEB makes two key assumptions: i) the mean single pool mass $m_r$ scales proportionally to $M$, both within and between species; ii) the number of pools $n_r$ is the same in a given species as $M$ varies during growth, but potentially changes between species.

To derive the scaling of $B$ between different species from DEB theory, we start with the argument that basal metabolism solely involves the maintenance of the structural component (growth is excluded in the usual definition of $B$), so it should like scale with the structural mass, 
\begin{equation}\label{deb1}
    B = a_s M_s,
\end{equation}
where $a_s$ is a constant. The energy required for this maintenance is taken from the reserve, and the energy flow is assumed to the be proportional to the total surface area of the pools. Since each pool has surface area proportional to $m_r^{2/3}$ and there are $n_r$ pools, we can write $B \propto n_r m_r^{2/3} = M_r / m_r^{1/3}$. Using assumption i, namely $m_r \propto M_s$, we have the following relation:
\begin{equation}\label{deb2}
    B = a_r M_r / M_s^{1/3},
\end{equation}
where $a_r$ is another constant. Equating Eqs.~\eqref{deb1}-\eqref{deb2} for $B$ gives us a scaling relation between the reserve and structural mass:
\begin{equation}
    \label{deb3}
    M_r = \frac{a_s}{a_r} M_s^{4/3}.
\end{equation}
Plugging Eqs.~\eqref{deb1}-\eqref{deb3} into the definition $M = M_s + M_r$ we can rewrite it as
\begin{equation}
    \label{deb4}
    M = \frac{1}{a_s} B + \frac{1}{a_r a_s^{1/3}} B^{4/3}.
\end{equation}
This expresses $M$ in terms of linear combinations of power law terms in $B$, rather than the reverse case seen in Eqs.~\eqref{cascade}-\eqref{twoexp}, where $B$ is expressed through power law terms in $M$. Eq.~\eqref{deb4} predicts $B \sim M^{3/4}$ for sufficiently large $M$, where the second term on the right 
dominates, and $B \sim M$ for small $M$. The crossover depends on the constants $a_s$ and $a_r$, but the 3/4 behavior at large $M$ asymptotically approaches the Kleiber scaling. Note the difference with Eq.~\eqref{twoexp}, which predicts $B \sim M^{2/3}$ at small $M$ and $B \sim M$ at large $M$. Eq.~\eqref{deb4} has the same form as the WBE theory prediction in Eq.~\eqref{wbe43}, up to different constants in front of each term. As a result, it also exhibits the same curvature misprediction relative to empirical data that was discussed below Eq.~\eqref{wbe43}.

Another DEB prediction is the increase in reserve relative to structural mass as organism sizes get larger, shown in Eq.~\eqref{deb3}. At first glance, this seems to be qualitatively borne out by the planarian experiments~\cite{thommen2019body} described in Sec.~\ref{sec:tissue}, where larger individuals have a significantly higher proportion of triglycerides and glycogen relative to the total mass. However even here there are discrepancies with DEB theory. Technically speaking the prediction of Eq.~\eqref{deb3} is for comparisons between different species. Based on assumption ii above, DEB actually predicts $M_r \propto M_s$ for intraspecies comparisons, like the planarian case. Even if one were to drop assumption ii, the actual increase in reserve mass in planarians is larger than the 4/3 scaling in Eq.~\eqref{deb3}: if one uses total protein mass as a proxy for $M_S$ and the total mass of lipids / carbohydrates as a proxy for $M_r$, the scaling exponent observed in the data is roughly $M_r \sim M_s^{1.6}$.

\subsubsection{Evolutionary optimization using metabolic growth modeling}

As discussed in Sec.~\ref{sec:growth}, if one assumes metabolic scaling applies to changes in mass during growth of an individual organism (and not just to comparisons between adult animals), this leads to predictions for the mass growth curve $M(t)$ over time $t$ . The model of \cite{white2022metabolic} uses this connection between scaling and growth to argue that the scaling exponent may have been shaped by evolutionary optimization of a third, related quantity: energy allocated to reproduction over the organism lifetime. To tie together scaling, growth, and reproduction, they introduced the following growth model, a variant of the ones considered in Sec.~\ref{sec:growth} (note we have simplified the notation from the original paper for ease of presentation):
\begin{equation}
    \label{white}
\begin{split} &\frac{dM(t)}{dt}\\
&= \begin{cases} g M^{\alpha}(t) & M(t) < M_\text{mat}\\
     g M^{\alpha}(t) - r M^{\gamma}(t) & M_\text{mat} \le M(t) \le M_\text{max}
     \end{cases}
\end{split}
\end{equation}
The first line on the right-hand side of Eq.~\eqref{white} describes the early growth of the organism, when the mass $M(t)$ is smaller than a threshold mass $M_\text{mat}$, corresponding to reproductive maturity. During this first phase, energy is allocated to maintenance and biomass synthesis, just as in the growth model of Eq.~\eqref{wbe}. However in that model we had $dM/dt = (B_0/E_m) M^\alpha - (B_m/E_m) M$, with an exponent $\alpha$ for the basal metabolism and an exponent of 1 for the maintenance term. In contrast, here the metabolism and maintenance are assumed to share the same scaling exponent $\alpha$, and the two terms can be combined into a single net growth term (related to the energy left over for growth after maintenance is taken into account), with some constant prefactor $g$. This is similar to the model posited by \cite{day1997bertalanffy} to describe pre-maturity animal growth.

The second line on the right-hand side of Eq.~\eqref{white} reflects the energy allocation after reproductive maturity, with the $r M^{\gamma}$ term (where $r$ is a constant) representing the energy needed for reproductive processes, leaving less energy for growth. The assumption is that typically the exponent $\gamma > \alpha$, so that the reproductive term rises more quickly with mass than the net growth term. The organism reaches a maximum mass $M_\text{max} = (g/r)^{1/(\gamma - \alpha)}$ when the two terms cancel each other and $dM/dt \to 0$ for large $t$. Thus in this model reproduction plays the analogous role to maintenance in the model of Eq.~\eqref{wbe}: it provides a counterbalancing term that scales faster with $M$ than metabolism, resulting in an asymptotic limiting mass.

To relate the growth curve $M(t)$ from Eq.~\eqref{white} to lifetime reproduction, the organism is assumed for simplicity to have a constant Poisson death rate $\lambda$, such that the probability of surviving until time $t$ and dying between $t$ and $t+dt$ is $\lambda e^{-\lambda t} dt$. The proposed measure of mean reproductive success ${\cal R}$ is just the energy allocated to reproduction integrated over the organism lifetime and weighted by the chances of survival,
\begin{equation}
    \label{white2}
    {\cal R} = \int_0^{t_\text{max}} dt\, \lambda e^{-\lambda t} r M^{\gamma}(t). 
\end{equation}
Note the integration is up until some maximum longevity $t_\text{max}$, which provides the final parameter in the model. If we fix all parameters other than $\alpha$ in Eqs.~\eqref{white}-\eqref{white2}, there will be a particular value of $\alpha$ that maximizes ${\cal R}$. Since larger ${\cal R}$ should in principle translate to more offspring, there would be evolutionary forces that favor this optimization.

\cite{white2022metabolic} estimate some of the parameters based on earlier biological measurements, and provide guesses for others (like $t_\text{max}$) that ultimately yield an optimal value of $\alpha$ close to 3/4, compatible with Kleiber scaling. However they reiterate that this value is not unique, since it depends on the remaining parameters in the model. Through a sensitivity analysis, varying the parameters over ranges that are argued to be biologically plausible, they show that the optimal values of $\alpha$ are typically $\lesssim 1$, giving a possible explanation for the ubiquity of sublinear metabolic scaling. 

Connecting metabolism, growth, and reproduction is an intriguing proposal, but the details of the model in \cite{white2022metabolic} have attracted criticism. The biological realism of the growth model in Eq.~\eqref{white} has been questioned in relation to data from specific examples like chicken or fish~\cite{froese2023comment,kearney2023comment}. More generally, even if one accepts the model assumptions, any conclusion about optimal $\alpha$ depends heavily on the other parameters (and what ranges of these parameters are considered plausible). With different choices of the ranges, sublinear optimal $\alpha$ would not necessarily be favored. Thus while \cite{white2022metabolic} helped draw attention to evolutionary / reproductive aspects of the metabolic scaling question, there is still work to be done to make the model more realistic and perhaps bolster its predictive value.

\subsection{Concluding remarks}

The above survey of models cover some potential connections between metabolic scaling, physiology, and evolution. However, there are caveats in these phenomenological approaches: model assumptions either rest on very specific biological features that may not be universal (i.e. the geometries of transport networks) while aiming to explain interspecies scaling or else are not well motivated by the underlying biology (i.e. the number of terms and associated exponents in Eq.~\eqref{twoexp}, or assumptions i and ii of DEB theory). In turn, a lack of consensus is apparent upon reviewing the literature. Overall there has been a strong emphasis on explaining a 3/4 exponent (or behavior that effectively looks like a 3/4 exponent), which is not surprising given the centrality of Kleiber scaling in the biological literature. Given the increasing comprehensiveness of modern data surveys, and the availability of new experimental approaches for metabolic measurements at all scales, there is an opportunity for future models to grapple with the subtleties of the complex empirical picture that is emerging. Earlier theoretical works offer useful conceptual ideas: the biological rationales emphasized by the different models---transport geometries, mechanical stability, energy reserves, reproductive success---all seem potentially relevant, and the heterogeneity in scaling across different classes of organisms may in fact arise from a heterogeneity in mechanisms.



\section{Metabolism in development}\label{section5}
The metabolic scaling discussed so far is mostly based on observations in adult animals at rest. Metabolism also plays a critical role during embryo development where the embryo grows in cell number and size and acquires robust tissue identity and morphology \cite{ghosh2022developmental}. This is a complex process where metabolic rate evolves dynamically in space and time. Does metabolic rate of the embryo scale with its size or other aspects of its morphology? Current technology allows to extend the investigations of metabolic scaling to embryo development with single cell and even subcellular resolution. We discuss recent works particularly on metabolic scaling and control in embryonic development.
 \vspace{-1em}
 \begin{figure*}[t]
 \includegraphics[width=1.0\textwidth]{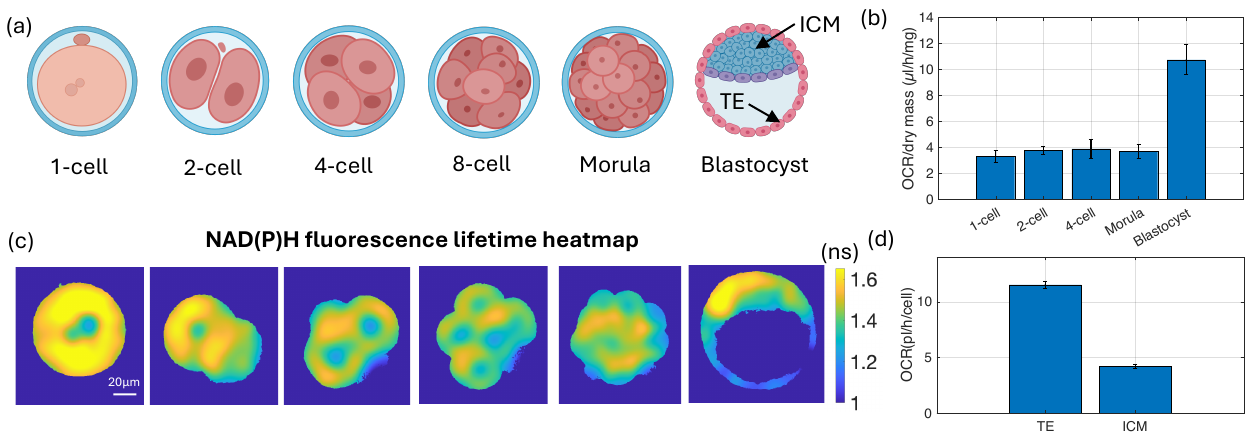}
 \caption{{\bf Spatiotemporal metabolic variations during embryo development}. (a) Schematic of mammalian preimplantation embryo development. (b) OCR normalized by dry mass of the embryo (specific OCR) throughout development. Specific OCR increases significantly at blastocyst stage despite a constant dry mass (Figure adapted from \cite{houghton1996oxygen}). (c) Heatmaps of averaged NAD(P)H fluorescence lifetime displaying complex spatiotemporal metabolic patterns during embryo development (Figure adapted from \cite{venturas2023noninvasive}). (d) Trophectoderm (TE) cells display higher OCR than inner cell mass (ICM) cells (Figure adapted from \cite{houghton2006energy}).}
 \label{fig:spatiotemporal_metab}
 \end{figure*}
\subsection{Metabolic rate in embryo development}\label{sec_embryo}
Metabolic activities are highly dynamic and display complex spatiotemporal patterns during embryo development \cite{ghosh2022developmental}. These dynamic metabolic patterns are accompanied by elaborate morphogenesis processes involving self-organization across scales, from proliferation and migration of cells to subcellular cytoskeleton rearrangements. It remains an open question as to what controls metabolic rate in embryo development. How does metabolic rate scale with the embryo size? What are the cellular and subcellular processes underlying the control of metabolic rate in embryos?  These are open questions that call for a combined effort of high-resolution experimental measurements with theories incorporating cell metabolism and self-organization.

As early as during the growth of an unfertilized egg, called oocyte, metabolic rate is regulated to accommodate the developmental needs. In early oocytes of both human and \textit{Xenopus}, the
oxygen consumption rate (OCR) of the oocytes is small because of the absence of functional complex I in the mitochondria \cite{rodriguez2022oocytes}. Complex I catalyzes the oxidation of NADH to donate electrons to oxygen through the ETC. The absence of functional complex I in early oocytes contributes to the reduced production of reactive oxygen species (ROS) to protect the oocytes from ROS damage. 

After fertilization of the oocyte, embryo development begins. Early mouse embryos develop from a single-cell zygote to a blastocyst consisting of hundreds of cells over a course of 3-4 days (Fig. \ref{fig:spatiotemporal_metab}(a)). During this period, the cellular mass of the embryo stays almost the same, but OCR of the embryo increases by over 3 times (Fig. \ref{fig:spatiotemporal_metab}(b)) \cite{mills1967oxygen, houghton1996oxygen}. This implies a composition change of the embryo, which can be characterized by a time-dependent metabolic flux in Eq.~\ref{calorimetry2main}. It remains an open question as to what accounts for the changes of metabolic fluxes over time. At the blastocyst stage, the first cell differentiation event happens, where some of the cells become inner cell mass (ICM) that will ultimately develop into the embryo proper, while the rest of the cells differentiate into trophectoderm (TE) that will develop into the placenta. By isolating cells from ICM and TE, it is observed that TE cells display a higher OCR than the ICM cells (Fig. \ref{fig:spatiotemporal_metab}(d)) \cite{houghton2006energy}, but the mechanism remains unknown. 

In addition to the dynamic changes of OCR, early embryos go through metabolic reprogramming where early cleavage embryos depend primarily on mitochondrial respiration until blastocyst stage, when glycolysis is upregulated \cite{krisher2012role}. This type of metabolic reprogramming has been observed in a broad biological context including cancer progression \cite{vander2009understanding} and T cell activation \cite{slack2015t}. Different mechanisms have been proposed to explain this metabolic reprogramming. For example, glycolysis is often associated with fast proliferating and growing cells, because it produces metabolite intermediates to support biosynthesis \cite{vander2009understanding}. Alternatively, it has been proposed that glycolysis is a faster process in terms of ATP production compared to mitochondrial respiration, hence cells transition to glycolysis for the maximization of ATP production rate \cite{kukurugya2024warburg}. In embryo development, glycolysis is critical in signaling the differentiation of TE by facilitating the nuclear localization of transcription factors, a mechanism independent of energy metabolism but highlights the importance of metabolism in cell signaling and cell fate specification \cite{chi2020glycolysis}.

To understand the control of metabolic rate, it is important to know the energetic costs in development \cite{rolfe1997cellular}. Heat release rate of developing zebrafish embryos is measured using calorimetry. The early zebrafish embryo undergoes a series of fast cell divisions to develop from a single cell to 1024 cells without a significant change of total embryo volume. Remarkably, the heat release rate of the zebrafish embryo increases significantly during this period and scales not with the number of cells but with the total surface area of the cells \cite{rodenfels2020contribution}. It is argued that the energetic costs from the maintenance and assembling of the plasma membranes could contribute appreciably to the energy budget of the embryo. Consistently, triploid \textit{Xenopus} embryos have fewer, larger cells than diploids embryos, resulting in reduced total cell surface area and a lower oxygen consumption rate compared to diploid embryos \cite{cadart2023polyploidy}. The heat release rate of the early zebrafish embryo also oscillates in synchrony with the cell cycle, which is associated with the energetic costs of cell-cycle signalling \cite{rodenfels2019heat}. Calorimetry measurement has revealed a significant increase of heat release rate during \textit{drosophila} embryogenesis. It is demonstrated that 10mJ of energy is released by the oxidation of glycogen and triacylglycerol stores during embryogenesis. Interestingly, an estimation implies that protein, RNA and DNA polymerization collectively require less than 10\% of the ATPs produced in the early embryo \cite{song2019energy}. Calorimetry is applied to measure the amount of heat released by frog embryo from cleavage stage to tailbud stage. Despite of a strong dependence of growth rate of the embryo on temperature, the amount of energy expenditure during each developmental stage is constant over the optimal temperature range \cite{nagano2014temperature}.

Relation between metabolic rate, quantified as the $\mathrm{CO_2}$ release rate, and mass of the \textit{Drosophila} larva is measured during the first, second and third instar stage of development. The average scaling exponent is 1.42, 1.04 and 0.85 corresponding to the first, second and third instar stage, respectively 
\cite{matoo2019genetic}. The origin of these scaling exponents remains unknown. Trachea is a network of internal tubes that supplies oxygen to the insect tissue without using blood. The tracheal volume in tobacco hornworm does not grow within each instar despite the increase of the larvae mass, but increases in size discretely at each molting stage
\cite{callier2011control}. This could suggest a potential limitation of oxygen supply that contributes to the sublinear scaling of hornworm larvae metabolic rate in Fig.\ref{intraspecies}.

An interesting question related to metabolic scaling is that do speeds of other biological processes depend on metabolic rates? Segmentation clock is a biological oscillator that characterizes the periodic formation of somites during the embryo development of vertebrates. The periods of the segmentation clock of diverse mammalian species are shown to be correlated with the length of embryogenesis but not with the metabolic rates or body weights of the animals \cite{lazaro2023stem}. However, perturbing metabolism, specifically NAD+/NADH ratio, changes the segmentation clock period \cite{diaz2023metabolic}. 

The complex metabolic dynamics during development is calling for an in-depth study of metabolic control beyond allometric metabolic scaling, where metabolic regulations are intertwined with energy and biosynthesis demand, cell fate specification and cell stress alleviation. However, it would be useful to systematically measure and characterize the relations between metabolic rate and embryo size to see how allometric scaling could breakdown in development. To further understand the metabolic scaling in development, it is useful to characterize the metabolic scaling of different organs in development, which can help identify new metabolic constraints.

\subsection{Spatial variations of metabolism in development}


As a result of the development of high spatial resolution metabolic profiling techniques, spatial patterns of metabolic activities are discovered across multiple stages of development. The existence of spatial variations of metabolism suggests that metabolic scaling could vary in space, and a comprehensive understanding of the metabolic scaling requires the consideration of spatial heterogeneities.

Fluorescence lifetime imaging (FLIM) of NADH combined with biophysical modeling of NADH redox cycles enables the inference of mitochondrial electron transport chain flux (ETC flux), which is proportional to OCR, with subcellular resolution. Using this technique, it is observed that a subcellular spatial gradient of ETC flux exists within a single mouse oocyte, where the ETC flux is higher in mitochondria closer to the cell membrane \cite{yang2021coarse}. ATP synthesis and proton leakage are two processes that are coupled to the ETC flux. Decoupling ATP synthesis from proton leakage through pharmaceutical perturbations reveals that the ETC flux gradient is due to heterogeneous proton leakage rather than ATP synthesis. This result suggests potential metabolism-driven self-organization of mitochondria within a single cell. 

FLIM of NAD(P)H also reveals spatial heterogeneities of NAD(P)H metabolic states throughout mouse preimplantation embryo development (Fig. \ref{fig:spatiotemporal_metab}(c)) \cite{venturas2023noninvasive}. Are these metabolic heterogeneities results of intracellular or intercellular variabilities? How do mitochondrial respiration and glycolysis contribute to these metabolic heterogeneities? These remain to be open questions for future research.

Glycolysis gradients exist through multiple stages of embryo development. During gastrulation of mouse embryo, tissue-level glycolysis gradients guide cell fate acquisition in the epiblast and mesoderm migration \cite{cao2024selective}. During the presomitic mesoderm (PSM) development of mouse and chicken embryos, a spatial gradient of glycolysis is discovered through 13C-glucose tracing experiments and spatial profiling of glycolytic enzymes \cite{bulusu2017spatiotemporal}. It is shown that glycolytic activities are higher at the posterior region as compared to the anterior region. This glycolytic gradient is correlated with cell differentiation and migration facilitating the body axis elongation \cite{oginuma2017gradient}. These results highlight the role of cell metabolism in coordinating with signaling gradients to facilitate morphogenesis.

Finally, spatial metabolic gradients can also be related to localized energy demand. During the anchor cell invasion of the basement membrane in \textit{C. elegans}, glucose transporters and mitochondria enrich at the site of invasion to power the invasive protrusions that breach basement membrane. The enrichment of glucose transporters and mitochondria lead to an enhanced ATP:ADP ratio at the invasion site \cite{garde2022localized}.

To understand the spatial control of metabolic rate in development, biophysical models involving reaction-diffusion, self-organization of mitochondria and interaction between metabolism and signaling gradients are needed.

\vspace{1em}
\section{Conclusions and Outlook}\label{section6}
Metabolism is a complex system with many interacting chemical species and reactions.  We have discussed the fundamentals of metabolism in the context of energy flows, coarse-grained metabolic biochemistry, and metabolic rate. We presented a framework based on irreversible thermodynamics of spatially heterogeneous systems, allowing us to decompose metabolic rate into separate contributions: from internal compositional changes, exchange of molecules with the environment, and growth as well as spatial heterogeneities. This can serve as a reference for future studies under different circumstances, as it applies beyond basal metabolic rate measurements (but excludes locomotive activity). The thermodynamic formalism naturally leads to a macrochemical equation---encapsulating the net effect of all reactions on the enthalpy change in the system---which is useful in estimating the heat release rate.  On one hand, this makes it simple to estimate the enthalpy change without the knowledge of the full chemical reaction network. On the other hand,  this hides the information about the reaction kinetics which are necessary to understand the metabolic rate. A somewhat related work studies a forward coarse-graining approach from the full chemical reaction network in well-mixed systems \cite{avanzini2023circuit}. This can become relevant when one needs to model the reaction kinetics accurately, since the macrochemical equation does not contain this information. It would be interesting to explore whether the two approaches can be combined.

We reviewed comprehensive data sets on metabolic scaling with mass, including interspecies and intraspecies organism-level measurements, as well as more fine-grained tissue- and cell-level  measurements. The picture that emerges is that ``metabolic scaling'' is not a single phenomenon, but really a family of different scaling behaviors, depending on which subsets of the data are analyzed. The only trend that can plausibly be claimed as universal is that basal metabolic rate $B$ scales linearly with organism mass $M$ over 22 orders of magnitude in $M$, from the smallest bacterial cells to the largest mammals. This is a striking illustration of Jacques Monod's famous claim that ``Anything that is true of E. coli must be true for elephants, except more so.''~\cite{morange2010scientific}. Behind this trend are deep similarities in cellular biochemistry across organisms, leading to a universal metabolic time scale (at least for aerobic respiration).

Because of this universal linear trend, the mass-specific metabolic rate, $b = B/M$, is a more useful observable to plot versus $M$, since it captures the residual scaling behavior after the overall isometric trend is factored out. Analyzed in this way, we observe sublinear scaling patterns in smaller subsets of the data, for example mammals, ectotherms, and plants separately scale with exponents close to (but not exactly) the Kleiber law expectation of $b \sim M^{-1/4}$. This sublinearity is echoed at other levels of biological organization: the trends in $b$ over distinct growth stages in organisms like insects, or in the $b$ values estimated for different organs as a function of organism mass. Notably, there is heterogeneity in the sublinear scaling exponents: it is certainly not the case that a single exponent governs all these phenomena.  In this sense the criticisms of elevating the Kleiber exponent into a ``law'' are completely justified~\cite{dodds2001re,hulbert2014sceptics}.

Despite these caveats, the evidence of different sublinear metabolic trends is compelling, and has implications for other aspects of animal biology. Metabolism is closely tied to growth, and its scaling can be used to constrain dynamical models for organism mass over time. We show how such models can be grounded in the macrochemical reaction framework, and related to the energy cost of biomass synthesis (another quantity that exhibits a roughly consistent order of magnitude across different organism sizes). Metabolic scaling also has consequences at the population level, potentially giving insights into the energy usage of whole ecosystems.

For all these reasons, seeking out the biophysical mechanisms behind metabolic scaling remains an important line of inquiry. We summarized a range of theoretical models that aim to describe how conserved scaling behavior with body size can emerge. To date, none of these theories have been conclusively validated, and only some of the hypotheses provide testable predictions. One difficulty in creating a comprehensive theoretical model is that data sets often range over a wide diversity of species. For example the data covering the morphological characteristics of organs, cell number, and metabolic rates may not all come from the same organisms. Thus, it will be crucial in the future to investigate these aspects within the same model organism. Additionally, models should tackle the complexity of organismal development, as physiological features and anatomical structures change during this process. There is room for new theoretical frameworks that can incorporate the interplay between metabolism and morphogenesis:  whole organism and individual tissue/organ metabolic rates,  as well as the morphology of internal anatomy, all display allometric scaling relations with body size~\cite{schmidt1984scaling}. The ideal model would capture at least some of this complexity, since the scaling of anatomy and metabolism are likely to be causally linked.

In recent years, planarians have emerged as a promising model to collect multiple observables for the same species. These include metabolic rate and body mass composition measurements \cite{thommen2019body} and internal anatomy such as the morphometric scaling of the gut---the primary transport organ in planarians--- spanning a 1000-fold change in body mass~\cite{hanauer2026model}. As studies in developmental biology increasingly turn toward metabolism, we expect more examples to emerge in other model organisms. It is an exciting era because new experimental techniques have enabled measurement of metabolic rates with single-cell and even subcellular resolution. These techniques open up opportunities to study metabolic scaling across different scales in development. How does the size of cell organelles scale with the metabolic rate of the cell? Can we get more precise estimates for how the metabolic rate of a tissue scales with its size during development? What are the biological implications of these scaling relations in morphogenesis? Answering these questions will help uncover new mechanisms responsible for the robustness of development.


In particular, studying metabolic scaling at the organelle level can help uncover molecular mechanisms underlying the scaling of organelle sizes. During reductive cleavage division in zebrafish embryo development, 
the mitotic spindle volume scales with the cell surface area \cite{rieckhoff2020spindle} while the heat release rate of the embryo increases in proportion to the total cell surface area \cite{rodenfels2020contribution}, implying that the mitotic spindle volume scales with the metabolic rate of the cell. 
It remains to be explored what biological mechanisms contribute to this correlation. 

Controllable developmental systems such as organoids, which are organs grown \textit{in vitro} \cite{clevers2016modeling}, may provide new insights into metabolic rate studies. The advance of organoid technology has greatly expanded our capabilities to study organ development by lifting many technological and ethical constraints. Meanwhile, organoids represent systems that are freed from \textit{in vivo} biological constraints, enabling the exploration of a much more expanded morphospace to understand how organs develop robustly. Organoids mimic the 3D structures of organs, which provides a natural system to study how metabolic rate scales with organoid size. Taking brain organoids as an example, 3D brain organoids can be cultured from iPSC cells and develop into a single layer of neuroepithelium surrounding the lumen acquiring apical-bascal polarity. The neuroepithelium will futher develop into layers of radial glial cells and intermediate neuron progenitors before differentiating into neurons at the outer layers \cite{lancaster2013cerebral}. In fact, one of the original motivations to develop brain organoids is to study brain disorders such as microcephaly, where abnormal brain development leads to smaller than normal brains \cite{lancaster2013cerebral}. Metabolic disorders have been shown to be associated with microcephaly \cite{xing2024role}. Therefore, understanding how metabolic rate scales with brain organoid size can potentially provide insights into how metabolism regulates neurogenesis. This requires not only the measurement of the global metabolic rate of the organoids, but also the spatiotemporal mapping of the metabolic rate throughout organoid development during cell proliferation and differentiation. In addition, metabolic rate normalized by mass, by cell number and by mitochondrial mass needs to be measured in the same biological system to provide insights into how metabolic rates are controlled at different levels.

Studies of metabolic scaling have predominantly focused on aerobic organisms so far. In development, anaerobic pathways can have significant contributions. Early embryos upregulate glycolysis to support rapid cell proliferation, a phenomenon reminiscent of the Warburg effect \cite{krisher2012role}. However, the oxygen consumption rate also increases as development progresses \cite{houghton1996oxygen}. Hence, the study of metabolic scaling in development requires the consideration of both aerobic and anaerobic pathways. Thornton's rule is expected to fail when the contribution of anaerobic pathways to heat release is significant, in which case a respirometer cannot be used to infer the total heat release rate of the organism. While a calorimeter can be used instead, combined measurements of mitochondrial respiration rate and glycolytic fluxes are needed to understand the scaling of the heat release rate during development.

Identifying the contributions from different cellular processes could help unravel metabolic scaling.
Recently, there has been  a growing interest in the study of entropy production, energetic cost (e.g., ATP consumption), and heat release in active biophysical processes \cite{yang2021physical,tu_nonequilibrium_2026}. Examples cover a variety of systems: synthesis and maintenance of biomolecules \cite{lynch2015bioenergetic,mahmoudabadi2019defining,mori2023functional},  cytoskeletal gels \cite{foster2023dissipation,sakamoto2024f,fodor2016nonequilibrium}, 
posttranscriptional regulation \cite{ilker2024bioenergetic},
biochemical sensing \cite{govern2014energy,sartori2015free,mehta2012energetic}, error correction \cite{schilling_why_2026,sartori2015thermodynamics} and  signaling \cite{wang2020price,ouldridge2017thermodynamics,song2021cost}. These works provide tools to relate cellular processes to dissipation, enabling a bottom-up approach for the energetics of living systems. 

The emergence of metabolic scaling with different
apparent exponents raises the question whether there are broad, underlying physical principles that govern these scaling laws. This connects to interesting questions in the physics of nonequilibrium systems more generally.  
For example, the metabolic rate is a nonequilibrium signature of living matter. This motivates exploring  analogous quantities in active matter, such as the entropy production rate.  As we discussed in Section \ref{sec:iib}, the entropy production rate is closely related to the metabolic rate, and can  be theoretically estimated for a given system. Metabolic scaling then corresponds to the question of how the entropy production rate scales with system size in active matter. 
In general, the entropy production rate is an extensive quantity that should scale linearly with volume. However, when a system is close to a critical point it may exhibit nontrivial scaling. There is a well defined notion of ``universality'' in statistical mechanics---complex systems with different microscopic details can yield the same macroscopic behavior---but whether metabolic scaling fits into this idea remains to be seen.

 Recent works have explored scaling of entropy production rate in other contexts. These include scaling upon coarse-graining for chemical reaction and biological kinetic networks \cite{yu2021inverse}, and power law scaling near criticality in active field theories (active models A and B \cite{caballero2020stealth,paoluzzi2022scaling}). While the entropy production rate naturally depends on microscopic details, it remains an open question whether universality‑like classes exist for its scaling, based on distinguishing the dominant contributions. A macroscopic stochastic thermodynamics perspective \cite{falasco2025macroscopic} could also represent a novel frontier in studying this problem. Systems with self-organized criticality are another example of complex systems where power law scaling appears \cite{bak1988self,jensen1998self}. Overall, nonequilibrium statistical physics approaches may provide insights for understanding metabolic scaling. 

\acknowledgments{We thank many colleagues for stimulating discussions including Jonathan Rodenfels, Jochen Rink, Christian Hanauer, Omar Adame-Arana, Jean-François Joanny, Thomas Lecuit, Jacqueline Janssen, Stefano Bo, Pablo Sartori, Anthony Hyman, Miki Ebisuya, and Joshua Holmes. We also thank Christian Massino, Mounir El Hankouri, Junjie Liu for a critical reading of the manuscript. This work received support from the French
government under the France 2030 investment plan, as part of the Initiative d’Excellence d’Aix-Marseille Université - Amidex (AMX-23-CEI-064). Figure 1, Figure 2 and Figure 10(a) are created with BioRender.com.}



\appendix

\section{Thermodynamic relations}
\label{appendixa}
We consider the Gibbs free energy $G=E-TS+pV$ as thermodynamic potential for given pressure $p$, which depends on the numbers of molecules $N_i$ of component $i=0,\dots,K$, and temperature T:
$G=G(N_0,\dots,N_K,p,T)$. We then define
the chemical potentials $\mu_i=\partial G/\partial N_i$, the entropy $S=-\partial G/\partial T$ and we obtain the volume as 
$V=\partial G/\partial p$. The enthalpy is
defined as $H=G+TS$. From the extensitivity of $G$ for a homogeneous phase, with $\alpha G=G(\alpha N_0,\dots,\alpha N_K,p,T)$ it follows that
$G=\sum_{i=0}^K \mu_i N_i$. We thus have
\begin{eqnarray}
    S&=&\sum_{i=0}^K s_i N_i \\
    H&=&\sum_{i=0}^K h_i N_i \\
    V&=&\sum_{i=0}^K \nu_i N_i \quad,
\end{eqnarray}
with entropy per molecule $s_i=-\partial \mu_i/\partial T$, enthalpy per molecule
$h_i=\mu_i+Ts_i$ and molecular volume $\nu_i=\partial \mu_i/\partial p$.

When considering spatially heterogeneous systems at local thermodynamic equilibrium, we introduce densities $g=G/V$, $s=S/V$, $h=H/V$ and $e=E/V$ of extensive quantities and concentrations $n_i=N_i/V$, where $V$ is a small volume element. These densities obey 
\begin{eqnarray}
    g&=&\sum_{i=0}^K \mu_i n_i \label{eq:g}\\
    s&=&\sum_{i=0}^K s_i n_i \\
    h&=&\sum_{i=0}^K h_i n_i \quad \label{eq:heq}.
 \end{eqnarray}
We can now express the time evolution of the Gibb free energy density at constant pressure $p$ as $\partial_t g =\partial_t (G/V)$
with
\begin{eqnarray}
    \partial_t g 
    &=& \frac{1}{V}\left (\sum_{i=0}^K\frac{\partial G}{\partial N_i}\partial_t N_i+\frac{\partial G}{\partial T}\partial_t T -\frac{G}{V}\partial_t V\right ) \nonumber\\
    &=& \sum_{i=0}^K \mu_i \partial_t n_i
    -s \partial_t T
\end{eqnarray}
where $\partial_t n_i=(\partial_t N_i)/V - (\partial_t V)N_i/V^2$. Furthermore, we
find
\begin{eqnarray}
    \partial_t s 
    &=& \frac{1}{V}\left (\sum_{i=0}^K-\frac{\partial \mu_i}{\partial T}\partial_t N_i+\frac{\partial S}{\partial T}\partial_t T -\frac{S}{V}\partial_t V\right ) \nonumber\\
    &=& \sum_{i=0}^K s_i \partial_t n_i
    +\frac{c_p}{T} \partial_t T \quad ,
    \label{eq:dts}
\end{eqnarray}
where we have defined the volume specific heat at constant pressure $c_p=(T/V) \partial S/\partial T$. Finally, we obtain
for $h=g+Ts$ the time evolution
$\partial_t h=\partial_t g+T\partial_t s
+s\partial_t T$ with
\begin{equation}
 \partial_t h = \sum_{i=0}^K h_i \partial_t n_i
    +c_p \partial_t T \quad . \label{eq:dth}
\end{equation}
Since energy is conserved and $e=h-p$, we 
have for contant pressure 
\begin{equation}
    \partial_t h=-\nabla\cdot {\bf j}_e \label{eq:dht}
\end{equation} where ${\bf j}_e$
denotes the energy flux. We thus obtain an
equation for the temperature dynamics
\begin{equation}
c_p \partial_t T + \nabla \cdot {\bf j}_e
= -\sum_{i=0}^K h_i \partial_t n_i \quad .
\end{equation}
Using the balance equations $\partial_t n_i +\nabla\cdot {\bf j}_i=r_i$ of 
molecule numbers (described in \eqref{eqchm1}), we obtain Eq. (\ref{eqheatbalance}), with heat flux
\begin{equation}
    {\bf j}_q={\bf j}_e-\sum_{i=0}^K h_i {\bf j}_i \quad .\label{eq:heatjq}
\end{equation}
Similarly, combining Eqns. (\ref{eq:dts})
and (\ref{eq:dth}), we have
\begin{equation}
T\partial_t s = -\sum_{i=0}^K \mu_i\partial_t n_i -\nabla\cdot {\bf j}_e
\quad ,
\end{equation}
from which we obtain
Eqns. (\ref{eqs1}-\ref{epr}) with
entropy flux 
\begin{equation}
    {\bf j}_s = \frac{1}{T}({\bf j}_e-\sum_{i=0}^K {\bf j}_i\mu_i) \quad, \label{eq:js1}
\end{equation}
or ${\bf j}_s={\bf j}_q/T+\sum_{i=0}^K  {\bf j}_is_i $. Thus entropy flux is associated with heat flux plus the entropy carried 
by chemical potentials of transported molecules.




\section{Derivation of the total heat release rate}\label{appendixb}
In this section, we show the detailed derivation of the total heat release rate given in Eq.~\eqref{calorimetry0main}. The heat leaving the system defines the
total heat release rate
\begin{equation}
    \dot Q= \oint_{\partial V_S}  {\bf j}_q \cdot {\bf dA}
\end{equation}
where ${\bf dA}={
\bf\hat n} dA$ is the surface element $dA$ pointing in the direction of the outward surface normal vector ${\bf\hat n}$.  In order to take into account the volume changes of the system, we first derive the fluxes across a moving interface. 
\subsection{Fluxes across a moving interface}
In general, a given system may exhibit growth associated with motion of the outer surface with velocity ${\bf u}=u {\bf \hat n}$ and hence the system volume can change according to  $\partial_t V_{\mathcal{S}}=\oint_{\partial V_{\mathcal{S}}} {\bf u}\cdot d{\bf A}$. The system is in contact with an \qmark{environment} and can exchange energy and matter through the system-environment interface, while the system and environment together (${\mathcal S}\cup\mathcal{E}$) may form an isolated system. 

In order to obtain the fluxes of local densities across the system-environment interface we first write the time evolution equations of densities. The local density $y$ of any extensive property $Y=yV$ in a volume element $V$, e.g., $y\in\{n_i, e, s, h\}$, evolves in the form:
\begin{equation}
    \frac{\partial y}{\partial t}+\nabla \cdot {\bf j}_y=\dot{\theta}_y \label{dydt}
\end{equation}
where ${\bf j}_y$ is the flux of $y$ by transport, and the associated local source term $\dot{\theta}_y\in\{r_i, 0, \dot{\theta}_s, \partial_t p \}$. For instance, first and second laws of thermodynamics demand $\dot{\theta}_\ed=0$ and  $\dot{\theta}_s>0$ respectively. We can discuss the net exchange flux of $y$ at the system-environment interface using the relation
\begin{equation}
    \frac{dY^\mathcal{S}}{dt}+\frac{dY^\mathcal{E}}{dt}=\int _{V_{\mathcal{S}} \cup V_{\mathcal{E}}} \dot{\theta}_y dV 
    \quad , \label{eq:constot1}
\end{equation}
where $Y^{\mathcal{S}/\mathcal{E}}=\int_{V_{\mathcal{S}/\mathcal{E}}} ydV$. Given the  interface velocity ${\bf u}$, the integrated quantities follow:
\begin{equation}
\frac{dY^{\mathcal{S}/\mathcal{E}}}{dt}=\int_{V_{\mathcal{S}/\mathcal{E}}}\frac{\partial y}{\partial t}dV\pm\oint_{\partial V_{\mathcal{S}/\mathcal{E}}}y^{\rm in/out}{\bf u}\cdot d{\bf A}\ ,
\end{equation}
where \qmark{in} and \qmark{out} represent inside and outside with respect to the system. Using Eq.~\eqref{dydt} yields:
\begin{equation}
\frac{dY^{\mathcal{S}/\mathcal{E}}}{dt}=\mp\oint_{\partial V_{\mathcal{S}/\mathcal{E}}}\left({\bf j}_y^{\rm in/out}- y^{\rm in/out}{\bf u}\right)\cdot d{\bf A} + \int _{V_{\mathcal{S}/\mathcal{E}} }\dot{\theta}_y dV 
\end{equation}
These expressions used in  Eq.~\eqref{eq:constot1} imply the fluxes across the interface
\begin{equation}
    {\bf j}_y^{\rm int}\equiv {\bf j}_y^{\rm in/out}-y^{\rm in/out}{\bf u}\
\end{equation}
where the superscripts \qmark{int} denote the net fluxes with respect to the moving interface. As the
current ${\bf j}^{\rm int}_y$ is continuous across the interface, we have a jump
in current
\begin{equation}
    {\bf j}_y^{\rm out}-{\bf j}_y^{\rm in}={\bf u}(y^{\rm out}-y^{\rm in})
\end{equation}
across the interface.

Applying this relation to molecular concentrations, energy density, and entropy density, we get:
\begin{subequations}
\begin{align}
      {\bf j}_i^{\rm int}&\equiv {\bf j}_i^{\rm in/out}-n_i^{\rm in/out}{\bf u}\ ,\label{eqjiint}\\
      {\bf j}_{\ed}^{\rm int}&\equiv{\bf j}_{\ed}^{\rm in/out}-{\ed}^{\rm in/out}{\bf u}\ \label{eqjeint},\\
      {\bf j}_s^{\rm int}&\equiv{\bf j}_{s}^{\rm in/out}-{s}^{\rm in/out}{\bf u}\ \label{jsint},\\
    {\bf j}_{h}^{\rm int}&\equiv{\bf j}_{h}^{\rm in/out}-{h}^{\rm in/out}{\bf u}\ .
    \end{align}
\end{subequations}
The heat flux behaves differently. Using Eq. (\ref{eq:heatjq}), we find 
\begin{equation}
    {\bf j}_q^{\rm int}={\bf j}^{\rm in/out}_q +p{\bf u} \quad .
\end{equation}

\subsection{Total heat release rate}\label{appb2}
The total heat release rate from the system is given by:
\begin{equation}
     \dot{Q}=\oint_{\partial V_{\mathcal{S}}}{\bf j}_q^{\rm out}\cdot d{\bf A} \label{qintflow}
\end{equation}
We have
\begin{eqnarray}
    {\bf j}_q^{\rm out}&=&{\bf j}_e^{\rm out}-\sum_{i=0}^K h_i^{\rm out}{\bf j}_i^{\rm out}\\
    &=&{\bf j}_e^{\rm int}-\sum_{i=0}^K h_i^{\rm out}{\bf j}_i^{\rm int}+(e^{\rm out}-h^{\rm out}){\bf u} 
\end{eqnarray}
where we used Eqs.~\eqref{eqjiint} and \eqref{eqjeint} in the second line. 
We can then write

\begin{eqnarray}
    \frac{dE_{\mathcal{S}}}{dt}&=&-\oint_{\partial V_{\mathcal{S}}} {\bf j}_e^{\rm int}\cdot {\bf dA}\nonumber \\
    &=&-\dot Q+ \oint_{\partial V_{\mathcal{S}}} (e^{\rm out}-h^{\rm out}){\bf u}\cdot{\bf dA}\nonumber \\
    &&- \sum_{i=0}^K \oint_{\partial V_{\mathcal{S}}} h_i^{\rm out}{\bf j}_i^{\rm int}\cdot {\bf dA} \quad. \label{eq:des1}
\end{eqnarray}
Using Eqs.~\eqref{eq:dht} and \eqref{eqjeint} we also have: 
\begin{equation}
   \frac{dE_{\mathcal{S}}}{dt}=\int_{V_\mathcal{S}} \partial_t h \ dV+\oint_{\partial V_{\mathcal{S}}}  e^{\rm in} {\bf u}\cdot{\bf dA}\quad .\label{eq:des2}
\end{equation}
Equating Eqs.~\eqref{eq:des1} and \eqref{eq:des2} with $e^{\rm in/out}-h^{\rm in/out}=p$ we find Eq.~\eqref{calorimetry0main}.

\section{Element conservation in macrochemical reactions}\label{app:c}
In order to express the overall chemical transformation in terms of macrochemical reactions where the spatial information is integrated on compartments of system $\mathcal{S}$, $\mathcal{E}$, and to their interface, we use the element conservation. From Eq.~\eqref{eq:constot1}, the total number of $i$-molecules $N_i^{\rm tot}=N_i^{\mathcal{S}}+N_i^{\mathcal{E}}$ evolve according to:
\begin{equation}
        \frac{dN_i ^{\rm tot}}{dt}=\int _{V_{\mathcal{S}} \cup V_{\mathcal{E}}} r_i dV 
\end{equation}
Assuming there are no metabolic reactions in environment $\mathcal{E}$, and using Eq. \eqref{eqchm1} of main text and Eq. \eqref{eqjiint} in appendix, we get:
\begin{equation}
        \frac{dN_i ^{\rm tot}}{dt}=\int _{V_{\mathcal{S}} } \frac{\partial n_i}{\partial t} dV +\oint_{\partial V_{\mathcal{S}}} {\bf j}_i^{\rm int}\cdot d{\bf A}+\oint_{\partial V_{\mathcal{S}}} {n}_i^{\rm in} {\bf u} \cdot d{\bf A}
\end{equation}
We define the net exchange fluxes as $\Phi_i=\oint_{\partial V_{\mathcal{S}}}{\bf j}_i^{\rm int}\cdot d{\bf A}$, the rate of net change of molecules in the system $\Gamma_i=\int_{V_{\mathcal{S}}}\partial_t n_idV$. Accordingly, we get:
\begin{equation}
        \frac{dN_i ^{\rm tot}}{dt}=\Gamma_i+\Phi_i+\oint_{\partial V_{\mathcal{S}}} {n}_i^{\rm in} {\bf u} \cdot d{\bf A} \quad. \label{eqdnitot}
\end{equation}
Now, we aim to write the last term in a similar form to the fluxes $\Gamma_i, \Phi_i$. To achieve this, we define biomass growth where biomass is an effective composition of chemical elements. As discussed in Section \ref{sec:IIa4}, each molecule $i$ is composed of $\eta_{ji}$ number of $j$-elements, whereas the effective species is composed of $\epsilon_{j}=\sum_{i=0}^K \eta_{ji}n_i$ number of $j$-elements.  The total number of element $j$ in  $V_{\mathcal{S}} \cup V_{\mathcal{E}}$ is $\sum_{i=0}^K\eta_{ji}N_i^{\rm tot}$. The element conservation states that $\sum_{i=0}^K \eta_{ji}\frac{dN_i}{dt}=0$ for each $j$.  This  can be written using Eq.~\eqref{eqdnitot} as
\begin{eqnarray}
    \sum_{i=0}^K \eta_{ji}\left(\Gamma_i+\Phi_i\right)+\bar{\epsilon}_j\Pi_b =0 \quad  \label{eqelement1}
\end{eqnarray}
where $\Pi_b=\oint_{\partial V_{\mathcal{S}}} {n}_b {\bf u} \cdot d{\bf A}$, $n_b$ is the concentration of a reference element, which we chose to be carbon and  ${\epsilon_j}/{{n}_b}$ is the number of atoms of element $j$ in an effective biomass molecule with one carbon atom. We also defined $\bar{\epsilon}_j$ which is the relative space-averaged growth of number of $j$ elements as
\begin{equation}
\bar{\epsilon}_j=\frac{\oint_{\partial V_{\mathcal{S}}} (\epsilon_j/{n}_b^{\rm in}) {n}_b^{\rm in} {\bf u} \cdot d{\bf A}}{\oint_{\partial V_{\mathcal{S}}} {n}_b^{\rm in} {\bf u} \cdot d{\bf A}} \quad.
\end{equation}
Since we take carbon as the reference element, $\bar{\epsilon}=1$ for carbon.

As in Section  \ref{sec:IIa4}, we restrict our attention to a subset of chemical species with nonzero $\Gamma_i$ and $\Phi_i$ and redefine the chemical species array as in Eqs.~\eqref{eq:xtilde},\eqref{rvect}. This reduced set contains $K_{\rm net}$ components. Accordingly, we can also define the number of elements in this reduced set as ${\mathbb{E}}_{\rm net}$. Then, the new element composition matrix $\tilde{\eta}_{ij}$ has dimensions $K_{\rm net}\times \mathbb{E}_{\rm net}$.  Eq.~\eqref{eqelement1} can be written as:
\begin{equation}
     \sum_{i=1}^{K_{\rm net}} \tilde{{\eta}}_{ji} \mathcal{I}_i=0  \label{eqelement3}
\end{equation}
where $\tilde{\eta}_{ji}$ is the number of $j$ elements in $i$ molecule in chemical species array Eq.~\eqref{eq:xtilde} for $i<K_{\rm net}$ and for the biomass, i.e., $i=K_{\rm net}$, $\tilde{\eta}_{ji}=\bar{\epsilon}_j$. Thus, using the above relation with Eq.~\eqref{eqnetreaction1}, we derive Eq.~\eqref{eqelement2}.

\section{Entropy production and free energy dissipation rates in the system}
\subsection{Entropy production rate}\label{app:epr}
The local entropy production rate is given in main text by Eq.~\eqref{epr}. We can integrate this over the system volume $V_{\mathcal{S}}$ to obtain the total entropy production rate in the system $\dot{\Theta}_{s} ^\mathcal{S}$. This results 
\begin{eqnarray}
    \dot{\Theta}_{s} ^\mathcal{S}=&-&\sum_{i=0}^K\int_{V_{\mathcal{S}}} r_i \frac{\mu_i}{T} dV\nonumber \\ &-& \sum_{i=0}^K\int_{V_{\mathcal{S}}}\left(\frac{{\bf j}_i}{T}\cdot \nabla \mu_i \right) dV\nonumber \\ &-& \int_{V_{\mathcal{S}}}\left(\frac{{\bf j}_s}{T}\cdot \nabla T\right) dV \quad .
\end{eqnarray}
We can rewrite this equation by using Eq.~\eqref{eqchm1} in the main text and Eq.~\eqref{eq:js1}:
\begin{eqnarray}
    \dot{\Theta}_{s} ^\mathcal{S}=&-&\sum_{i=0}^K\int_{V_{\mathcal{S}}} \frac{\partial n_i}{\partial t} \hat{\mu_i}^{\rm in}  dV\nonumber \\ &-& \sum_{i=0}^K\oint_{\partial V_{\mathcal{S}}} \hat{\mu_i}^{\rm in} {\bf j}_i^{\rm in}\cdot d{\bf A}\nonumber \\ &-&\int_{V_{\mathcal{S}}}\left(\frac{{\bf j}_e}{T^2}\cdot \nabla T\right) dV  \label{totaleprsys}
\end{eqnarray}
where we introduced $\hat{\mu_i}^{\rm in/out}=\mu_i^{\rm in/out}/T^{\rm in/out}$. Using Eq.~\eqref{eqjiint} for the particle flux across the interface, we can rewrite Eq.~\eqref{totaleprsys} as:
\begin{eqnarray}
    \dot{\Theta}_{s} ^\mathcal{S}=&-&\sum_{i=0}^K\int_{V_{\mathcal{S}}} \hat{\mu_i}^{\rm in}\frac{\partial n_i}{\partial t}  dV\nonumber \\ &-& \sum_{i=0}^K\oint_{\partial V_{\mathcal{S}}} \hat{\mu_i}^{\rm in} {\bf j}_i^{\rm int}\cdot d{\bf A}\nonumber \\ &-& \sum_{i=0}^K\oint_{\partial V_{\mathcal{S}}} \hat{\mu_i}^{\rm in} {n}_i^{\rm in} {\bf u}\cdot d{\bf A}\nonumber \\ &-& \int_{V_{\mathcal{S}}}\left(\frac{{\bf j}_e}{T^2}\cdot \nabla T\right) dV \quad .
\end{eqnarray}
Now we use an analog transformation as we did for the heat release rate in in Section \ref{sec:IIa4}. This leads to
\begin{eqnarray}
    \dot{\Theta}_{s} ^\mathcal{S}=&-&\sum_{i=0}^K \hat{\mu}_i ^{\rm in}\Gamma_i -\sum_{i=0}^K \hat{\mu}_i ^{\rm in}\Phi_i- \hat{\mu}_b ^{\rm }\Pi_b\nonumber 
    \\&+& \delta\dot{\Theta}_{s} ^\mathcal{S} \label{eq:totalepr}
\end{eqnarray}
where $\bar{\hat{\mu}}_i ^{\rm in}=V_{\mathcal{S}}^{-1}\int_{V_{\mathcal{S}}} \hat{\mu}_i \ dV$,  $\bar{\hat{\mu}}_b = A^{-1}\oint_{\partial V_{\mathcal{S}}} \hat{\mu
}_b \ d{A}$ and $\hat{\mu
}_b=\sum_{i=0}^K \hat{\mu
}_i n_i/n_b$. The last term $\delta\dot{\Theta}_{s} ^\mathcal{S}$ is due to the spatial heterogeneities in chemical potentials $\delta \hat{\mu}_{i}=\hat{\mu}_{i}-\bar{\hat{\mu}}_{i}^{\rm in}$, $\delta \hat{\mu}_{b}=\hat{\mu}_{b}-\bar{\hat{\mu}}_{b}$ and in temperature. It is given by:
\begin{eqnarray}
    \delta\dot{\Theta}_{s} ^\mathcal{S}=&-&\sum_{i=0}^K \int_{V_{\mathcal{S}}}\delta \hat{\mu}_i\frac{\partial n_i}{\partial t}dV-\sum_{i=0}^K\oint_{\partial V_{\mathcal{S}}} \delta \hat{\mu}_i{\bf j}_i^{\rm int}\cdot d{\bf A} \nonumber \\&-&\oint_{\partial V_{\mathcal{S}}} \delta \hat{\mu}_b {n}_b {\bf u} \cdot d{\bf A} \nonumber \\&-& \int_{V_{\mathcal{S}}}\left(\frac{{\bf j}_e}{T^2}\cdot \nabla T\right) dV  \label{eq:deldepr}
\end{eqnarray}

 We finally apply the transformations for the net reaction following Sections \ref{sec:IIa4}, \ref{sec:macro}, i.e., introduce Eq.~\eqref{rvect} reduced to $K_{\rm net}$ non-zero components and the array $\hat{\bm{\mu}}=(\bar{\hat{\bm{\mu}}}^{\rm in},\hat{\mu}_b) $,  we reach an analog of Eq.~\eqref{qdotmacrochemical2} for the total entropy production rate in the system:

\begin{equation}
    \dot{\Theta}_{s} ^\mathcal{S}=-\sum_{\beta=1}^{\tilde{R}}\mathcal{R}_{\beta}\Delta \hat{G}_{\beta}+ \delta\dot{\Theta}_{s} ^\mathcal{S}\quad \label{eq:totaleprv2}
\end{equation}
where $\Delta \hat{G}_{\beta}=\sum_{i=1}^{K_{\rm net}} 
\tilde{\sigma}_{i\beta} \bar{\hat{\mu}}_i$. 
A final remark is that the system also induces entropy production rate in the environment. This additional amount is determined by integrating Eq.~\eqref{epr} of the main text in environment volume $V_\mathcal{E}$ as
\begin{equation}
    \dot{\Theta}_{s} ^{\mathcal{E}\leftarrow\mathcal{S}}=- \sum_{i=0}^K\int_{V_{\mathcal{E}}}\left(\frac{{\bf j}_i}{T}\cdot \nabla \mu_i \right) dV- \int_{V_{\mathcal{E}}}\left(\frac{{\bf j}_s}{T}\cdot \nabla T\right) dV \label{extraeprsys}
\end{equation}
since no metabolic reactions occur outside the system and we used \qmark{$\mathcal{E}\leftarrow\mathcal{S}$} to denote the marginal entropy production rate in environment induced by the system of interest. The majority of the contribution comes near the interface, while in general it is assumed that environment is a large reservoir, thus the contributions from $\nabla {\mu}_i$, $\nabla T$ terms are negligible.

\subsection{Free energy dissipation rate for isothermal systems}
For isothermal systems, it is more convenient to study free energy dissipation rate. Given the Gibbs free energy density in Eq.~\eqref{eq:g}, the free energy balance is written as:
\begin{equation}
    \frac{\partial g}{\partial t}+\nabla {\cdot}\sum_{i=0}^K{\bf j}_i\mu_i=-\dot{\theta}_g
\end{equation}
where $\dot{\theta}_g$ is the free energy dissipation rate. We can determine $\dot{\theta}_g$ from previous relations. Thus, using Eqs.~\eqref{eq:g} and \eqref{epr}, we reach:
\begin{equation}
\dot{\theta}_g=T\dot{\theta}_{s}+{\bf j}_s\nabla T+s\frac{\partial T}{\partial t}.
\end{equation}
Then, we follow the similar lines as before (see Sections  \ref{sec:IIa4}, \ref{sec:macro}, and \ref{app:epr}) for expressing the total free energy dissipation rate in the system in terms of macrochemical reactions. As a result, we obtain when $\partial_t T=0$ and $\nabla T=0$:
\begin{equation}
    \dot{\Theta}_{g} ^\mathcal{S}=\int_{V_{\mathcal{S}}}\dot{\theta}_g \ dV=-\sum_{\beta=1}^{\tilde{R}}\mathcal{R}_{\beta}\Delta G_{\beta} + \delta\dot{\Theta}_{g} ^\mathcal{S}\quad\quad 
\end{equation}
where $\Delta {G}_{\beta}=\sum_{i=1}^{K_{\rm net}} 
\tilde{\sigma}_{i\beta} \bar{{\mu}}_i$ and
\begin{eqnarray}
    \delta\dot{\Theta}_{g} ^\mathcal{S}=&-&\sum_{i=0}^K \int_{V_{\mathcal{S}}}\delta {\mu}_i\frac{\partial n_i}{\partial t}dV-\sum_{i=0}^K\oint_{\partial V_{\mathcal{S}}} \delta {\mu}_i {\bf j}_i^{\rm int}\cdot d{\bf A} \nonumber \\&-&\oint_{\partial V_{\mathcal{S}}} \delta {\mu}_b {n}_b {\bf u} \cdot d{\bf A}  \quad \label{eq:deldg}
\end{eqnarray}
with $\delta {\mu}_{i}={\mu}_{i}-\bar{{\mu}}_{i}^{\rm in}$, $\delta {\mu}_{b}={\mu}_{b}-\bar{{\mu}}_{b}$. The averaged quantities are $\mu_i ^{\rm in}=V_{\mathcal{S}}^{-1}\int_{V_{\mathcal{S}}} \mu_i \ dV$,  $\bar{\mu}_b = A^{-1}\oint_{\partial V_{\mathcal{S}}} \hat{\mu
}_b \ d{A}$ and $\mu
_b=\sum_{i=0}^K \mu
_i n_i/n_b$.

\section{Heat release rate per oxygen consumed for different metabolic pathways}\label{appendixd}
The macrochemical equation for respiration using glucose as a carbon source is discussed in the main text Section \ref{thorntons_rule} leading to the enthalpy change per oxidation $\Delta h_{\rm ox}^{\rm glu}=-468.6 \text{ kJ/mol}$. Similarly, we can estimate $\Delta h_{\rm ox}$ for when lipid or proteins are used as a carbon source. Let us consider Eq.~\eqref{qdotmacrochemical2} with $\partial_tT=0$, $\delta \dot{Q}=0$:
\begin{equation}
\dot{Q}=-\sum_{\beta=1}^{\tilde{R}} \mathcal{R}_{\beta}\Delta H_{\beta}  \quad. \label{qdotmacrochemical3}  
\end{equation}
Dividing $\dot{Q}$
by oxygen consumption rate $\Phi_{\text{O}_2}$ gives the heat released per oxygen consumed:
\begin{equation}
\Delta h_{\rm ox}=\frac{\dot{Q}}{\Phi_{\text{O}_2}}\quad .\label{eq:hox}  
\end{equation}

For the particle enthalpies $h_i$, we use the values of standard enthalpy of formation from \cite{domalski1972selected,heijnen1999bioenergetics}.

\textit{Lipid as a carbon source:} Lipids are typically used in the form of fatty acids in energy metabolism. For instance, for full oxidation of palmitic acid (${\rm C}_{16} {\rm H}_{32} {\rm O}_2$) producing ${\rm CO}_2$ and ${\rm H}_2{\rm O}$, we have the macrochemical reaction
\begin{equation}
 {\rm C}_{16} {\rm H}_{32} {\rm O}_2+23{\rm O}_2\rightleftharpoons 16{\rm CO}_2+ 16{\rm H}_2{\rm O}  \quad \label{eq:palmitic}
\end{equation}
where the coefficients satisfy element conservation. The enthalpy change per oxidation  $\Delta h_{\rm ox}^{\rm pal}=\frac{16}{23}\left(h_{{\rm CO}_2}+h_{{\rm H}_2{\rm O}}-\frac{h_{\rm pal}}{16} \right)$. At $T=298 ^{\circ}$K, the enthalpy of formation values are given as $h_{{\rm pal}}=-892 \text{ kJ/mol}$, $h_{{\rm CO}_2}=-393.5 \text{ kJ/mol}$, $h_{{\rm H}_2{\rm O}}=-285.8 \text{ kJ/mol}$, and  $h_{{\rm O}_2}=0$. This leads to $\Delta h_{\rm ox}^{\rm pal}=-433.8 \text{ kJ/mol}$.

\textit{Protein as a carbon source:} Proteins are polymers composed of molecular units called amino acids. While the proteins will vary in amino acid composition, we can get an estimate on enthalpy of oxidation of amino acids. Full oxidation of an amino acid typically produces ${\rm CO}_2$, ${\rm H}_2{\rm O}$, and urea (${\rm C}{\rm H}_4 {\rm N}_2 {\rm O}$). Let us take an example amino acid alanine (${\rm C}_3{\rm H}_7 {\rm N} {\rm O}_2$) leading to the macrochemical reaction
\begin{equation}
 {\rm C}_3{\rm H}_7 {\rm N} {\rm O}_2+3{\rm O}_2\rightleftharpoons \frac{5}{2}{\rm CO}_2+ \frac{5}{2}{\rm H}_2{\rm O} + \frac{1}{2}{\rm C}{\rm H}_4 {\rm N}_2 {\rm O} \quad .\label{eq:alanine}
\end{equation} 
The enthalpy change per oxidation $\Delta h_{\rm ox}^{\rm ala}=\frac{1}{3}\left(\frac{5}{2}(h_{{\rm CO}_2}+h_{{\rm H}_2{\rm O}})-h_{\rm ala}+ \frac{1}{2}h_{\rm urea}\right)$. At $T=298 ^{\circ}$K, the enthalpy of formation values are given as $h_{{\rm ala}}=-560 \text{ kJ/mol}$ and  $h_{{\rm urea}}=-339.2 \text{ kJ/mol}$. This leads to $\Delta h_{\rm ox}^{\rm ala}=-436 \text{ kJ/mol}$.

\begin{figure*}[t]
\includegraphics[width=.85\textwidth]{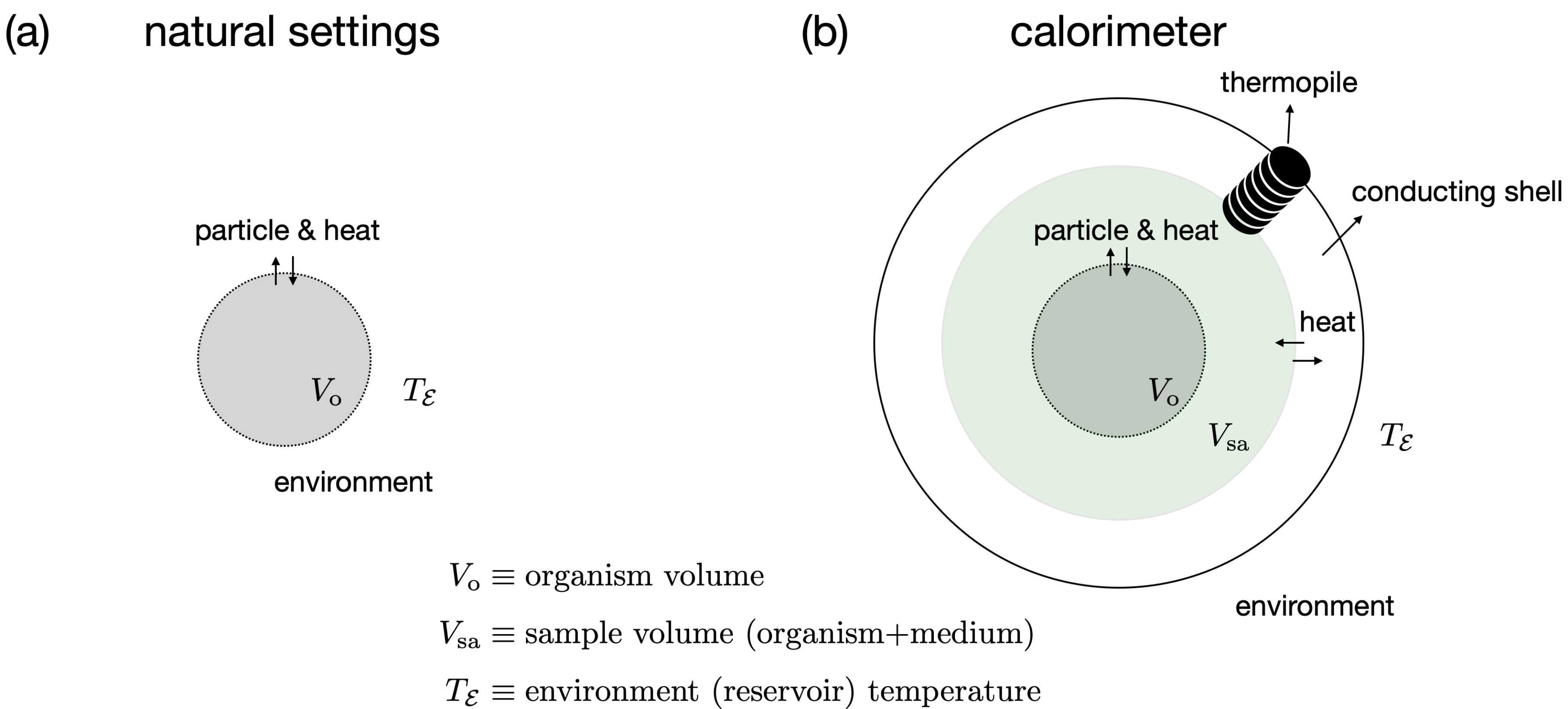}\\
\caption{Sketch of enclosed volumes and exchanges between each region.}\label{fig:11}
\end{figure*}
\textit{Deviations from the Thornton's rule in different metabolic pathways:}
Next, we illustrate deviations from Thornton factor for enthalpy change per oxygen consumed ($\sim$ -450 kJ/mol) in different metabolisms. Cells can use alternative metabolic pathways than respiration to generate ATP. One such pathway is lactate fermentation that differs from respiration after the glycolysis step, which is followed by using pyruvate to generate lactate completing the cycle. The net balanced reaction using this pathway is given in Eq.~\eqref{eq:glyc_ferm_steady} which is:
\begin{equation}
 {\rm C}_6 {\rm H}_{12} {\rm O}_6 \rightleftharpoons 2 {\rm C}_3 {\rm H}_{6}  {\rm O}_3\quad . \label{eq:fermentation}
\end{equation}
We can decompose the right hand side of Eq.~\eqref{qdotmacrochemical3}  into two net reactions with varying fluxes $\mathcal{R}_{\rm res}$ and $\mathcal{R}_{\rm lac}$ leading to:
\begin{equation}
    \dot{Q}=-\mathcal{R}_{\rm res} \Delta H_{\rm res}-\mathcal{R}_{\rm {lac}} \Delta H_{\rm lac}
\end{equation}
where we defined reaction enthalpies for the net reactions of respiration $\Delta H_{\rm res}$, and lactate fermentation $\Delta H_{\rm lac}$. Then the enthalpy change per oxygen consumed becomes:
\begin{equation}
    \frac{\dot{Q}}{\Phi_{{\rm O}_2}}=-\Delta h_{\rm ox}-\frac{\mathcal{R}_{\rm {lac}}}{6\mathcal{R}_{\rm {res}}} \Delta H_{\rm lac}
\end{equation}
where we used that oxygen consumption rate $\Phi_{{\rm O}_2}=6 \mathcal{R}_{\rm {res}}$ and $\Delta h_{\rm ox}=\Delta h_{\rm res}/6$. We can express the flux ratio $\mathcal{R}_{\rm {lac}}/\mathcal{R}_{\rm {res}}$ in terms of fraction of ATP produced by each processes. The net ATP generated per glucose differs for cellular respiration and lactate fermentation, which are 32 ATP/glucose and 2 ATP/glucose respectively. Hence, the fraction of ATP produced by lactate fermentation is $\zeta_{\rm lac}^{\rm ATP}=2 \mathcal{R}_{\rm lac}/(2 \mathcal{R}_{\rm lac}+32 \mathcal{R}_{\rm res})$. This leads to:
\begin{equation}
    \frac{\dot{Q}}{\Phi_{{\rm O}_2}}=-\Delta h_{\rm ox}-\frac{16\zeta_{\rm lac }^{\rm ATP}}{6(1-\zeta_{\rm lac }^{\rm ATP})} \Delta H_{\rm lac} \quad.
\end{equation}
We can now calculate the value of $\Delta H_{\rm lac}=2 h_{\rm lactate}-h_{\rm glu}$, using  $h_{\rm lactate}=-687 \text{ kJ/mol}$ and $h_{\rm glu}=-1264.2 \text{ kJ/mol}$ at $T=298 ^{\circ}$K, which lead to $\Delta H_{\rm lac}=-109.8 \text{ kJ/mol}$. Using the Thornton value $\Delta h_{\rm ox}=-450 \text{ kJ/mol}$, we finally have
\begin{equation}
    \frac{\dot{Q}}{\Phi_{{\rm O}_2}}=-\Delta h_{\rm ox}\left(1+0.62\frac{\zeta_{\rm lac}^{\rm ATP}}{(1-\zeta_{\rm lac}^{\rm ATP})} \right) \quad.
\end{equation}
which illustrates the deviation from the Thornton value for enthalpy change per oxygen consumed as a function of share of each lactate fermentation in energy metabolism (i.e. fraction of ATP produced $\zeta_{\rm lac}^{\rm ATP}$). This can also be expressed in terms of glucose consumption ratios. We can define the fraction of glucose consumed by lactate fermentation as $\zeta_{\rm lac}^{\rm glu}=\mathcal{R}_{\rm lac}/(\mathcal{R}_{\rm res}+\mathcal{R}_{\rm lac})$. Then, carrying out the calculations, the equivalent of the above equation becomes:
\begin{equation}
    \frac{\dot{Q}}{\Phi_{{\rm O}_2}}=-\Delta h_{\rm ox}\left(1+0.04\frac{\zeta_{\rm lac}^{\rm glu}}{(1-\zeta_{\rm lac}^{\rm glu})} \right) \quad.
\end{equation}
This means 90$\%$ use of glucose on lactate fermentation already reaches  $\dot{Q}/\Phi_{{\rm O}_2}>600 \text{ kJ/mol}$.

\section{Measuring the total heat release rate}\label{appendixc}
The organism is an open system in its natural setup as portrayed in Fig.\ref{fig:11}(a). In Fig.\ref{fig:11}(b), we sketch the calorimeter setup  with four domains: organism, medium, conducting shell, and surrounding environment that is kept at fixed temperature $T_{\mathcal{E}}$. We consider the sample as the system that contains organism and culture medium. The sample constitutes a closed system with no particle exchange which means vanishing of ${\bf j}_i^{\rm int}\cdot d{\bm A}=0$. Accordingly, Eq.~\eqref{calorimetry1main} becomes
\begin{eqnarray}
   \dot{Q}_{\rm sa}+\int_{V_{\rm sa}}c_p \frac{\partial T}{\partial t}dV=&-&\sum_{i=0}^K \int_{V_{\rm sa}}h_i\frac{\partial n_i}{\partial t}dV\nonumber\\&-&\oint_{\partial V_{\rm sa}}h^{\rm in}{\bf u}\cdot d{\bf A} \label{dotqcalorimetry}
\end{eqnarray}
where $V_{\rm sa}$ is the sample volume including organism and medium, and $\dot{Q}_{\rm sa}$ is the total rate of heat release from the sample to its surroundings. The heat flow from the sample can be obtained using Newton's law of cooling which can be derived from local heat balance for our case. We model the sample and the measurement unit---comprising the sample and surrounding conducting layer---as two concentric cylinders with radii $R_{\rm sa}$ and $R$ and height $h$. We assume that along the conducting shell there is no heat source and the temperature profile relaxes quickly to steady state such that $\partial_t T=0$. Then, the heat balance equation Eq.~\eqref{eqheatbalance} becomes in the conducting shell region:
\begin{equation}
    \nabla\cdot {\bf j}_{q}=0 \quad \label{eqheatss}.
\end{equation}
Without chemical reactions and molecular transport, the heat flux vector should follow Fourier's law ${\bf j}_{q}=- k\nabla T $ to ensure non-negativity of entropy production rate Eq.~\eqref{epr}, where $k>0$ is the thermal conductivity of the medium. Replacing heat flux vector with this expression, and solving along the conducting shell in Fig.\ref{fig:11}(b) and setting temperatures $T_{\rm sa}$ and $T_{\mathcal{E}}$ respectively at the sample-shell and the shell-environment interfaces, we get $T(r)=T_{\rm sa}+(T_{\mathcal{E}}-T_{\rm sa})\ln(r/R_{\rm sa})/\ln(R/R_{\rm sa})$ for $R\geq r\geq R_{\rm sa}$ where $r$ is the displacement from the center of the sample. Then, we can calculate the integral resulting:
\begin{equation}
    \dot{Q}_{\rm sa}=\oint_{\partial V_{\rm sa}}{\bf j}_q^{\rm out}\cdot d{\bf A}=\kappa (T_{\rm sa}-T_{\mathcal{E}})\label{qdotappout}
\end{equation}
where we absorbed the constant terms in $\kappa=2\pi kh/\ln(R/R_{\rm sa})$ which is the heat transfer coefficient across the conducting shell. This equation corresponds to Newton's law of cooling, which is more general than the specific geometry considered here, with the form of 
$\kappa$ depending on the geometry. The calorimeter measures the difference in temperature $T_{\rm sa}-T_{\mathcal{E}}$ through voltage difference in thermopile. Since $T_{\mathcal{E}}$ is set by the calorimeter, $T_{\rm sa}$ can also be inferred from this voltage measurement.


We can assume that the only net reaction fluxes in the sample is due to metabolic reactions of the organism and $\partial_t T \approx 0$ in the medium leading to:
\begin{equation}
   \dot{Q}_{\rm sa}\approx \dot{Q} \quad.
   \label{dotqequivalence}
\end{equation}

Thus, Eq.~\eqref{dotqequivalence} together with Eq.~\eqref{qdotappout} leads to:
\begin{equation}
    \dot{Q} = \kappa (T_{\rm sa}-T_{\mathcal{E}}) \quad. \label{eq:qnewton}
\end{equation}

We can now explicitly write a relation between measurements and chemical processes, by using the enthalpy change due to macrochemical equation given in Eq.~\eqref{qdotmacrochemical2} when $\delta \dot{Q}=0$:
\begin{equation}
    -\sum_{\beta=1}^{\tilde{R}} \mathcal{R}_{\beta}\Delta H_{\beta} =\kappa (T_{\rm sa}-T_{\mathcal{E}})
\end{equation}
where the sum is over the net reactions.

It is finally possible to correct the heat flow rate measurements by assuming a space-independent quasi-static temperature change inside the sample such that $\partial_t T\approx \partial_t T_{\rm sa}$, and obtain the calorimetry equation using Eq.~\eqref{qdotmacrochemical2} in Eq.~\eqref{eq:qnewton} when $\delta \dot{Q}=0$:
\begin{eqnarray}
       -\sum_{\beta=1}^{\tilde{R}} \mathcal{R}_{\beta}\Delta H_{\beta}=\kappa (T_{\rm sa}-T_{\mathcal{E}})+\bar{C}_p\frac{\partial {T}_{\rm sa}}{\partial t}\nonumber\\
   \label{dotqcalorimetry2}
\end{eqnarray}
where $\bar{C}_p=\int_{V_{\rm sa}} c_p dV$.
\\

 \vspace{1em}
\bibliography{main}
\end{document}